\documentclass[11pt,english]{article}
\usepackage[latin9]{inputenc}
\usepackage{color}
\usepackage{amsmath}
\usepackage{graphicx}
\usepackage{setspace}
\usepackage{subscript}
\makeatletter

\providecommand{\tabularnewline}{\\}

\@ifundefined{date}{}{\date{}}
\usepackage{esint}
\usepackage{hyperref}
\usepackage{latexsym}
\usepackage{graphicx}\usepackage{bm}\usepackage{longtable}

\usepackage{xcolor}

\@addtoreset{equation}{section}

\makeatother

\usepackage{babel}
\begin{document}
\title{Correlation function of flavored fermion in holographic QCD}
\maketitle
\begin{center}
Si-wen Li\footnote{Email: siwenli@dlmu.edu.cn}, Yi-peng Zhang\footnote{Email: ypmahler111@dlmu.edu.cn},
Hao-qian Li\footnote{Email: lihaoqian@dlmu.edu.cn},
\par\end{center}

\begin{center}
\emph{Department of Physics, School of Science,}\\
\emph{Dalian Maritime University, }\\
\emph{Dalian 116026, China}\\
\par\end{center}

\vspace{12mm}

\begin{abstract}
By using the gauge-gravity duality, we investigate the correlation
function of flavored fermion in the $\mathrm{D}_{p}/\mathrm{D}_{p+4}$
model as top-down approaches of holographic QCD for $p=4,3$. The
bulk spinor, as the source of the flavored fermion in QCD, is identified
to the worldvolume fermion on the flavor $\mathrm{D}_{p+4}$-branes
and the standard form of its action can be therefore obtained by the
T-duality rules in string theory. Keeping this in hand, we afterwards
generalize the prescription for two-point correlation function in
AdS/CFT dictionary into general D-brane backgrounds and apply it to
the case of $p=4,3$, i.e. the D4/D8 and D3/D7 approach respectively.
Resultantly, our numerical calculation with the bubble background
always displays discrete peaks in the correlation functions which
imply the bound states created by the flavored fermions as the confinement
in QCD. With the black brane background, the onshell condition illustrated
by the correlation function covers basically the dispersion curves
of fermion obtained by the hard thermal loop approximation in the
hot medium. Finally, we interpret the flavored fermions in the bubble
background as baryons by taking into account a baryon vertex, then
find the two-point correlation function is able to fit the lowest
baryon spectrum.\textcolor{blue}{{} }In this sense, we conclude remarkably
that our top-down approach in this work could reveal the fundamental
properties of QCD both in the confined and deconfined phase.
\end{abstract}
\newpage{}

\tableofcontents{}

\section{Introduction}

The gauge-gravity duality and AdS/CFT correspondence have become a
very powerful tool to investigate strongly coupled quantum field theory
(QFT) by analyzing the corresponding classical gravity system \cite{key-1,key-2,key-3}
in holography. Particularly, the real-time two-point correlation functions
of (composite) operators are significant to understand the QFT at
finite temperature or with dense matter since the information about
the collective behavior e.g. the properties of transport, existence
of (quasi-) particles, are encoded in them. The prescription to compute
the two-point retarded Green functions in AdS/CFT has been justified
in many different ways e.g. in literature \cite{key-4,key-5,key-6,key-7,key-8,key-9,key-10,key-11,key-12,key-13,key-14,key-15}
which have instrumentally extracted many important insights about
strongly coupled QFT systems from AdS/CFT. On the other hand, quantum
chromodynamics (QCD) is the underlying theory to describe the property
of strong interaction, however it is extremely complex to solve in
the low-energy (strongly coupled) region due to its asymptotic freedom,
especially at finite temperature with dense matter \cite{key-15,key-16}.
Therefore investigating QCD through gauge-gravity duality naturally
becomes an interesting topic. Although there are several bottom-up
models e.g. \cite{key-17,key-18,key-19} attempting to give a holographic
version of QCD, the top-down approaches based on D4/D8 (the Witten-Sakai-Sugimoto
model \cite{key-20,key-21,key-22}) and D3/D7 \cite{key-14} models
are the most famous and successful achievements in holography since
almost all the elementary features of QCD are included in these models
in a very simple way \cite{key-23,key-24,key-25}. 

Moreover, it is known that the self-energy of fermion has been studied
with a very long history and attracted many interests since fermion
is one of the fundamental elements in QFT as well as in nuclear physics.
For example, in weakly coupled system, the self-energy of fermion
can be analyzed by the perturbation method of QFT or be characterized
by the thermal mass in the hot medium \cite{key-26,key-27} while
the results may not hold in the strong coupling regime \cite{key-28,key-29}.
Hence there are many researches about fermionic correlations in order
to investigate the properties of the strongly coupled fermionic system
\cite{key-11,key-12,key-30,key-31,key-32,key-33} in holography. As
we can see, the presented works about holographic fermion are based
on bottom-up approaches or with minimal coupled fermions, and the
bulk fermionic field is identified to the fundamental representation
of a $U\left(N\right)$ group when the chemical potential is taken
into account especially. However, such a fermionic field in bulk is
less clear in the top-down models e.g. D4/D8 and D3/D7 approaches.

Motivated by this issue, in this work we attempt to find a holographic
correspondence to evaluate the two-point correlation function of the
flavored fermion in $\mathrm{D}_{p}/\mathrm{D}_{p+4}$ model (for
$p=4,3$, i.e. the D4/D8 and D3/D7 approach) as two top-down approaches
to holographic QCD. Specifically, the $\mathrm{D}_{p}/\mathrm{D}_{p+4}$
model holographically dual to QCD consists of $N_{c}$ $\mathrm{D}_{p}$-branes
as colors and $N_{f}$ probe $\mathrm{D}_{p+4}$-branes as flavors
which are intersected to each other. In the large $N_{c}$ limit,
the dynamics of the bulk geometry is described approximately by the
type II supergravity. Due to the supersymmetry on the $N_{f}$ probe
$\mathrm{D}_{p+4}$-branes, we identify the bulk spinor (as the source
of the flavored fermion) to the worldvolume fermionic field on the
probe $\mathrm{D}_{p+4}$-branes and this may be the only consistent
correspondence by analyzing the bulk spectrum in this holographic
system. In this sense, the action of the bulk fermion can be obtained
by the T-duality rules in string theory \cite{key-34,key-35} and
it reveals the bulk fermion is not minimal coupled. Keeping this in
hand, we further generalize the prescription for two-point correlation
function in AdS/CFT dictionary into general D-brane backgrounds with
respect to the flavored fermion in the dual theory. Afterwards, we
apply our method to the D4/D8 and D3/D7 models as tests in order to
evaluate the two-point retarded Green functions variously. 

Our numerical calculation displays that the retarded Green functions
behave similarly in the D4/D8 and D3/D7 approaches. That is, with
bubble background, the Green functions always contain many discrete
peaks representing various bound states in QCD and it is consistent
with the QCD confinement since the bubble background corresponds to
confined phase of QCD in holography \cite{key-14,key-20,key-21,key-23,key-25,key-36,key-38}.
Remarkably, the bound energies obtained in the Green function also
agree with the numerical evaluation in \cite{key-36} for $p=4$ quantitatively.
With the black brane background, the particle onshell condition obtained
from the Green functions covers the dispersion curves obtained from
the hard thermal loop approximation and the effective thermal mass
of fermion can be also determined numerically while the chemical potential
is not turned on in our current work. We notice the behaviors of confined
Green functions in our top-down approaches are very different from
them in the bottom-up approaches or it with minimal coupled fermion
\cite{key-30,key-31,key-32,key-33}, while the deconfined Green functions
(obtained from the black brane background) behave similarly. Thus
it illustrates seemingly that the property of QCD confinement is less
clear in the previous works with bottom-up models or with minimal
coupled fermion e.g. \cite{key-30,key-31,key-32,key-33}. Finally,
we discuss how to interpret the fermionic bound states in confined
phase as baryon states by including a baryonic brane as baryon vertex
\cite{key-45} with the framework of $\mathrm{D}_{p}/\mathrm{D}_{p+4}$
model which may support that the open strings on the D-brane behaves
somehow as baryons \cite{key-37,key-37+1}.

The outline of this manuscript is as follows. In Section 2, we review
the $\mathrm{D}_{p}/\mathrm{D}_{p+4}$ model as holographic QCD briefly
for the case of $p=4,3$. In Section 3, we discuss how to identify
the bulk spinor, its associated action and the holographic prescription
to compute the two-point correlation function in D-brane background.
In Section 4, 5, we apply our method to D4/D8 and D3/D7 model respectively
in order to evaluate the retarded Green functions for flavored fermions
then analyze the numerical results. Summary and discussion are given
in Section 6.

\section{The $\mathrm{D}_{p}/\mathrm{D}_{p+4}$ model as holographic QCD}

In this section, let us review the holographic QCD through gauge-gravity
duality based on the type II string theories. In order to obtain a
dual field theory close to QCD, we start with the D-brane system consisted
of $N_{c}$ coincident $\mathrm{D}_{p}$-branes and $N_{f}$ coincident
$\mathrm{D}_{p+4}$-branes. 

Taking the near-horizon limit $\alpha^{\prime}\rightarrow0$ and the
large $N_{c}$ limit by keeping $N_{f}$ finite, the bulk dynamic
is described approximately by the type II supergravity (SUGRA) for
$N_{c}$ coincident $\mathrm{D}_{p}$-branes whose action takes the
following form as \cite{key-38,key-39},

\begin{equation}
S_{\mathrm{II}}=\frac{1}{2\kappa_{10}^{2}}\int d^{10}x\sqrt{-g}\left[e^{-2\phi}\left(\mathcal{R}+4\partial_{M}\phi\partial^{M}\phi\right)-\frac{g_{s}^{2}}{2}\left|F_{p+2}\right|^{2}\right],\label{eq:1}
\end{equation}
where $\mathcal{R},\phi,C_{p+1}$ is respectively the 10-dimensional
(10d) scalar curvature, the dilaton and the Ramond-Ramond (R-R) $p+1$-form
field with its strength $F_{p+2}=dC_{p+1}$. The constant $\kappa_{10}$
is the gravity coupling constant related to the 10d Newtonian constant
$G_{10}$ as $2\kappa_{10}^{2}=16\pi G_{10}=\left(2\pi\right)^{7}l_{s}^{8}g_{s}^{2}$
where the string coupling constant and string length is denoted by
$g_{s},l_{s}$. Since we will discuss the D3/D7 and D4/D8 approach
based on IIB and IIA string theory respectively, the cases for $p=3,4$
are only considered here. Vary the action (\ref{eq:1}) with respect
to the metric, the dilaton and the R-R field, then the associated
equations of motion can be obtained accordingly and they can be solved
in general as the black brane solution \cite{key-39},

\begin{align}
ds^{2} & =H_{p}^{-\frac{1}{2}}\left[f\left(r\right)dt^{2}+d\vec{x}\cdot d\vec{x}\right]+H_{p}^{\frac{1}{2}}\left[\frac{dr^{2}}{f\left(r\right)}+r^{2}d\Omega_{8-p}^{2}\right],\nonumber \\
e^{\phi} & =H_{p}^{-\frac{p-3}{4}},C_{01...p}=g_{s}^{-1}H_{p}^{-1},F_{r01...p}=\frac{\left(7-p\right)g_{s}^{-1}h_{p}^{7-p}}{r^{8-p}H_{p}^{2}},\label{eq:2}
\end{align}
where the functions $f\left(r\right),H_{p}\left(r\right)$ are given
respectively as,

\begin{equation}
f\left(r\right)=1-\frac{r_{H}^{7-p}}{r^{7-p}},H_{p}\left(r\right)=1+\frac{r_{p}^{7-p}}{r^{7-p}}.
\end{equation}
Here $r$ is the radial coordinate of the bulk which corresponds usually
to the holographic direction. $r_{H},r_{p}$ are two constants related
respectively to the position of the horizon and the R-R charge carried
by the $\mathrm{D}_{p}$-branes\footnote{In our solution, the constant $h_{p}$ is defined as $\left(h_{p}^{7-p}\right)^{2}=\left(r_{p}^{7-p}\right)^{2}+r_{p}^{7-p}r_{H}^{7-p}$.}.
We note that in the near-horizon limit, the harmonic function $H_{p}\left(r\right)$
becomes $H_{p}\left(r\right)\rightarrow r_{p}^{7-p}/r^{7-p}$ .

The dual field theory associated to the bulk geometry (\ref{eq:2})
in the large $N_{c}$ limit can be examined by taking into account
a probe $\mathrm{D}_{p}$-brane located at the holographic boundary
$r\rightarrow\infty$. And as it is known, the resultantly dual field
theory is respectively the $\mathcal{N}=4$ super Yang-Mills (YM)
theory \cite{key-14} for $p=3$ and the type $\mathcal{N}=\left(2,0\right)$
super conformal field theory (SCFT) reduced on a circle \cite{key-20,key-38}
for $p=4$, however both of them are less close to QCD due to the
presence of supersymmetry. Thus Witten proposed a scheme \cite{key-20}
to construct the bulk geometry presented in (\ref{eq:2}) in order
to obtain a dual field theory close to QCD with confinement. Specifically,
the first step is to perform the double Wick rotation $\left\{ x^{0}\rightarrow-ix^{p},x^{p}\rightarrow-ix^{0}\right\} $
on the $\mathrm{D}_{p}$-brane where $x^{0}=t,x^{p}$ refer to the
time and the $p$-th spacial direction of the $\mathrm{D}_{p}$-brane.
So the geometry (\ref{eq:2}) becomes a bubble configuration as,

\begin{align}
ds^{2} & =H_{p}^{-\frac{1}{2}}\left[\eta_{\mu\nu}dx^{\mu}dx^{\nu}+f\left(r\right)\left(dx^{p}\right)^{2}\right]+H_{p}^{\frac{1}{2}}\left[\frac{dr^{2}}{f\left(r\right)}+r^{2}d\Omega_{8-p}^{2}\right],\nonumber \\
f\left(r\right) & =1-\frac{r_{KK}^{7-p}}{r^{7-p}},H_{p}\left(r\right)=1+\frac{r_{p}^{7-p}}{r^{7-p}},\ \ \mu,\nu=0,1...p-1.\label{eq:4}
\end{align}
which is defined only for $r\in\left[r_{KK},\infty\right)$. We have
replace $r_{H}$ by $r_{KK}$ in (\ref{eq:4}) because there is not
a horizon in the bubble configuration. Notice that in the black brane
solution (\ref{eq:2}), $x^{0}=t$ is compactified on a circle which
implies in the bubble case $x^{p}$ becomes now periodic as,

\begin{equation}
x^{p}\sim x^{p}+\delta x^{p},\delta x^{p}=\frac{2\pi}{M_{KK}},\label{eq:5}
\end{equation}
where $M_{KK}$ is the Klein-Kaluza (KK) energy scale. So the second
step to found a QCD-like theory is to get rid of all massless field
other than the gauge fields by imposing the periodic and anti-periodic
boundary condition to bosons and supersymmetric fermions respectively
along the periodic direction $x^{p}$. Therefore the supersymmetric
fermion acquires mass of order $M_{KK}$ which would be decoupled
to the low-energy theory, while the gauge boson remains to be massless.
Accordingly, the dual field theory on the $\mathrm{D}_{p}$-brane
is $p$-dimensional pure $U\left(N_{c}\right)$ Yang-Mills theory
in the large $N_{c}$ limit below the energy scale $M_{KK}$. We note
that the wrap factor $H_{p}\left(r\right)$ in (\ref{eq:4}) never
goes to zero, hence the behavior of the Wilson loop in this geometry
obeys the area law which implies the dual field theory will exhibit
confinement. Altogether, these $\mathrm{D}_{p}$-branes compactified
on a circle correspond holographically to the color sector of QCD
and its number $N_{c}$ is identified to the color number. The $U\left(N_{c}\right)$
gauge field on the $\mathrm{D}_{p}$-branes is therefore identified
as the gluon field and in the large $N_{c}$ limit, the bulk geometry
is described by (\ref{eq:4}) with confinement. To further obtain
a deconfined phase of QCD at finite temperature\footnote{Identifying the black brane solution (\ref{eq:2}) to the deconfined
phase of QCD may contain some issues \cite{key-40,key-41}. Nonetheless
we will continue our discussion by this identification as most works
with the $\mathrm{D}_{p}/\mathrm{D}_{p+4}$ system. The reason is
that the black brane solution corresponds to the dual theory at finite
temperature which may include some properties of hot QCD, close to
the deconfinement in QCD.}, we can campactify the $p$-th direction in the black brane solution
(\ref{eq:2}) as (\ref{eq:5}) then follow the same discussion as
it is in the confined geometry to construct the dual theory \cite{key-14,key-20,key-38}.
Afterwards the dual theory could be identified as deconfined QCD at
finite temperature since there is a horizon in (\ref{eq:2}) introducing
the Hawking temperature in the dual theory.

As QCD also includes fermions as flavors, the existing $N_{f}$ coincident
$\mathrm{D}_{p+4}$-branes in this holographic model can just play
the role of flavor. The D-brane configuration of the $\mathrm{D}_{p}/\mathrm{D}_{p+4}$
system is given in Table \ref{tab:1} for the D3/D7 and D4/D8 approach
in this manuscript. 
\begin{table}
\begin{centering}
\begin{tabular}{|c|c|c|c|c|c|}
\hline 
 & $x^{\mu}$ & $x^{p}$ & $x^{p+1}\left(r\right)$ & $x^{p+2},...x^{p+5}$ & $x^{p+6},...x^{9}$\tabularnewline
\hline 
\hline 
$\mathrm{D}_{p}$ & - & - &  &  & \tabularnewline
\hline 
$\mathrm{D}_{p+4}$ & - &  & - & - & \tabularnewline
\hline 
\end{tabular}
\par\end{centering}
\caption{\label{tab:1} The D-brane configuration of the $\mathrm{D}_{p}/\mathrm{D}_{p+4}$
system for $p=3,4$. ``-'' represents that the D-branes extend along
this direction and $x^{p}$ is the periodic direction. }

\end{table}
 Since $N_{f}$ is set to be finite, the $\mathrm{D}_{p+4}$-branes
become the probes embedded into (\ref{eq:2}) (\ref{eq:4}) and $N_{c}\rightarrow\infty$
implies $\mathrm{D}_{p}$-brane dominates the bulk geometry. By analyzing
the spectrum of the open string connecting the $N_{c}$ $\mathrm{D}_{p}$-branes
and $N_{f}$ $\mathrm{D}_{p+4}$-branes in R-sector, the low-energy
theory includes fermions in the fundamental representation of $U\left(N_{c}\right)$
and $U\left(N_{f}\right)$ which can be identified nicely as the fundamental
quarks in this model \cite{key-14,key-21,key-22}. In this sense,
the dual theory contains both colors and flavors, thus it is expected
to be the holographic version of QCD.

\section{The holographic correspondence with flavored fermion}

As our goal is to work out the two-point correlation function of the
flavored fermions through gauge-gravity duality, in this section,
let us attempt to found the associated holographic correspondence
in the top-down approach first, then discuss how to generalize the
prescription in AdS/CFT dictionary into a non-conformal background
to compute the Green function.

\subsection{Identification of the bulk spinor}

To begin with, let us recall the principle of the AdS/CFT which means
in general the partition function of the dual field theory $Z_{QCD}$
is equal to its gravitational partition function $Z_{gravity}$ in
the bulk. Since the dual theory living in the boundary $\partial\mathcal{M}$
of (\ref{eq:2}) (\ref{eq:4}) would be expected to be QCD as it is
reviewed in Section 1, we use $Z_{QCD}$ here to refer to its partition
function. For a spinor $\psi$ in the bulk $\mathcal{M}$ whose boundary
value is $\psi_{0}=\lim_{r\rightarrow\infty}\psi$, we can write down,

\begin{equation}
Z_{QCD}\left[\bar{\psi}_{0},\psi_{0}\right]=Z_{gravity}\left[\bar{\psi},\psi\right],\label{eq:6}
\end{equation}
with

\begin{align}
Z_{QCD}\left[\bar{\psi_{0}},\psi_{0}\right] & =\left\langle \exp\left\{ \int_{\partial\mathcal{M}}\left(\bar{\chi}\psi_{0}+\bar{\psi}_{0}\chi\right)d^{D}x\right\} \right\rangle ,\nonumber \\
Z_{gravity}\left[\bar{\psi},\psi\right] & =\exp\left\{ \int_{\mathcal{M}}\mathcal{L}_{gravity}^{ren}\left[\bar{\psi},\psi\right]d^{D+1}x\right\} .\label{eq:7}
\end{align}
And $\mathcal{L}_{gravity}^{ren}$ refers to the classical renormalized
Lagrangian of the bulk field $\psi$. In our concern, $\chi$ refers
to the flavored fermion in the dual field theory (QCD). Then the two-point
retarded Green function $G_{R}$ for the spinor $\chi$ can be evaluated
by following \cite{key-10,key-13,key-14} as,

\begin{equation}
\left\langle \chi\left(\omega,\vec{k}\right)\right\rangle =G_{R}\left(\omega,\vec{k}\right)\psi_{0}\left(\omega,\vec{k}\right),\label{eq:8}
\end{equation}
and,

\begin{align}
\left\langle \bar{\chi}\left(\omega,\vec{k}\right)\right\rangle  & =-\frac{\delta S_{gravity}^{ren}}{\delta\psi_{0}}=\Pi_{0}\left(\omega,\vec{k}\right),\nonumber \\
S_{gravity}^{ren} & =\int_{\mathcal{M}}\mathcal{L}_{gravity}^{ren}\left[\bar{\psi}_{0},\psi_{0}\right]d^{D+1}x.\label{eq:9}
\end{align}
Here $\omega,\vec{k}$ refers respectively to the frequency and 3-momentum
of the associated Fourier modes. We note that the relation of the
Euclidean Green function $G_{E}$ and the real-time Green function
$G_{R}$ is given by $G_{E}\left(\omega_{E},\vec{k}\right)=G_{R}\left(\omega,\vec{k}\right)$
with $\omega_{E}=-i\omega$.

The next step is to identify the bulk spinor field $\psi$ in order
to investigate its action $S_{gravity}^{ren}\left[\psi\right]$ in
holography. We notice that, in several bottom-up or phenomenological
holographic approaches e.g. \cite{key-12,key-30,key-31,key-32,key-33},
the bulk field $\psi$ is usually identified to a fundamental fermion
of the $U\left(N\right)$ group whose action is the minimally coupled
Dirac action. However such a bulk field $\psi$ is less clear in the
$\mathrm{D}_{p}/\mathrm{D}_{p+4}$-brane system with a bosonic background
(\ref{eq:2}) (\ref{eq:4}) since in general, the bulk modes created
by the strings in the type II string theory do not include a $U\left(N\right)$
fundamental fermion. To figure out this problem, it is worth retaking
into account the interaction of the open strings in the $\mathrm{D}_{p}/\mathrm{D}_{p+4}$-brane
system according to the string theory. For example, it is known the
AdS/CFT dictionary for the approach of vector field 

\begin{equation}
\left\langle \exp\left\{ \int_{\partial\mathcal{M}}J^{\mu}A_{\mu}^{\left(0\right)}d^{D}x\right\} \right\rangle =\exp\left\{ -S_{bulk}\left[A_{\mu}\right]\right\} ,\ \ A_{\mu}^{\left(0\right)}=\lim_{r\rightarrow\infty}A_{\mu}.\label{eq:10}
\end{equation}
illustrates that the correlator of the flavored current operator $J^{\mu}$
living on a probe $\mathrm{D}_{p}$-brane at the boundary could be
evaluated by varying the action of the worldvolume vector $A_{\mu}$
on the $\mathrm{D}_{p+4}$-branes as the bulk action $S_{bulk}\left[A_{\mu}\right]$
\cite{key-5,key-14,key-10}, since the $\mathrm{D}_{p+4}$-branes
as flavors extend along the holographically radial direction of the
bulk according to Table \ref{tab:1}. In this sense, the source term
$J^{\mu}A_{\mu}^{\left(0\right)}$ presented in (\ref{eq:10}) implies
the interaction of the $\left(p,p\right)$ and $\left(p+4,p+4\right)$
strings\footnote{Here we use $\left(p,q\right)$ to refer to an open string connecting
$\mathrm{D}_{p}$- and $\mathrm{D}_{q}$-brane.}. 

\begin{figure}
\begin{centering}
\includegraphics[scale=0.3]{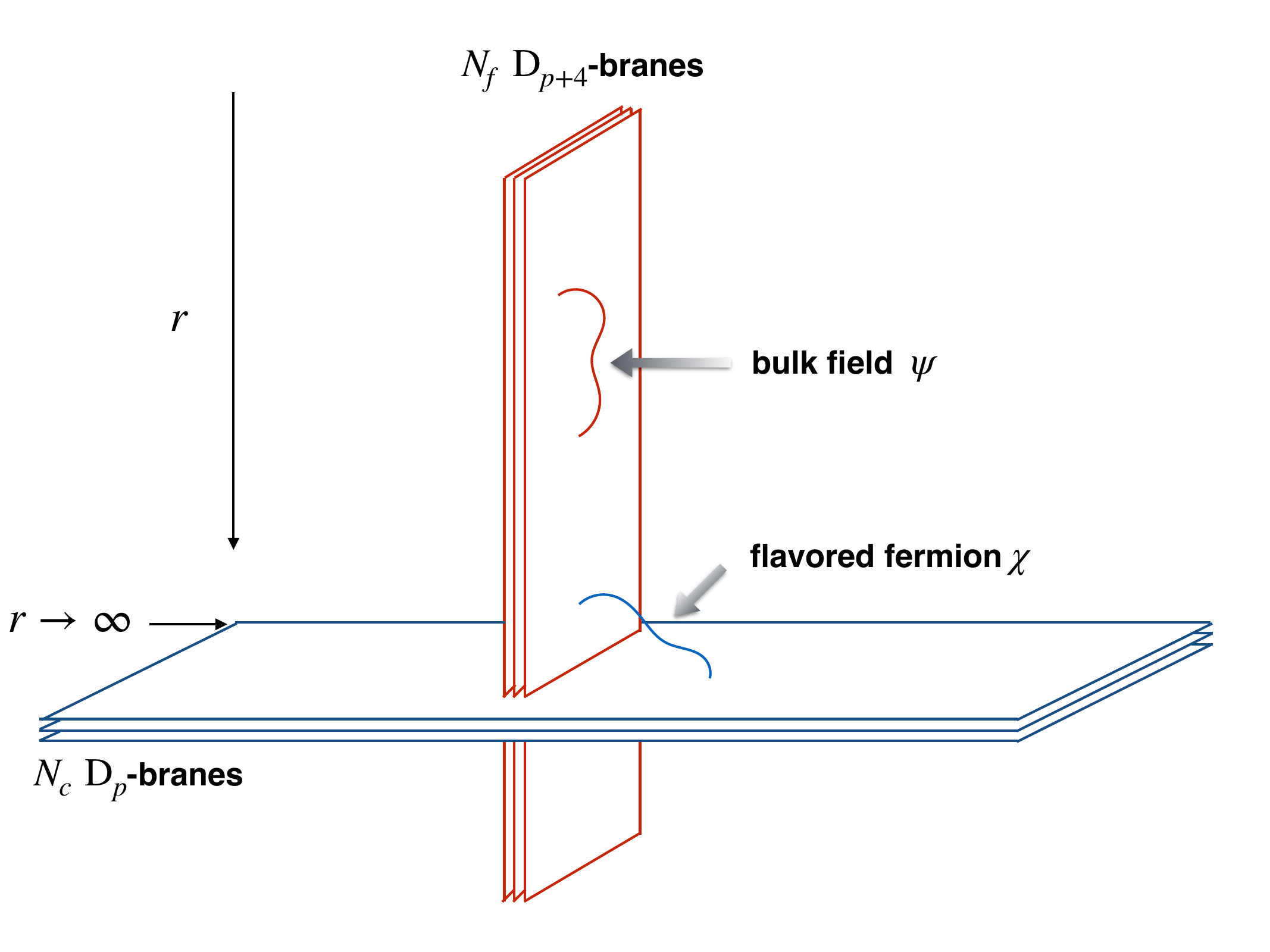}
\par\end{centering}
\caption{\label{fig:1} The correspondence of the bulk field $\psi$ and boundary
operator $\chi$ in the $\mathrm{D}_{p}/\mathrm{D}_{p+4}$ system.
Here $r$ refers to the radial direction as the holographic direction.
$\chi$ is a hadronic field as a gauge-invariant operator produced
by multiple fundamental quark fields, or equivalently by multiple
$\left(p,p+4\right)$ strings.}
\end{figure}

Keeping the above in mind and turning to our case, once the spinor
$\chi$ in (\ref{eq:6}) (\ref{eq:7}) is identified to the flavored
fermion living on the $\mathrm{D}_{p}$-brane at boundary, we must
find its fermionic dual field $\psi$ by analyzing the low-energy
spectrum of type II string theories. Among the fields in the massless
spectrum of the $\mathrm{D}_{p}/\mathrm{D}_{p+4}$ model, the only
possible choice for $\psi$ should be the fermionic field on the $\mathrm{D}_{p+4}$-branes\footnote{Since we are working in the bosonic background given in (\ref{eq:2})
(\ref{eq:4}) , the fermions in the sector of close string can be
neglected.}. In this sense the source term $\bar{\chi}\psi_{0}+\bar{\psi}_{0}\chi$
presented in (\ref{eq:7}) corresponds to the interaction of $\left(p,p+4\right)$
and $\left(p+4,p+4\right)$ strings at boundary, which reveals the
interaction to boundary theory involving the fermionic part to $A_{\mu}$
presented in (\ref{eq:10}), as it is displayed in Figure \ref{fig:1}.
 We note that, in the D-brane setup given in Table \ref{tab:1} on
the bubble background (\ref{eq:4}), the $\left(p+4,p+4\right)$ string
remains to be supersymmetric \cite{key-37+1,key-38,key-42} since
there is not any mechanism to break down the supersymmetry on the
$\mathrm{D}_{p+4}$-branes in principle while the $\mathrm{D}_{p}$-brane
is not supersymmetric due to the boundary condition on its compactified
direction. Therefore the fermionic field $\psi$ in the worldvolume
of $\mathrm{D}_{p+4}$-branes must exist in the low-energy theory,
and in our $\mathrm{D}_{p}/\mathrm{D}_{p+4}$ model, it is seemingly
to be the suitable candidate uniquely as the source of the flavored
fermion $\chi$ living in the boundary which may also be the only
consistent interpretation in this system. However in the black brane
solution (\ref{eq:2}), the worldvolume fermion on the $\mathrm{D}_{p+4}$-branes
may acquire thermal mass due to the contribution from loop diagrams.
This means the supersymmetry may break down so that the supersymmetric
fermion could decouple to the low-energy theory. Nonetheless, it would
be interesting to investigate the fermionic correlation in the black
brane background (as the deconfined phase of QCD) as a comparison
to several bottom-up approaches with deconfined fermions \cite{key-30,key-31,key-32,key-33}.
Therefore we will identify the bulk field $\psi$ to the supersymmetric
fermion on worldvolume of the $\mathrm{D}_{p+4}$-brane in order to
continue our discussion. 

In addition, we also need an interpretation of the dual operator $\chi$
in terms of hadron physics because it is presented in the dual theory
which is very close to QCD in holography. As the bulk field $\psi$
and its dual operator $\chi$ are usually gauge-invariant operators
and share same quantum numbers in gauge-gravity duality or AdS/CFT
\cite{key-14,key-38,key-39}, it implies $\chi$ must be a color singlet
in the language of QCD. On the other hand, the bulk field $\psi$
is in the adjoint representation of the flavor group (just as its
bosonic superpartner), it means dual operator $\chi$ is also the
adjoint representation of the flavor group thus it must be a fermionic
hadronic field. Mathematically, $\chi$ is not the fundamental representation
of the color group, instead it must be the singlet irreducible representation
obtained by decomposing the tensor product of the fundamental representation
of the color group. So $\chi$ could be a mesino (the supersymmetric
fermionic meson) \cite{key-36,key-37,key-37+1} or baryon produced
by multiple fundamental quark fields, or equivalently by multiple
$\left(p,p+4\right)$ strings. This would be clear by recalling the
construction of the hadronic state in QCD with $SU\left(3\right)$
color group. For example, the two-quarks color singlet can be obtained
by decomposing the tensor product of irreducible representation of
$SU\left(3\right)$ in QCD as $\mathbf{3}\otimes\mathbf{3}=\mathbf{8}\oplus\mathbf{1}$,
where $\mathbf{1}$ refers to the two-quarks color singlet i.e. meson.
Similarly, for the three-quarks color singlet, we have $\mathbf{3}\otimes\mathbf{3}\otimes\mathbf{3}=\boldsymbol{10}\oplus\mathbf{8}\oplus\mathbf{8}^{*}\oplus\mathbf{1}$
where $\mathbf{1}$ refers to the three-quarks color singlet i.e.
baryon. Therefore $\chi$ refers to such a color singlet as a hadronic
operator in our holographic approach with the $\mathrm{D}_{p}/\mathrm{D}_{p+4}$
model. However, interpreting $\chi$ as mesino is less close to the
realistic hadron physics since mesino is always absent in QCD, thus
we attempt to interpret $\chi$ as baryon by taking into account a
baryon vertex in this work and we will discuss it in details in the
end of this paper. 

\subsection{Action for the bulk spinor}

In the last section, since the bulk field dual to the flavored fermion
is identified to the worldvolume fermion on the $N_{f}$ coincident
$\mathrm{D}_{p+4}$-branes, let us take a look at the action for such
a fermionic field on the worldvolume of the D-brane. 

As it is known that the action for the worldvolume fields on a D-brane
is in principle obtained under the rule of T-duality \cite{key-38,key-39}
in string theory which includes a bosonic part consisted of Dirac-Born-Infeld
(DBI) term plus Wess-Zumino (WZ) term, and a fermionic part. Since
our concern is the worldvolume fermion on the D-brane, let us focus
the fermionic part of the $\mathrm{D}_{p}$-brane action which on
the bosonic background is given by \cite{key-34,key-35},

\begin{equation}
S_{f}^{\mathrm{D}_{p}}=\frac{iT_{p}}{2}\int d^{p+1}xe^{-\phi}\sqrt{-\left(g+\mathcal{F}\right)}\bar{\Psi}\left(1-\Gamma_{\mathrm{D}_{p}}\right)\left(\Gamma^{\alpha}\hat{D}_{\alpha}-\Delta+\mathrm{L}_{\mathrm{D}_{p}}\right)\Psi,\label{eq:11}
\end{equation}
where, for the type IIA string theory ($p$ is even number), 

\begin{align}
\hat{D}_{\alpha} & =\nabla_{\alpha}+\frac{1}{4\cdot2!}H_{\alpha NK}\Gamma^{NK}\bar{\gamma}+\frac{1}{8}e^{\phi}\left(\frac{1}{2!}F_{NK}\Gamma^{NK}\Gamma_{\alpha}\bar{\gamma}+\frac{1}{4!}F_{KLNP}\Gamma^{KLNP}\Gamma_{\alpha}\right),\nonumber \\
\Delta & =\frac{1}{2}\left(\Gamma^{M}\partial_{M}\phi+\frac{1}{2\cdot3!}H_{MNK}\Gamma^{MNK}\bar{\gamma}\right)+\frac{1}{8}e^{\phi}\left(\frac{3}{2!}F_{MN}\Gamma^{MN}\bar{\gamma}+\frac{1}{4!}F_{KLNP}\Gamma^{KLNP}\right),\nonumber \\
\Gamma_{\mathrm{D}_{p}} & =\frac{1}{\sqrt{-\left(g+\mathcal{F}\right)}}\sum_{q}\frac{\epsilon^{\alpha_{1}...\alpha_{2q}\beta_{1}...\beta_{p-2q+1}}}{q!2^{q}\left(p-2q+1\right)!}\mathcal{F}_{\alpha_{1}\alpha_{2}}...\mathcal{F}_{\alpha_{2q-1}\alpha_{2q}}\Gamma_{\beta_{1}...\beta_{p-2q+1}}\bar{\gamma}^{\frac{p-2q+2}{2}},\nonumber \\
\mathrm{L}_{\mathrm{D}_{p}} & =\sum_{q}\frac{\epsilon^{\alpha_{1}...\alpha_{2q}\beta_{1}...\beta_{p-2q+1}}}{q!2^{q}\left(p-2q+1\right)!}\frac{\left(-\bar{\gamma}\right)^{\frac{p}{2}-q+1}}{\sqrt{-\left(g+\mathcal{F}\right)}}\mathcal{F}_{\alpha_{1}\alpha_{2}}...\mathcal{F}_{\alpha_{2q-1}\alpha_{2q}}\Gamma_{\beta_{1}...\beta_{p-2q+1}}^{\ \ \ \ \ \ \ \ \ \ \ \ \ \lambda}\hat{D}_{\lambda},\label{eq:12}
\end{align}
for the type IIB string theory ($p$ is odd number),

\begin{align}
\hat{D}_{\alpha}= & \nabla_{\alpha}+\frac{1}{4\cdot2!}H_{\alpha NK}\Gamma^{NK}\bar{\gamma}-\frac{1}{8}e^{\phi}\big(F_{N}\Gamma^{N}\Gamma_{\alpha}\bar{\gamma}+\frac{1}{3!}F_{KLN}\Gamma^{KLN}\Gamma_{\alpha}\nonumber \\
 & +\frac{1}{2\cdot5!}F_{KLMNP}\Gamma^{KLMNP}\Gamma_{\alpha}\bar{\gamma}\big),\nonumber \\
\Delta= & \frac{1}{2}\left(\Gamma^{M}\partial_{M}\phi+\frac{1}{2\cdot3!}H_{MNK}\Gamma^{MNK}\bar{\gamma}\right)+\frac{1}{2}e^{\phi}\left(F_{M}\Gamma^{M}\bar{\gamma}+\frac{1}{2\cdot3!}F_{KLN}\Gamma^{KLN}\right),\nonumber \\
\Gamma_{\mathrm{D}_{p}}= & \frac{1}{\sqrt{-\left(g+\mathcal{F}\right)}}\sum_{q}\frac{\epsilon^{\alpha_{1}...\alpha_{2q}\beta_{1}...\beta_{p-2q+1}}}{q!2^{q}\left(p-2q+1\right)!}\mathcal{F}_{\alpha_{1}\alpha_{2}}...\mathcal{F}_{\alpha_{2q-1}\alpha_{2q}}\Gamma_{\beta_{1}...\beta_{p-2q+1}}\bar{\gamma}^{\frac{p-2q+1}{2}},\nonumber \\
\mathrm{L}_{\mathrm{D}_{p}} & =\sum_{q}\frac{\epsilon^{\alpha_{1}...\alpha_{2q}\beta_{1}...\beta_{p-2q+1}}}{q!2^{q}\left(p-2q+1\right)!}\frac{\left(-i\sigma_{2}\right)\left(\bar{\gamma}\right)^{\frac{p-2q+1}{2}}}{\sqrt{-\left(g+\mathcal{F}\right)}}\mathcal{F}_{\alpha_{1}\alpha_{2}}...\mathcal{F}_{\alpha_{2q-1}\alpha_{2q}}\Gamma_{\beta_{1}...\beta_{p-2q+1}}^{\ \ \ \ \ \ \ \ \ \ \ \ \ \lambda}\hat{D}_{\lambda}.\label{eq:13}
\end{align}
Since the two-point Green function in holography is our concern, only
the quadratic term in fermionic action is given in (\ref{eq:11}).
Then let us clarify the notation presented in (\ref{eq:11}) - (\ref{eq:13}).
$T_{p}$ refers to the tension of a $\mathrm{D}_{p}$-brane given
as $T_{p}=g_{s}^{-1}\left(2\pi\right)^{-p}l_{s}^{-\left(p+1\right)}$.
The indices denoted by capital letters $K,L,M,N...$ run over the
10d spacetime and indices denoted by lowercase letters $a,b,...$
run over the tangent space of the 10d spacetime. We use Greek alphabet
$\alpha,\beta,\lambda$ to denote the indices running over the worldvolume
of the $\mathrm{D}_{p}$-brane. The metric is written in terms of
elfbein as $g_{MN}=e_{M}^{a}\eta_{ab}e_{N}^{b}$, so the gamma matrices
are defined by

\begin{equation}
\left\{ \gamma^{a},\gamma^{b}\right\} =2\eta^{ab},\left\{ \Gamma^{M},\Gamma^{N}\right\} =2g^{MN},
\end{equation}
with $e_{M}^{a}\Gamma^{M}=\gamma^{a}$. $\omega_{\alpha ab}$ refers
to the spin connection and $\nabla_{\alpha}=\partial_{\alpha}+\frac{1}{4}\omega_{\alpha ab}\gamma^{ab}$
is the covariant derivative for fermion. The gamma matrix with multiple
indices is defined by ranking alternate anti-symmetrically or symmetrically
the indices e.g. 
\begin{equation}
\gamma^{ab}=\frac{1}{2}\left[\gamma^{a},\gamma^{b}\right],\gamma^{abc}=\frac{1}{2}\left\{ \gamma^{a},\gamma^{bc}\right\} ,\gamma^{abcd}=\frac{1}{2}\left[\gamma^{a},\gamma^{bcd}\right]...
\end{equation}
$\Gamma^{MNK...}$ shares the same definition as $\gamma^{abc...}$.
$\bar{\gamma}$ is defined as $\bar{\gamma}=\gamma^{01...9}$ and
$\sigma_{2}$ refers to the associated Pauli matrix. The worldvolume
field $\mathcal{F}$ is given as $\mathcal{F}=B+\left(2\pi\alpha^{\prime}\right)f$
where $B$ is the NS-NS 2-form in type II string theory with $H=dB$
and $f$ is the Yang-Mills field strength. $F_{M},F_{MN},F_{KLM}...$
refer to the associated field strength of the massless R-R fields
presented in IIA and IIB string theory. We note that, in our $\mathrm{D}_{p}/\mathrm{D}_{p+4}$
approach, $p$ should be replaced by $p+4$ in the above formulas
since the worldvolume fermion on the probe $\mathrm{D}_{p+4}$-brane
is our concern. 

\subsection{The prescription for two-point correlation function}

In this section, let us attempt to generalize the prescription in
AdS/CFT dictionary for two-point correlation function into the background
presented in (\ref{eq:2}) and (\ref{eq:4}). We first summarize the
steps as follows then give some comments.
\begin{enumerate}
\item Since one dimension of the $\mathrm{D}_{p}$-brane is compactified
on a circle, the dual theory is effectively $p$ dimensional below
the energy scale provided by the circle size. So we need to simplify
the T-dualitized action (\ref{eq:11}) with respect to the background
fields given in (\ref{eq:2}) (\ref{eq:4})\footnote{When we recall the background (\ref{eq:2}) (\ref{eq:4}), it means
we will use their near-horizon version for holography.} by integrating out the dependence on $S^{8-p}$ in order to obtain
an effective $p+1$ dimensional bulk action involving the bulk spinor
$\psi\left(x,r\right)$ as 
\begin{align}
S_{p+1} & \propto i\int d^{D}xdr\bar{\psi}\gamma^{r}\partial_{r}\psi+...\nonumber \\
 & =i\int d^{D}x\left(\bar{\psi}\gamma^{r}\psi\right)\big|_{r=r_{KK,H}}^{r=\infty}-i\int d^{D}xdr\partial_{r}\bar{\psi}\gamma^{r}\psi+...\label{eq:16}
\end{align}
\item Derive the equation of motion for $\psi\left(x,r\right)$ by varying
the $p+1$ dimensional action obtained in Step 1, then solve it by
using the Fourier mode as the ansatz for the $\psi\left(x,r\right)$
as,
\begin{equation}
\psi\left(x,r\right)=e^{ik\cdot x}\beta\left(\omega,\vec{k},r\right),k_{\mu}=\left(-\omega,\vec{k}\right).\label{eq:17}
\end{equation}
\item Impose the solution for $\psi\left(x,r\right)$ obtained in Step 2
in to the $p+1$ dimensional action obtained in Step 1, so the $p+1$
dimensional fermionic action becomes onshell $S_{p+1}^{\mathrm{cl}}$,
then define the boundary value of $\psi$ as $\psi_{0}$, hence the
two-point correlation function is obtained as,
\begin{align}
G_{R} & =-\frac{1}{Z_{gravity}}\frac{\delta}{\delta\bar{\psi}_{0}}\frac{\delta}{\delta\psi_{0}}Z_{gravity}\left[\bar{\psi},\psi\right]|_{\bar{\psi}_{0},\psi_{0}=0}\nonumber \\
 & =-\frac{\delta}{\delta\bar{\psi}_{0}}\frac{\delta}{\delta\psi_{0}}S_{p+1}^{\mathrm{cl}}\left[\bar{\psi},\psi\right],
\end{align}
leading to 
\begin{equation}
\bar{\Pi}_{0}=G_{R}\psi_{0},
\end{equation}
where $\Pi_{0}$ is defined in (\ref{eq:9}). Accordingly, one can
obtained the correlation function by solving $\psi_{0}$ and $\Pi_{0}$
in gravity side.
\end{enumerate}
The above steps are based on the prescription in AdS/CFT dictionary
for two-point correlation function \cite{key-10,key-14,key-38}. Note
that the last term in (\ref{eq:16}) would become vanished once the
classical solution (\ref{eq:17}) is imposed since (as we will see)
this term is nothing but the Dirac equation. Accordingly, in the actual
calculation, we will introduce a ratio function defined as 
\begin{equation}
\bar{\Pi}=\xi\psi,\label{eq:20}
\end{equation}
 where 
\begin{equation}
\Pi=-\frac{\delta}{\delta\psi}S_{p+1}^{\mathrm{cl}}=-i\bar{\psi}\gamma^{r},
\end{equation}
so that the boundary value of $\xi$ is the correlation function.
Therefore, our goal is to derive and solve the equations for $\xi$
with the in-falling boundary conditions usually used in AdS/CFT.

Besides, in the AdS/CFT correspondence, the onshell action (\ref{eq:16})
would always include a finite part due to the isometry of AdS or equivalently
the conformal symmetry in the dual theory \cite{key-14}. Hence the
holographic renormalization could work consistently to remove the
divergence presented in action (\ref{eq:16}). However, in a general
background, the onshell action (\ref{eq:16}) might not include any
finite parts which could lead to a non-renormalizable dual theory.
Nevertheless, it is possible to define a finite Green function consistently
by rescaling the boundary value $\psi_{0}$ with respect to $\psi$
which, as we will see, is equivalent to exact the finite part of $\xi$
given in (\ref{eq:20}). Therefore in the following sections, we will
test our prescription in the top-down D4/D8 and D3/D7 approaches by
holography.

\section{Approach to the D4/D8 model}

In this section, let us apply our prescription to the D4/D8 model
(Witten-Salai-Sugimoto model) to compute the two-point Green function.
We will first introduce the D4/D8 model briefly, then test our prescription
and analyze the results numerically in the 1+3 dimensional QCD. In
Section 4.1, we perform the Step 1 in the prescription which is to
obtain a 5d effective bulk action for fermion with respect to the
bubble and black D4 background. In Section 4.2, we follow the Step
2 and Step 3, that is to solve the associated equations of motion
for the bulk field, define its boundary value then derive its onshell
action to evaluate the correlation function.

\subsection{The 5d action for the bulk spinor}

The D4/D8 model is based on type IIA string theory and it corresponds
to the case of $p=4$ as it is discussed in Section 2. For the readers'
convenience, let us give its supergravity solution which is to set
$p=4$ in the background geometry (\ref{eq:4}) for the confined phase
as \cite{key-21},

\begin{align}
ds_{c}^{2} & =\left(\frac{r}{R}\right)^{3/2}\left[\eta_{\mu\nu}dx^{\mu}dx^{\nu}+f\left(r\right)\left(dx^{4}\right)^{2}\right]+\left(\frac{R}{r}\right)^{3/2}\left[\frac{dr^{2}}{f\left(r\right)}+r^{2}d\Omega_{4}^{2}\right],\nonumber \\
e^{\phi} & =\left(\frac{r}{R}\right)^{3/4},\ F_{4}=dC_{3}=3R^{3}g_{s}^{-1}\epsilon_{4},\ f\left(r\right)=1-\frac{r_{KK}^{3}}{r^{3}},\mu,\nu=0,..3.\label{eq:22}
\end{align}
Here $R$ relates to the radius of the spacetime, $\epsilon_{4}$
is the volume form of a unit $S^{4}$ and we have taken the near-horizon
limit. In the language of QCD, the parameters presented in (\ref{eq:22})
can be expressed as,

\begin{equation}
R^{3}=\frac{1}{2}\frac{g_{\mathrm{YM}}^{2}N_{c}l_{s}^{2}}{M_{KK}},\ r_{KK}=\frac{2}{3}g_{\mathrm{YM}}^{2}N_{c}M_{KK}l_{s}^{2},\ g_{s}=\frac{1}{2\pi}\frac{g_{\mathrm{YM}}^{2}}{M_{KK}l_{s}},
\end{equation}
where $g_{\mathrm{YM}}$ refers to the Yang-Mills coupling constant
in the dual theory. For the deconfined phase, the bulk geometry in
the near-horizon limit is obtained from (\ref{eq:2}) with $p=4$
as,

\begin{align}
ds_{d}^{2} & =\left(\frac{r}{R}\right)^{3/2}\left[-f_{T}\left(r\right)dt^{2}+\delta_{ij}dx^{i}dx^{j}+\left(dx^{4}\right)^{2}\right]+\left(\frac{R}{r}\right)^{3/2}\left[\frac{dr^{2}}{f\left(r\right)}+r^{2}d\Omega_{4}^{2}\right]\nonumber \\
e^{\phi} & =\left(\frac{r}{R}\right)^{3/4},\ F_{4}=dC_{3}=3R^{3}g_{s}^{-1}\epsilon_{4},\ f_{T}\left(r\right)=1-\frac{r_{H}^{3}}{r^{3}},i,j=1,2,3.\label{eq:24}
\end{align}

\begin{figure}
\begin{centering}
\includegraphics[scale=0.3]{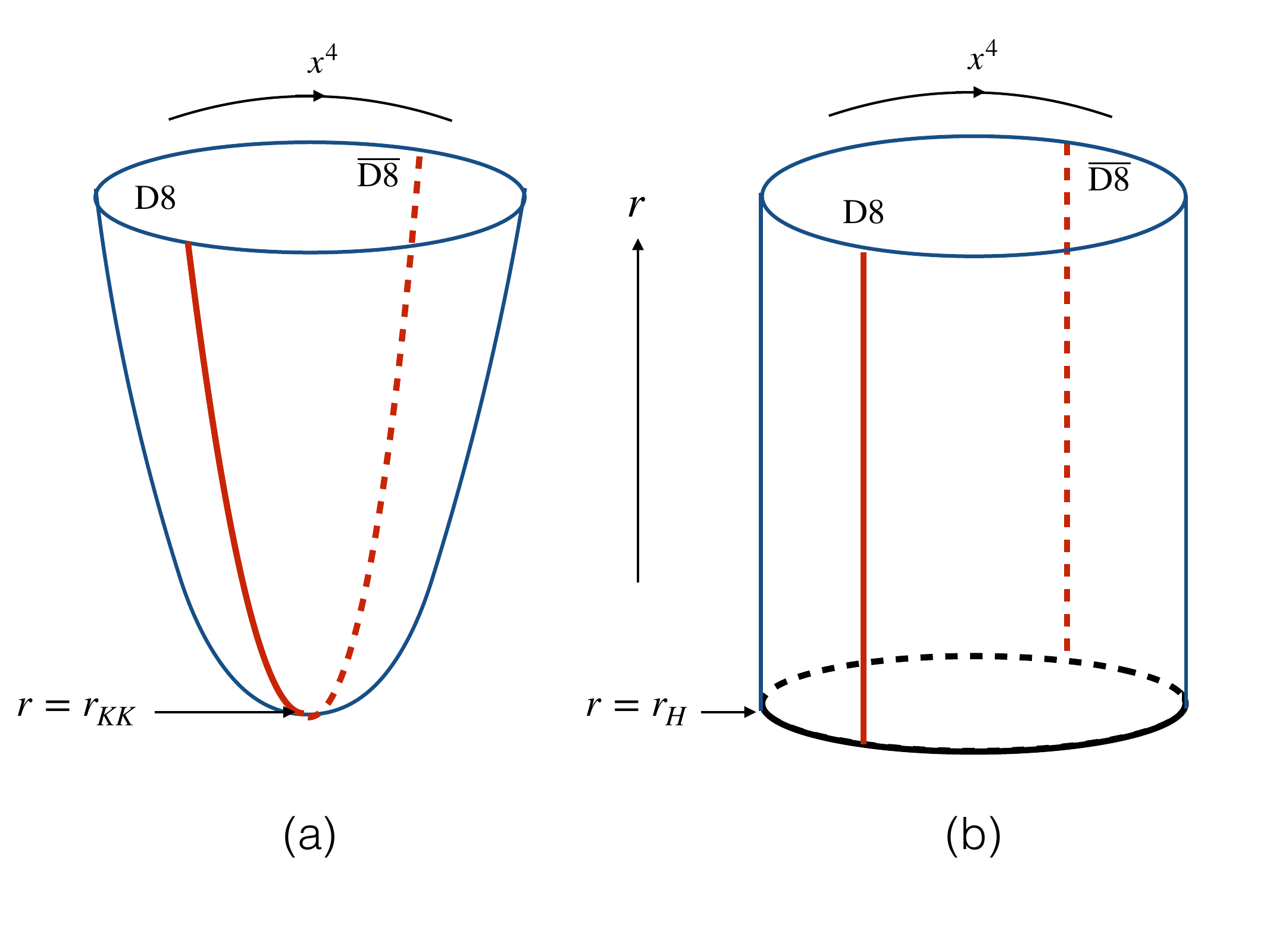}
\par\end{centering}
\caption{\label{fig:2} The configuration of the D4/D8 model in the bubble
(confined) background (a) and black brane (deconfined) background
(b) on $r-x^{4}$ plane. Red line refers to the D8-branes ($\mathrm{D8}$)
and anti D8-branes ($\mathrm{\overline{D8}}$). }

\end{figure}

\begin{figure}
\begin{centering}
\includegraphics[scale=0.3]{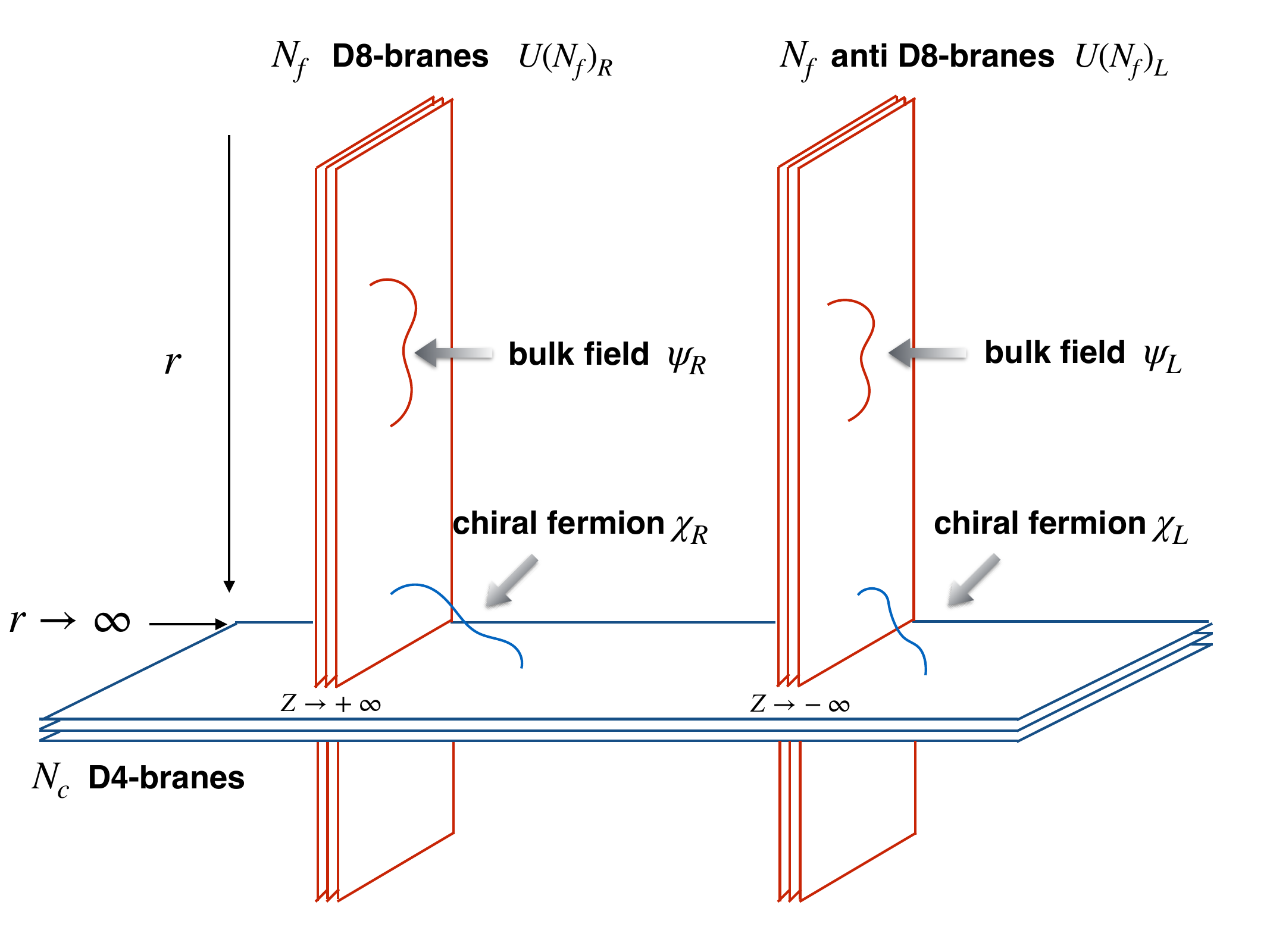}
\par\end{centering}
\caption{\label{fig:3} The correspondence of the bulk fields $\psi_{R,L}$
and flavored fermions $\chi_{R,L}$ consisted of chiral quarks in
the D4/D8 model. The $N_{f}$ pairs of D8- and anti D8-branes are
located at $Z\rightarrow\pm\infty$ at boundary providing $U\left(N_{f}\right)_{R}\times U\left(N_{f}\right)_{L}$
chiral symmetry in the dual theory. The bulk field $\psi_{R,L}$ as
source couples to the chiral quarks in $\chi_{R,L}$ respectively
as it is in QCD.}
\end{figure}

As it is discussed in Section 2, the bulk spinor is identified to
the fermionic field on the worldvolume of the probe D8-branes, we
need the induced metric on the worldvolume of the D8-branes which
according to Table \ref{tab:1} is given as,

\begin{align}
ds_{\mathrm{D8},c}^{2} & =\left(\frac{r}{R}\right)^{3/2}\eta_{\mu\nu}dx^{\mu}dx^{\nu}+\left(\frac{R}{r}\right)^{3/2}\left[\frac{dr^{2}}{f\left(r\right)}+r^{2}d\Omega_{4}^{2}\right],\nonumber \\
ds_{\mathrm{D8},d}^{2} & =\left(\frac{r}{R}\right)^{3/2}\left[-f_{T}\left(r\right)dt^{2}+\delta_{ij}dx^{i}dx^{j}\right]+\left(\frac{R}{r}\right)^{3/2}\left[\frac{dr^{2}}{f_{T}\left(r\right)}+r^{2}d\Omega_{4}^{2}\right],
\end{align}
for the confined and deconfined phase respectively. Considering the
most simple configuration to reveal the chirality in the dual theory,
we note that the D8- and anti D8-branes are embedded at the antipodal
position of $x^{4}$ i.e. $x^{4}=\mathrm{const}$ in the bubble D4-brane
background (\ref{eq:4}). For the black brane background (\ref{eq:2}),
the D8- and anti D8-branes are embedded parallel which also implies
$x^{4}=\mathrm{const}$ as it is illustrated in Figure \ref{fig:2}.
 Then the action for the worldvolume fermion on D8-branes can be collected
from (\ref{eq:11}) (\ref{eq:12}) as,

\begin{align}
S_{f}^{\mathrm{D8}} & =\frac{iT_{8}}{2}\int d^{9}xe^{-\phi}\sqrt{-g}\bar{\Psi}P_{-}\big[\Gamma^{\alpha}\nabla_{\alpha}+\frac{1}{8\cdot4!}e^{\phi}F_{KLNP}\left(\Gamma^{\alpha}\Gamma^{KLNP}\Gamma_{\alpha}-\Gamma^{KLNP}\right)\nonumber \\
 & \ \ \ \ \ \ \ \ \ \ \ \ \ \ -\frac{1}{2}\Gamma^{M}\partial_{M}\phi\big]\Psi.\label{eq:26}
\end{align}
Since our concern is the fermionic part, we have turned off the irrelative
bosonic fields by setting $\mathcal{F}=0$. Imposing the solution
for the dilaton $\phi$ and the R-R form $F_{4}$ presented in (\ref{eq:22})
or (\ref{eq:24}), the action (\ref{eq:26}), after some algebraic
calculations, can be simplified in the confined background as,

\begin{align}
S_{f,c}^{\mathrm{D8}}= & \frac{i\mathcal{T}_{c}}{\left(2\pi\alpha^{\prime}\right)^{2}\Omega_{4}}\int d^{4}xdZd\Omega_{4}\bar{\Psi}P_{-}\big(\frac{2}{3}M_{KK}K^{\frac{7}{12}}\gamma^{m}\nabla_{m}^{S^{4}}+K^{\frac{5}{12}}\gamma^{\mu}\partial_{\mu}+M_{KK}K^{\frac{13}{12}}\gamma^{Z}\partial_{Z},\nonumber \\
 & \ \ \ \ \ \ \ \ \ \ \ \ \ \ +\frac{13}{12}M_{KK}ZK^{\frac{1}{12}}\gamma^{4}\big)\Psi,\label{eq:27}
\end{align}
with,

\begin{equation}
K\left(Z\right)=1+Z^{2},\mathcal{T}_{c}=\frac{1}{2}\left(\frac{2}{3}\right)^{13/2}T_{8}\Omega_{4}\left(2\pi\alpha^{\prime}\right)^{2}\left(M_{KK}R\right)^{11/2}R^{5},P_{-}=\frac{1}{2}\left(1-\Gamma_{\mathrm{D8}}\right).
\end{equation}
In the deconfined background, the action (\ref{eq:26}) becomes,

\begin{align}
S_{f,d}^{\mathrm{D8}}= & \frac{i\mathcal{T}_{d}}{\left(2\pi\alpha^{\prime}\right)^{2}\Omega_{4}}\int d^{4}xdZd\Omega_{4}\bar{\Psi}P_{-}\bigg[\frac{4}{3}\pi TZK^{1/12}\gamma^{m}\nabla_{m}^{S^{4}}+K^{5/12}\gamma^{0}\partial_{0}+ZK^{-1/12}\gamma^{i}\partial_{i}\nonumber \\
 & \ \ \ \ \ \ \ \ \ \ \ \ \ \ +2\pi TZK^{7/12}\gamma^{Z}\partial_{Z}+\pi T\left(\frac{13}{6}Z^{2}K^{-5/12}+K^{7/12}\right)\gamma^{Z}\bigg]\Psi,\label{eq:29}
\end{align}
where 

\begin{equation}
\mathcal{T}_{d}=\frac{1}{2}\left(\frac{2}{3}\right)^{13/2}T_{8}\Omega_{4}\left(2\pi\alpha^{\prime}\right)^{2}\left(2\pi TR\right)^{11/2}R^{5},r_{H}=\frac{16}{9}\pi^{2}R^{3}T^{2},T=\beta_{t}^{-1}.
\end{equation}
Here $T,\beta_{t}$ refers to the Hawking temperature and the period
of the time direction in (\ref{eq:2}). We have used $\Psi$ to denote
the worldvolume fermion whose boundary value relates to $\psi_{0}$.
The index $m$ runs over $S^{4}$ and the unit volume of $S^{4}$
is given as $\Omega_{4}=8\pi^{2}/3$. We recall that in the D4/D8
model, the flavor branes are introduced as $N_{f}$ pairs of probe
D8- and anti D8-branes located at the antipodal points of $x^{4}$.
Hence the Cartesian coordinates $Z$ is usually used as,

\begin{equation}
K\left(Z\right)=1+Z^{2}=\frac{r^{3}}{r_{KK,H}^{3}},
\end{equation}
so that the D8- and anti D8-branes are located respectively at $Z\rightarrow\pm\infty$
representing approximately the chiral symmetry $U\left(N_{f}\right)_{R}\times U\left(N_{f}\right)_{L}$
at boundary. And the dual theory on the D4-branes at boundary takes
the chirally symmetric action as QCD as \cite{key-42},

\begin{align}
S= & \int_{\mathrm{D4}}d^{5}x\sqrt{-g}\left[\delta\left(Z-Z_{+\infty}\right)\chi_{L}^{\dagger}\bar{\sigma}^{\mu}\left(i\nabla_{\mu}+A_{\mu}\right)\chi_{L}+\delta\left(Z-Z_{-\infty}\right)\chi_{R}^{\dagger}\bar{\sigma}^{\mu}\left(i\nabla_{\mu}+A_{\mu}\right)\chi_{R}\right]\nonumber \\
 & -\frac{1}{4g_{\mathrm{YM}}^{2}}\int_{\mathrm{D4}}d^{5}x\sqrt{-g}\mathrm{Tr}F_{\mu\nu}F^{\mu\nu},\ \mu,\nu=0,1...3,
\end{align}
where we use Weyl spinor $\chi_{R,L}$ to denote the fundamental chiral
fermion and $F_{MN}$ is the gauge field strength of the gluon $A_{M}$.
Therefore the source term of spinor presented in (\ref{eq:7}) can
be written in terms of Weyl spinors as,

\begin{equation}
\left\langle \exp\left\{ \int_{\partial\mathcal{M}}\left(\chi_{L}^{\dagger}\psi_{L}+\chi_{R}^{\dagger}\psi_{R}+h.c.\right)d^{D}x\right\} \right\rangle ,
\end{equation}
once we focus on the Green function of chiral fermion, as it is illustrated
in Figure \ref{fig:3}. 

In order to compute the two-point Green function, let us reduce the
fermionic action (\ref{eq:27}) (\ref{eq:29}) to a 5d form as most
works about AdS/CFT. First, we decompose the 10d spinor into a 3+1
dimensional part $\psi\left(x,Z\right)$ with holographic coordinate
$Z$, an $S^{4}$ part $\varphi$ and a remaining 2d part $\beta$
as\footnote{We note that in $D$ dimension, a spinor has $\left[D/2\right]$ components
where $\left[D/2\right]$ refers to the integer part of $D/2$.},

\begin{equation}
\psi\left(x,Z\right)\otimes\varphi\left(S^{4}\right)\otimes\beta,\label{eq:34}
\end{equation}
so that the worldvolume fermion on D8-brane should be $\Psi=\psi\otimes\varphi$.
Then the associated 10d gamma matrices are chosen as,

\begin{align}
\gamma^{\mu} & =\sigma_{1}\otimes\boldsymbol{\gamma}^{\mu}\otimes\boldsymbol{1},\mu=0,1,2,3\nonumber \\
\gamma^{Z} & =\sigma_{1}\otimes\boldsymbol{\gamma}\otimes\boldsymbol{1},\nonumber \\
\gamma^{4} & =\sigma_{2}\otimes\boldsymbol{1}\otimes\tilde{\boldsymbol{\gamma}},\nonumber \\
\gamma^{m} & =\sigma_{2}\otimes\boldsymbol{1}\otimes\boldsymbol{\gamma}^{m},m=6,7,8,9,\nonumber \\
\boldsymbol{\gamma} & =i\boldsymbol{\gamma}^{0}\boldsymbol{\gamma}^{1}\boldsymbol{\gamma}^{2}\boldsymbol{\gamma}^{3},\nonumber \\
\tilde{\boldsymbol{\gamma}} & =i\boldsymbol{\gamma}^{6}\boldsymbol{\gamma}^{7}\boldsymbol{\gamma}^{8}\boldsymbol{\gamma}^{9},\label{eq:35}
\end{align}
where we use bold to denote the $4\times4$ gamma matrices. $\boldsymbol{\gamma}^{m},m=6,7,8,9$
refers to the gamma matrix on tangent space of $S^{4}$. We note that,
the 10d chirality matrix takes a very simple form as $\bar{\gamma}=\sigma_{3}\otimes\boldsymbol{1}\otimes\boldsymbol{1}$
in this decomposition. Choosing the $\sigma_{3}$ representation,
$\beta$ can therefore be decomposed by the eigenstates of $\sigma_{3}$
with

\begin{equation}
\sigma_{3}\beta_{\pm}=\beta_{\pm},\sigma_{1}\beta_{\pm}=\beta_{\mp},\sigma_{2}\beta_{\pm}=\pm i\beta_{\mp},\label{eq:36}
\end{equation}
where we use $\beta_{\pm}$ to denote the two eigenstates of $\sigma_{3}$.
Since the kappa symmetry fixes the condition $\bar{\gamma}\Psi=\Psi$\footnote{This condition could be opposite on the anti D8-branes.},
we have to chose $\beta=\beta_{+}$ on the D8-brane. In addition,
as $\varphi$ must satisfy the Dirac equation on $S^{4}$, we can
decompose it by using the spherical harmonic function with the eigenstates
of $\Gamma^{\underline{m}}\nabla_{m}^{S^{4}}$ as \cite{key-43},

\begin{equation}
\boldsymbol{\gamma}^{m}\nabla_{m}^{S^{4}}\varphi^{\pm l,s}=i\Lambda_{l}^{\pm}\varphi^{\pm l,s};\Lambda_{l}^{\pm}=\pm\left(2+l\right),l=0,1...\label{eq:37}
\end{equation}
Here $s,l$ represents the angular quantum numbers of the spherical
harmonic function.

Put (\ref{eq:34}) into (\ref{eq:27}) (\ref{eq:29}) with the decomposition
(\ref{eq:35}) - (\ref{eq:37}) for $\varphi$ and $\beta$, we could
obtain a 5d effective action for the fermion on the worldvolume of
the D8-branes after integrating the $S^{4}$ part as,

\begin{align}
S_{f,c}^{\mathrm{D8}}= & \frac{i\mathcal{T}_{c}}{\left(2\pi\alpha^{\prime}\right)^{2}}\int d^{4}xdZ\bar{\psi}\big(-\frac{2}{3}M_{KK}\Lambda_{l}K^{\frac{7}{12}}+K^{\frac{5}{12}}\boldsymbol{\gamma}^{\mu}\partial_{\mu}+M_{KK}K^{\frac{13}{12}}\boldsymbol{\gamma}\partial_{Z}\nonumber \\
 & \ \ \ \ \ \ \ \ \ \ \ \ \ +\frac{13}{12}M_{KK}ZK^{\frac{1}{12}}\boldsymbol{\gamma}\big)\psi,\label{eq:38}
\end{align}
and

\begin{align}
S_{f,d}^{\mathrm{D8}}= & \frac{i\mathcal{T}_{d}}{\left(2\pi\alpha^{\prime}\right)^{2}}\int d^{4}xdZ\bar{\psi}\bigg[-\frac{4}{3}\pi TZK^{1/12}\Lambda_{l}+K^{5/12}\boldsymbol{\gamma}^{0}\partial_{0}+ZK^{-1/12}\boldsymbol{\gamma}^{i}\partial_{i}\nonumber \\
 & \ \ \ \ \ \ \ \ \ \ \ \ \ \ +2\pi TZK^{7/12}\boldsymbol{\gamma}\partial_{Z}+\pi T\left(\frac{13}{6}Z^{2}K^{-5/12}+K^{7/12}\right)\boldsymbol{\gamma}\bigg]\psi,\label{eq:39}
\end{align}
where $\Lambda_{l}=\left|\Lambda_{l}^{\pm}\right|$. As we are going
to focus on the Green function in the dual theory according to the
Section 3.3, the action (\ref{eq:38}) can be further simplified by
rescaling $\psi\rightarrow2\pi\alpha^{\prime}K^{-13/24}\psi$ as,
\begin{equation}
S_{f,c}^{\mathrm{D8}}=i\mathcal{T}_{c}\int d^{4}xdZ\bar{\psi}\left(-\frac{2}{3}M_{KK}\Lambda_{l}K^{-1/2}+K^{-2/3}\boldsymbol{\gamma}^{\mu}\partial_{\mu}+M_{KK}\boldsymbol{\gamma}\partial_{Z}\right)\psi,\label{eq:40}
\end{equation}
while the action (\ref{eq:39}) is rescaled by $\psi\rightarrow2\pi\alpha^{\prime}K^{-7/24}Z^{-1/2}\psi$
as,

\begin{align}
S_{f,d}^{\mathrm{D8}} & =i\mathcal{T}_{d}\int d^{4}xdZ\bar{\psi}\bigg(-\frac{4}{3}\pi TK^{-1/2}\Lambda_{l}+Z^{-1}K^{-1/6}\boldsymbol{\gamma}^{0}\partial_{0}+K^{-2/3}\boldsymbol{\gamma}^{i}\partial_{i}\nonumber \\
 & \ \ \ \ \ \ \ \ \ \ \ \ \ +2\pi T\boldsymbol{\gamma}\partial_{Z}+\pi TZK^{-1}\boldsymbol{\gamma}\bigg)\psi.\label{eq:41}
\end{align}
And we will use these 5d fermionic actions (\ref{eq:40}) (\ref{eq:41})
to evaluate the two-point Green function.

\subsection{Use the prescription}

Since the 5d effective action is obtained in Section 4.1, in this
section, let us use Step 2 and Step 3 in the prescription to evaluate
the two-point Green function with confined (\ref{eq:4}) and deconfined
(\ref{eq:2}) geometry by using action (\ref{eq:40}) (\ref{eq:41})
respectively. 

\subsubsection*{Confined phase}

Let us start with the Step 2 in Section 3.3 i.e. solve the equation
of motion associated to action (\ref{eq:40}) which is derived as,

\begin{equation}
\left(-\frac{2}{3}M_{KK}\Lambda_{l}K^{-1/2}+K^{-2/3}\boldsymbol{\gamma}^{\mu}\partial_{\mu}+M_{KK}\boldsymbol{\gamma}\partial_{Z}\right)\psi=0.\label{eq:42}
\end{equation}
Picking up the ansatz in Weyl basis as,

\begin{equation}
\psi=e^{ik\cdot x}\left(\begin{array}{c}
\psi_{R}\\
\psi_{L}
\end{array}\right),\ k\cdot x=k_{\mu}x^{\mu}=-\omega t+\vec{k}\cdot\vec{x},\label{eq:43}
\end{equation}
the equation (\ref{eq:42}) becomes

\begin{align}
-\frac{2}{3}M_{KK}\Lambda_{l}K^{-1/2}\psi_{R}-K^{-2/3}\left(\sigma\cdot k\right)\psi_{L}+M_{KK}\partial_{Z}\psi_{R} & =0,\nonumber \\
-\frac{2}{3}M_{KK}\Lambda_{l}K^{-1/2}\psi_{L}-K^{-2/3}\left(\bar{\sigma}\cdot k\right)\psi_{R}-M_{KK}\partial_{Z}\psi_{L} & =0,\label{eq:44}
\end{align}
where we have chosen the 5d gamma matrices as,

\begin{equation}
\boldsymbol{\gamma}^{\mu}=i\left(\begin{array}{cc}
0 & \sigma^{\mu}\\
\bar{\sigma}^{\mu} & 0
\end{array}\right),\boldsymbol{\gamma}=\left(\begin{array}{cc}
1 & 0\\
0 & -1
\end{array}\right),
\end{equation}
with $\sigma^{\mu}=\left(1,-\sigma^{i}\right),\bar{\sigma}^{\mu}=\left(1,\sigma^{i}\right)$.
Solving (\ref{eq:44}), we can further obtain the decoupled equations
for $\psi_{R,L}$ as,

\begin{align}
\partial_{Z}^{2}\psi_{L}+\frac{2K^{\prime}}{3K}\partial_{Z}\psi_{L}+\left(\frac{\Lambda_{l}K^{\prime}}{9K^{3/2}}-\frac{4}{9}\frac{\Lambda_{l}^{2}}{K}-\frac{k^{2}}{M_{KK}^{2}}\frac{1}{K^{4/3}}\right)\psi_{L} & =0,\nonumber \\
\partial_{Z}^{2}\psi_{R}+\frac{2K^{\prime}}{3K}\partial_{Z}\psi_{R}-\left(\frac{\Lambda_{l}K^{\prime}}{9K^{3/2}}+\frac{4}{9}\frac{\Lambda_{l}^{2}}{K}+\frac{k^{2}}{M_{KK}^{2}}\frac{1}{K^{4/3}}\right)\psi_{R} & =0,\label{eq:46}
\end{align}
where ``$\prime$'' refers to the derivative with respect to $Z$.
The above equations can be analytically solve at $Z\rightarrow+\infty$
as,

\begin{align}
\psi_{R} & =AZ^{\frac{2}{3}\Lambda_{l}}+BZ^{-\frac{1}{3}-\frac{2}{3}\Lambda_{l}},\nonumber \\
\psi_{L} & =CZ^{-\frac{1}{3}+\frac{2}{3}\Lambda_{l}}+DZ^{-\frac{2}{3}\Lambda_{l}},\label{eq:47}
\end{align}
where $A,B,C,D$ are constant spinors depended on $\omega,\vec{k},\Lambda_{l}$
and furthermore one can find the following relations as,

\begin{equation}
C=\frac{3}{\left(1-4\lambda\right)M_{KK}}\left(\bar{\sigma}\cdot k\right)A,\ B=\frac{3}{\left(1+4\lambda\right)M_{KK}}\left(\sigma\cdot k\right)D,
\end{equation}
once the analytical solution (\ref{eq:47}) is plugged back into (\ref{eq:44}).
Besides, we notice that the solution at $Z\rightarrow-\infty$ is
equivalent to interchange $\psi_{R,L}$ in the solution (\ref{eq:47}).
$\Lambda_{l}$ is the eigenvalue of the $S^{4}$ part and it satisfies
$\Lambda_{l}\geq2$ according to (\ref{eq:37}). So we obtain the
boundary behavior of $\psi$ as,

\begin{equation}
\lim_{Z\rightarrow+\infty}\psi\rightarrow\left(\begin{array}{c}
\psi_{R}\\
0
\end{array}\right),\ \lim_{Z\rightarrow-\infty}\psi\rightarrow\left(\begin{array}{c}
0\\
\psi_{L}
\end{array}\right).
\end{equation}
This can be nicely interpreted as the source coupling to the chiral
fermion in each boundary theory, as it is illustrated in Figure \ref{fig:3}.
Hence we will focus on the calculation with respect to the $N_{f}$
D8-branes since the discussion would be exactly same on the anti D8-branes.
Afterwards, let us take care of the boundary value $\psi_{0}$ of
$\psi$ at $Z\rightarrow+\infty$ by picking up its dominantly finite
part in (\ref{eq:47}) as \cite{key-13,key-14},

\begin{equation}
\psi_{0}=\lim_{Z\rightarrow+\infty}Z^{-\frac{2}{3}\Lambda_{l}}\psi=\left(\begin{array}{c}
A\\
0
\end{array}\right).
\end{equation}
Next, the conjugate momentum $\Pi_{0}$ associated to $\psi_{0}$
can be evaluated by using (\ref{eq:9}). To this goal, let us take
a look at the action (\ref{eq:40}), 

\begin{align}
S_{f,c}^{\mathrm{D8}}= & \ i\mathcal{T}_{c}\int d^{4}xdZ\bar{\psi}\left(-\frac{2}{3}M_{KK}\Lambda_{l}K^{-1/2}+K^{-2/3}\boldsymbol{\gamma}^{\mu}\partial_{\mu}+M_{KK}\boldsymbol{\gamma}\partial_{Z}\right)\psi,\nonumber \\
= & \ i\mathcal{T}_{c}\int d^{4}x\left(\bar{\psi}M_{KK}\boldsymbol{\gamma}\psi\right)|_{-\infty}^{+\infty}\nonumber \\
 & -i\mathcal{T}_{c}\int d^{4}xdZ\left(\partial_{Z}\bar{\psi}M_{KK}\boldsymbol{\gamma}+\frac{2}{3}M_{KK}\Lambda_{l}K^{-1/2}\bar{\psi}-K^{-2/3}\partial_{\mu}\bar{\psi}\boldsymbol{\gamma}^{\mu}\right)\psi,\nonumber \\
\equiv & \int d^{4}x\left(\Pi\psi\right)|_{-\infty}^{+\infty}=\int d^{4}x\left(\Pi_{R}\psi_{R}+\Pi_{L}\psi_{L}\right)|_{-\infty}^{+\infty},\label{eq:51}
\end{align}
where

\begin{equation}
\Pi_{R}=-\mathcal{T}_{c}M_{KK}\psi_{L}^{\dagger},\ \Pi_{L}=\mathcal{T}_{c}M_{KK}\psi_{R}^{\dagger}.
\end{equation}
Notice that the term in the third line of (\ref{eq:51}) vanishes
since it is nothing but the conjugate equation to (\ref{eq:42}).
So once we impose the solution (\ref{eq:47}) to (\ref{eq:51}), it
implies the action (\ref{eq:51}) includes divergence at boundary
$Z\rightarrow\infty$ as\footnote{Since the solution at $Z\rightarrow-\infty$ can be obtained by interchanging
the roles of $\psi_{R,L}$ in (\ref{eq:47}), the action (\ref{eq:51})
has same asymptotic behavior at $Z\rightarrow-\infty$ on the anti
D8-branes. },

\begin{align}
S_{f,c}^{\mathrm{D8}} & \supseteq-\mathcal{T}_{c}M_{KK}\int d^{4}x\left[C^{\dagger}AZ^{-\frac{1}{3}+\frac{4}{3}\Lambda_{l}}+D^{\dagger}A+h.c\right]|_{Z\rightarrow+\infty}+...\label{eq:53}
\end{align}
Therefore we can see, it requests for a holographic boundary counterterm
$S_{ct}$ as,

\begin{equation}
S_{ct}=\mathcal{T}_{c}M_{KK}\int d^{4}x\left[C^{\dagger}AZ^{-\frac{1}{3}+\frac{4}{3}\Lambda_{l}}+h.c\right]|_{Z\rightarrow+\infty},
\end{equation}
to remove the divergence in (\ref{eq:51}). We note that action (\ref{eq:53})
contains a finite part which will survive after holographic renormalization.
And this should relate to the residual isometry of $\mathrm{AdS}_{7}$
since the D4/D8 model is based on the holographic correspondence on
$\mathrm{AdS}_{7}\times S^{4}$ \cite{key-20,key-38}. Then the conjugate
momentum on the D8-branes to $\psi_{0}$ is evaluated as

\begin{equation}
\Pi_{0}=-\frac{\delta S^{ren}}{\delta\psi_{0}}=\mathcal{T}_{c}M_{KK}\left(0,D^{\dagger}\right),\ S^{ren}=S_{f,c}^{\mathrm{D8}}+S_{ct},
\end{equation}
which implies the Green function $G_{R}\left(\omega,\vec{k}\right)$
is obtained by

\begin{equation}
\mathcal{T}_{c}M_{KK}D=G_{R}\left(\omega,\vec{k}\right)A,
\end{equation}
according to (\ref{eq:8}) (\ref{eq:9}). Using the representation

\begin{equation}
\psi_{R}=\left(\begin{array}{c}
\psi_{R}^{\left(1\right)}\\
\psi_{R}^{\left(2\right)}
\end{array}\right),\psi_{L}=\left(\begin{array}{c}
\psi_{L}^{\left(1\right)}\\
\psi_{L}^{\left(2\right)}
\end{array}\right),\label{eq:57}
\end{equation}
and introducing the ratio

\begin{equation}
\xi_{1}=\frac{\psi_{L}^{\left(1\right)}}{\psi_{R}^{\left(1\right)}},\xi_{2}=\frac{\psi_{L}^{\left(2\right)}}{\psi_{R}^{\left(2\right)}},\label{eq:58}
\end{equation}
the Green function can therefore been written as,

\begin{equation}
G_{R}^{\left(1,2\right)}=\mathcal{T}_{c}M_{KK}\lim_{Z\rightarrow\infty}Z^{\frac{4}{3}\Lambda_{l}}\xi_{1,2}\left(Z\right).\label{eq:59}
\end{equation}
satisfying the equation of motion,

\begin{equation}
\xi_{1,2}^{\prime}=-\frac{4}{3}\Lambda_{l}K^{-1/2}\xi_{1,2}-\frac{K^{-2/3}}{M_{KK}}\left(-\omega+h\cdot\mathrm{k}\right)-\frac{K^{-2/3}}{M_{KK}}\left(-\omega-h\cdot\mathrm{k}\right)\xi_{1,2}^{2},\label{eq:60}
\end{equation}
according to (\ref{eq:44}). Here we have set $k_{\mu}=\left(-\omega,\mathrm{k},0,0\right)$
with,

\begin{equation}
\frac{\psi_{R}^{\left(2\right)}}{\psi_{R}^{\left(1\right)}}=\frac{\psi_{L}^{\left(2\right)}}{\psi_{L}^{\left(1\right)}}=h.\label{eq:61}
\end{equation}
 On the other hand, since the dual theory is chirally symmetric, we
obtain

\begin{equation}
\psi_{R}|_{Z\rightarrow+\infty}=\psi_{L}|_{Z\rightarrow-\infty},
\end{equation}
which implies \cite{key-36}

\begin{equation}
\psi_{R}\left(0\right)=\pm\psi_{L}\left(0\right).\label{eq:63}
\end{equation}
Therefore we can set $h=\pm1$ in (\ref{eq:61}) for $\xi_{1,2}$
respectively. Altogether, (\ref{eq:60}) could be numerically solved
with the incoming wave boundary condition $\xi_{1,2}\left(0\right)=\pm1$.
\begin{figure}[h]
\begin{centering}
\includegraphics[scale=0.27]{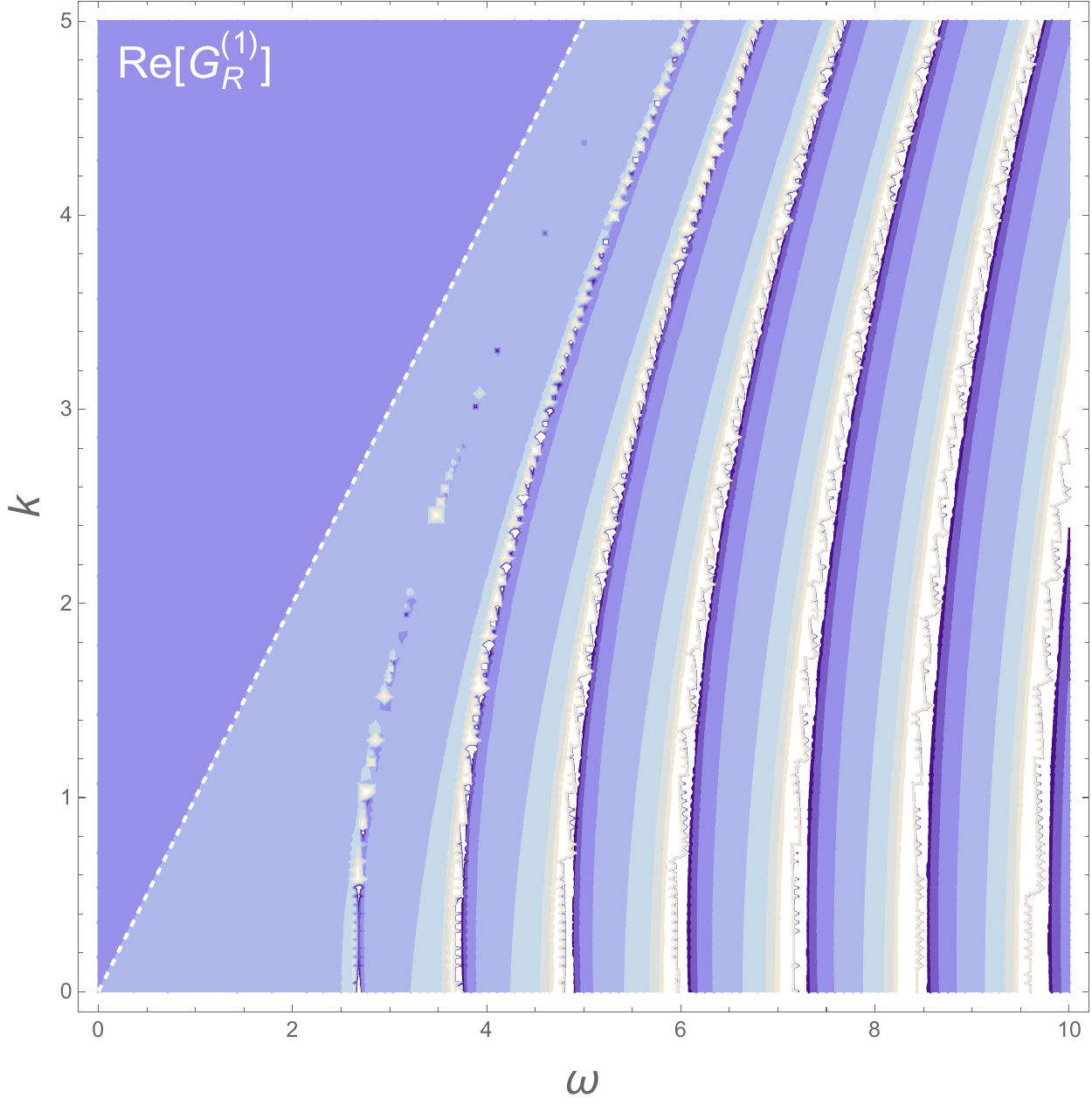}\includegraphics[scale=0.27]{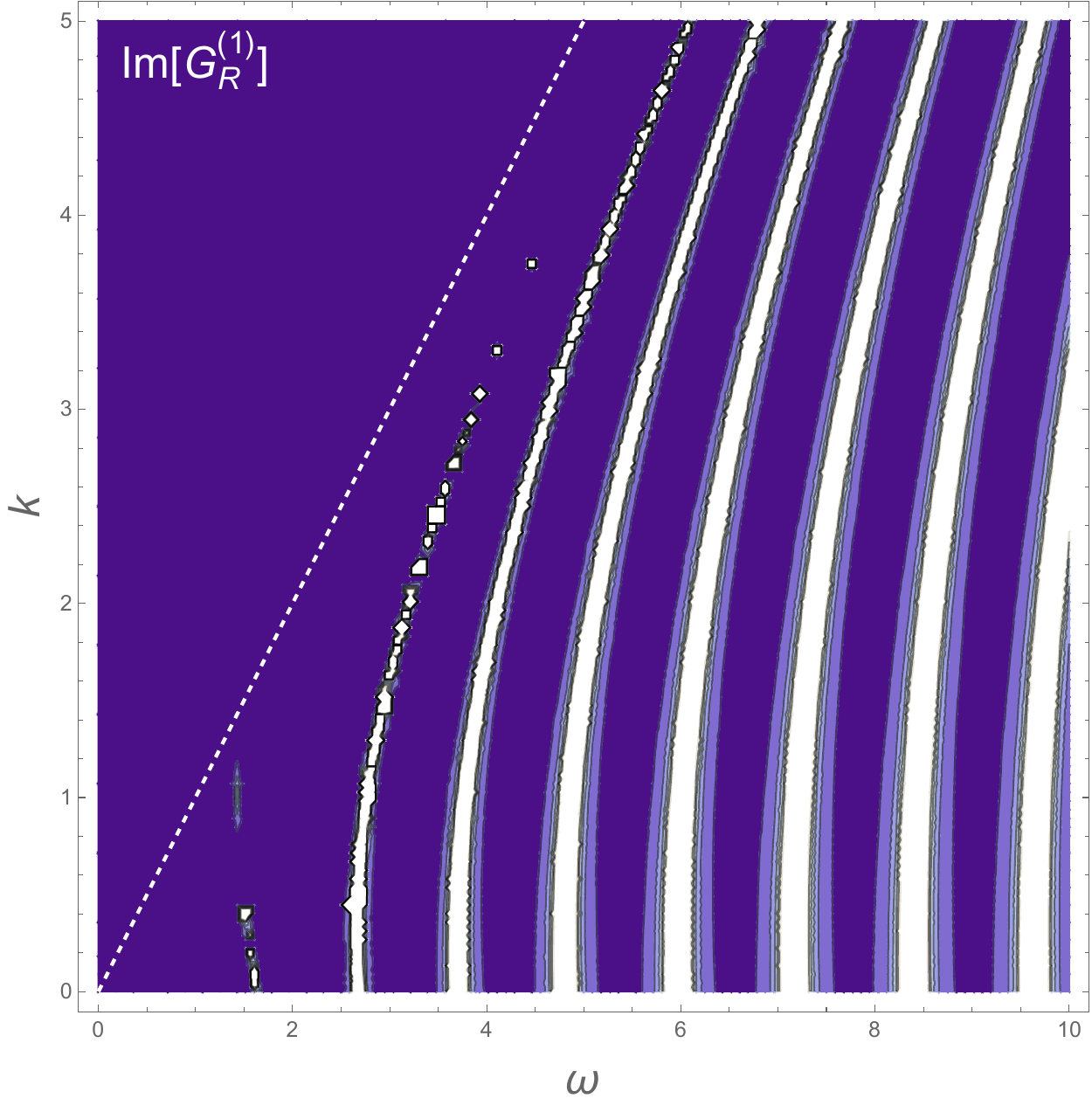}
\par\end{centering}
\begin{centering}
\includegraphics[scale=0.27]{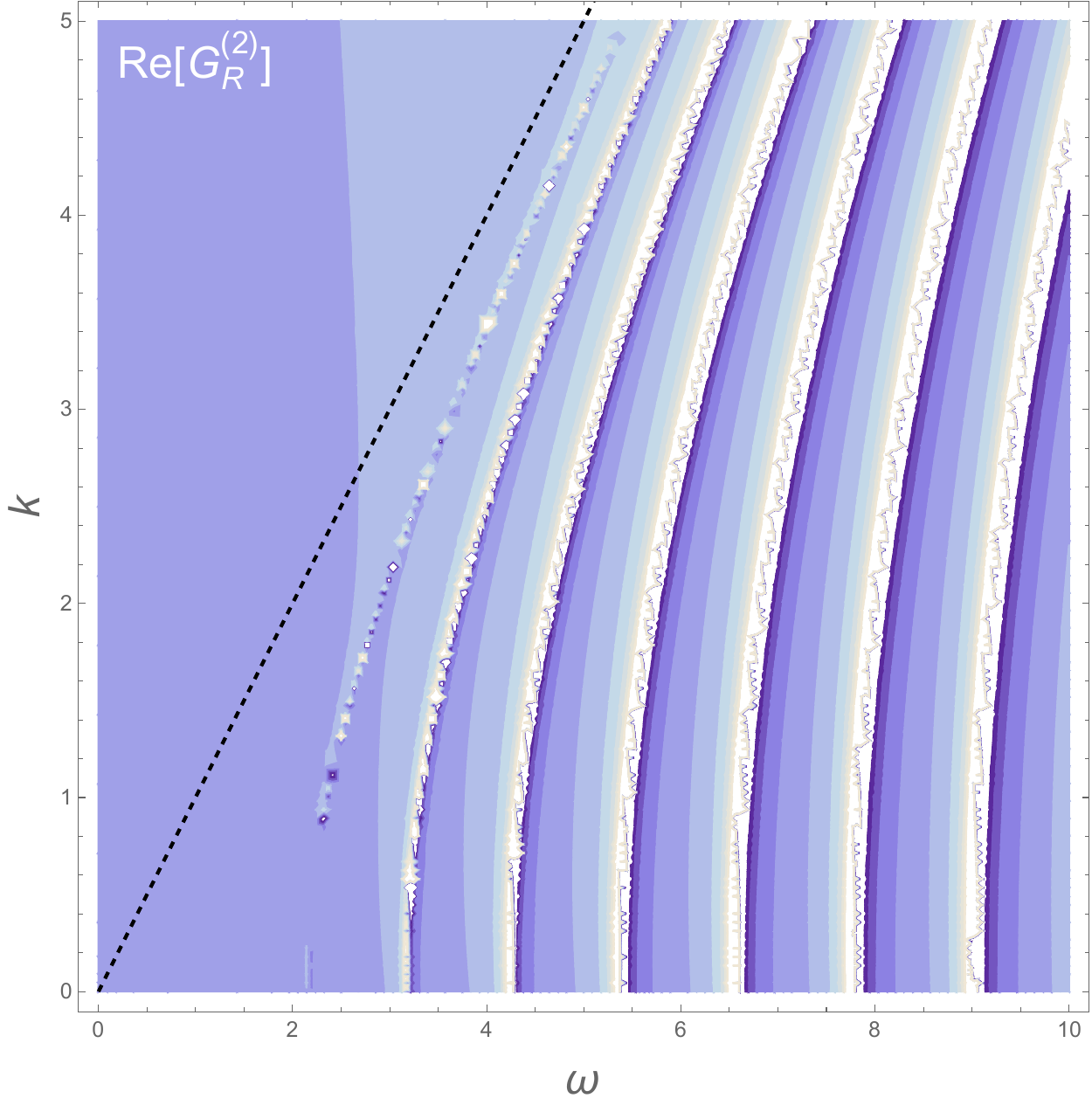}\includegraphics[scale=0.27]{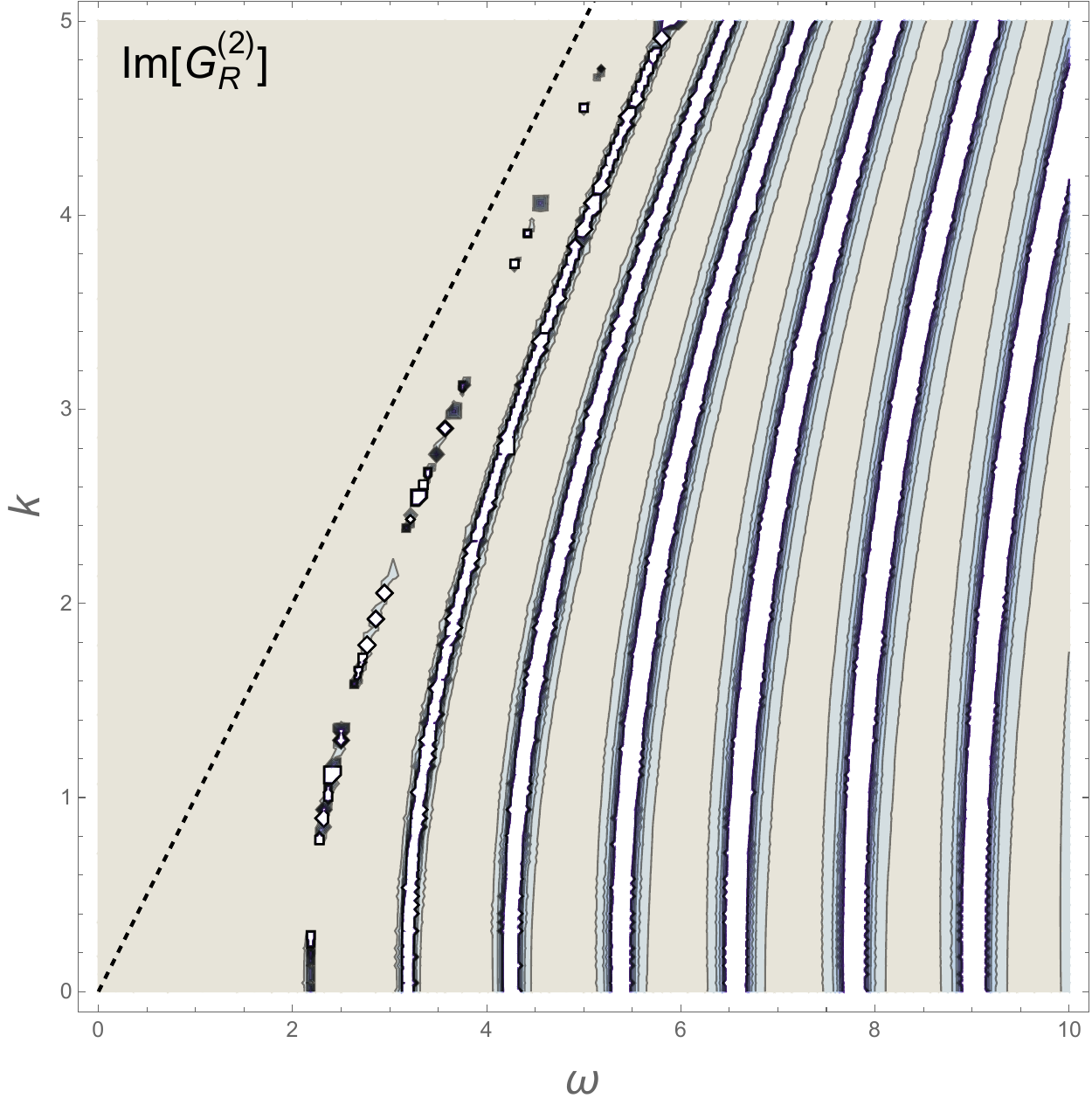}
\par\end{centering}
\caption{\label{fig:4} Density plot of the confined retarded Green function
$G_{R}^{\left(1,2\right)}$ as the spectral function of 1+3 dimensional
QCD from the D4/D8 model. The white regions refer to the peaks in
the Green function and the dashed lines refer to $\omega=\mathrm{k}$
as the light cones. The parameters are chosen as $\Lambda_{l}=2,l=0,\mathcal{T}_{c}=1$
in the unit of $M_{KK}=1$.}
\end{figure}
 
\begin{figure}[h]
\begin{centering}
\includegraphics[scale=0.3]{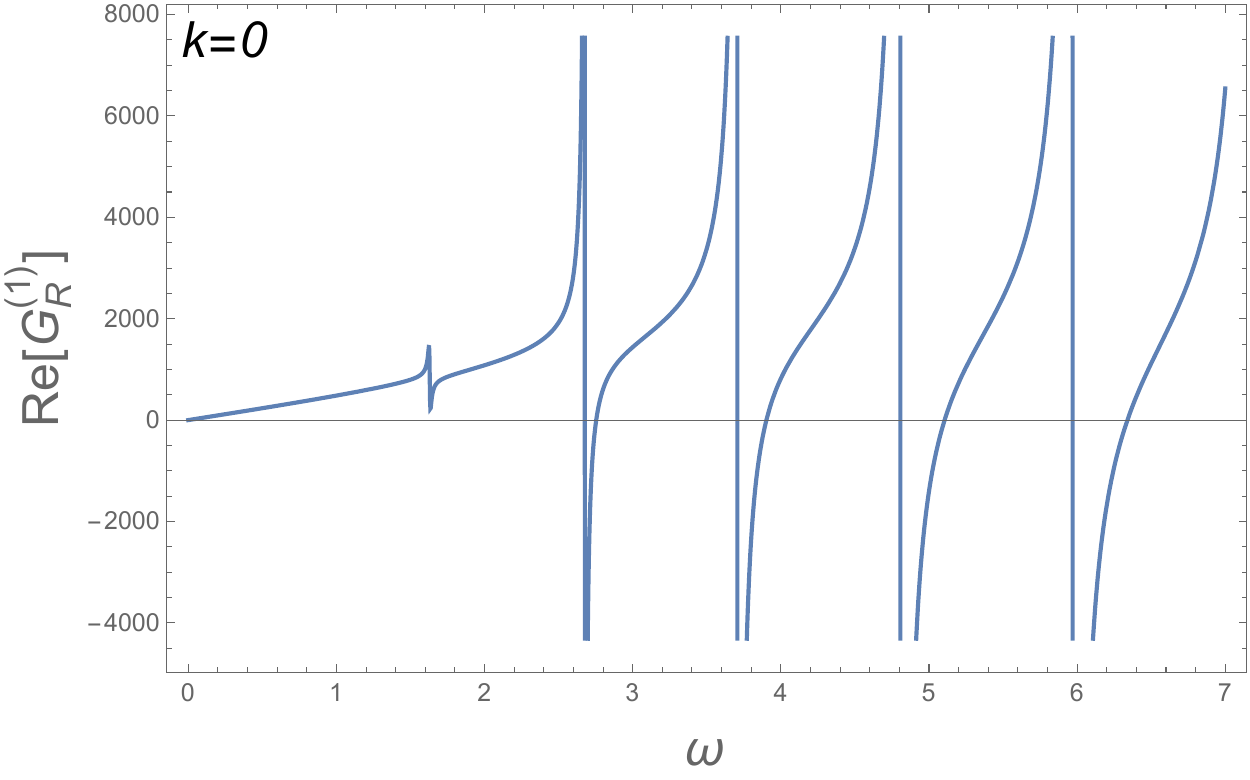}\includegraphics[scale=0.3]{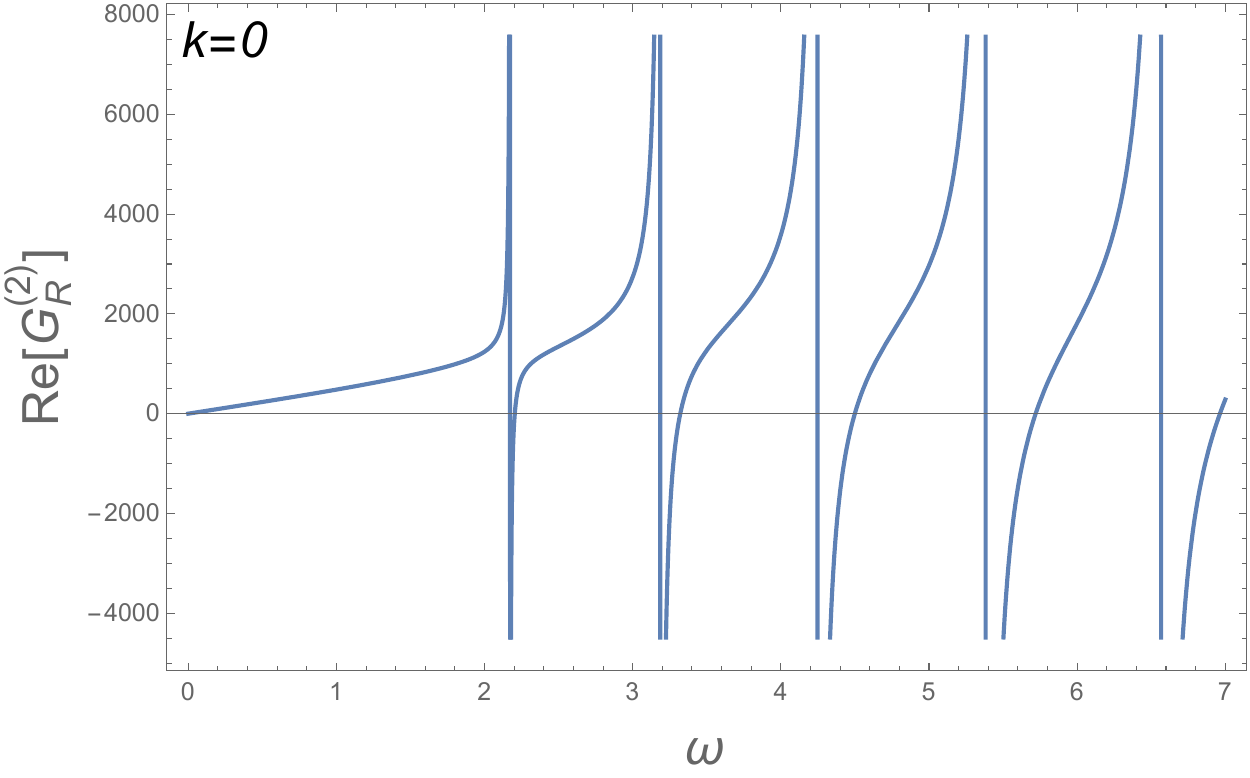}
\par\end{centering}
\begin{centering}
\includegraphics[scale=0.3]{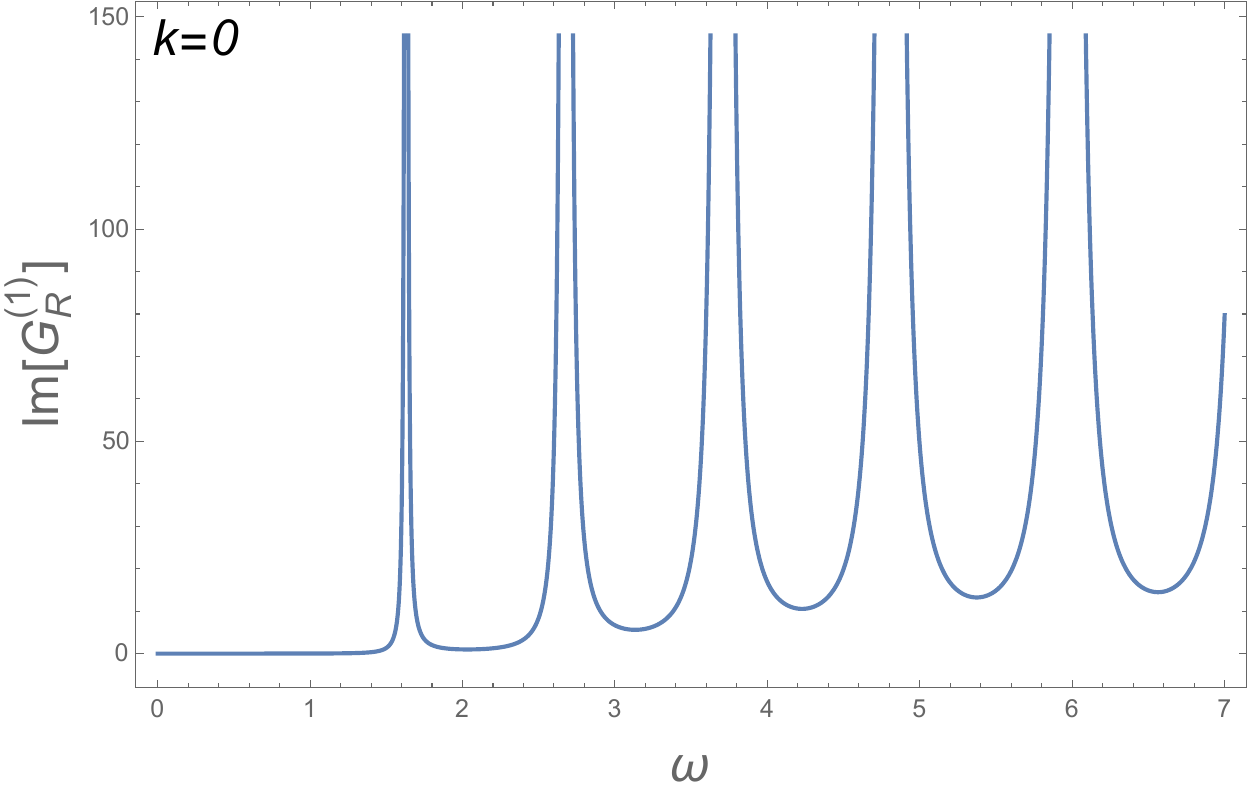}\includegraphics[scale=0.3]{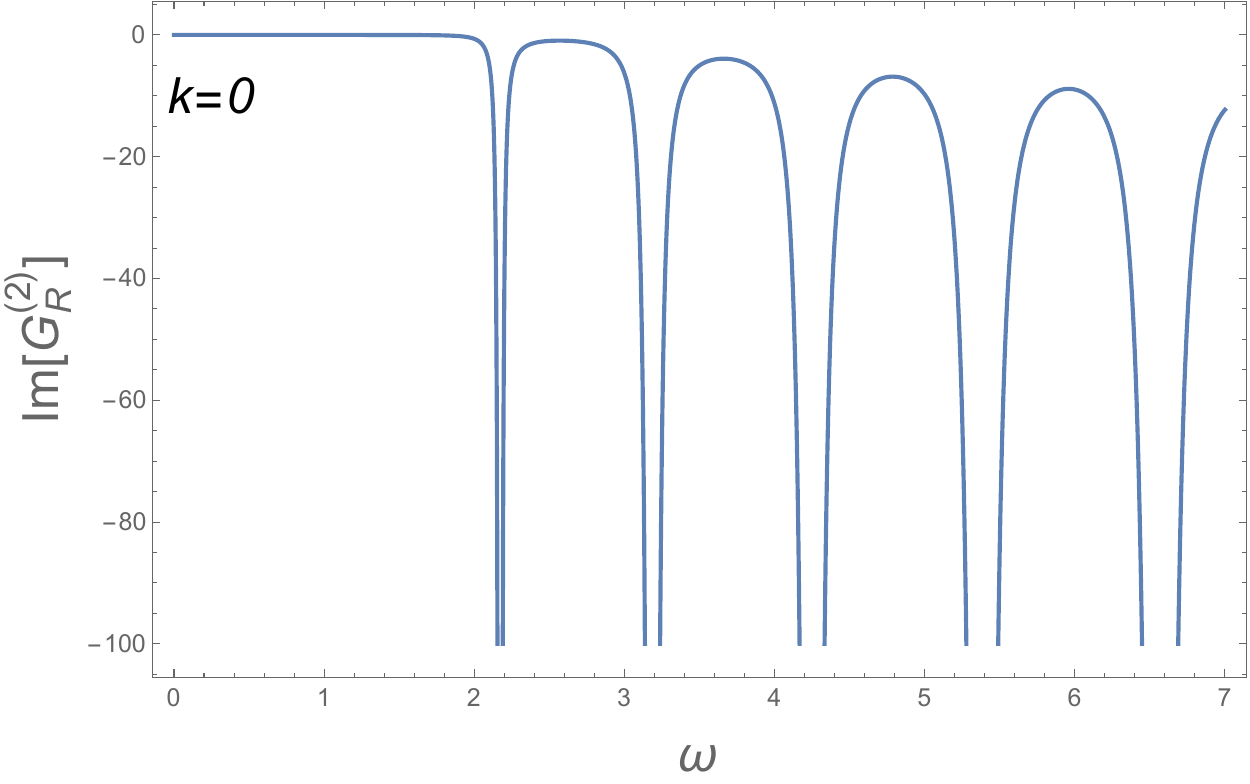}
\par\end{centering}
\caption{\label{fig:5} The confined retarded Green function $G_{R}^{\left(1,2\right)}$
as the spectral function of 1+3 dimensional QCD from the D4/D8 model.
The parameters are chosen as $\Lambda_{l}=2,\mathrm{k}=0,\mathcal{T}_{c}=1$
in the unit of $M_{KK}=1$.}
\end{figure}

\subsubsection*{Deconfined phase}

In the deconfined phase, the equation of motion for $\psi$ reads
from (\ref{eq:41}) as, 

\begin{equation}
\left(-\frac{4}{3}\pi TK^{-1/2}\Lambda_{l}+Z^{-1}K^{-1/6}\boldsymbol{\gamma}^{0}\partial_{0}+K^{-2/3}\boldsymbol{\gamma}^{i}\partial_{i}+2\pi T\boldsymbol{\gamma}\partial_{Z}+\pi TZK^{-1}\boldsymbol{\gamma}\right)\psi=0.\label{eq:64}
\end{equation}
Using the ansatz

\begin{equation}
\psi=e^{ik\cdot x}\left(\begin{array}{c}
\psi_{R}\\
\psi_{L}
\end{array}\right),\psi_{R,L}=\left(\begin{array}{c}
\psi_{R,L}^{\left(1\right)}\\
\psi_{R,L}^{\left(2\right)}
\end{array}\right),
\end{equation}
with $k_{\mu}=\left(-\omega,\mathrm{k},0,0\right)$, (\ref{eq:64})
reduces to

\begin{align}
\partial_{Z}\psi_{R}^{\left(1,2\right)}+\left(\frac{ZK^{-1}}{2}-\frac{2}{3}K^{-1/2}\Lambda_{l}\right)\psi_{R}^{\left(1,2\right)}+\left(\frac{\omega}{2\pi T}Z^{-1}K^{-1/6}+\frac{h\cdot\mathrm{k}}{2\pi T}K^{-2/3}\right)\psi_{L}^{\left(1,2\right)} & =0\nonumber \\
\partial_{Z}\psi_{L}^{\left(1,2\right)}+\left(\frac{ZK^{-1}}{2}+\frac{2}{3}K^{-1/2}\Lambda_{l}\right)\psi_{L}^{\left(1,2\right)}+\left(\frac{h\cdot\mathrm{k}}{2\pi T}K^{-2/3}-\frac{\omega}{2\pi T}Z^{-1}K^{-1/6}\right)\psi_{R}^{\left(1,2\right)} & =0,\label{eq:66}
\end{align}
which can be solved analytically at $Z\rightarrow\infty$ as,

\begin{align}
\psi_{R}^{\left(1,2\right)} & =AZ^{\frac{2}{3}\Lambda_{l}-\frac{1}{2}}+BZ^{-\frac{2}{3}\Lambda_{l}-\frac{5}{6}},\nonumber \\
\psi_{L}^{\left(1,2\right)} & =CZ^{-\frac{2}{3}\Lambda_{l}-\frac{1}{2}}+DZ^{\frac{2}{3}\Lambda_{l}-\frac{5}{6}}.\label{eq:67}
\end{align}
The asymptotics (\ref{eq:67}) leads to that the onshell action (\ref{eq:41})
takes form at $Z\rightarrow\infty$ as,

\begin{align}
S_{f,d}^{\mathrm{D8}} & \supseteq2\pi T\int d^{4}xdZi\bar{\psi}\boldsymbol{\gamma}\partial_{Z}\psi=-2\pi T\int d^{4}x\psi_{L}^{\dagger}\psi_{R}+...\nonumber \\
 & =-2\pi T\int d^{4}xD^{\dagger}AZ^{\frac{4}{3}\Lambda_{l}-\frac{4}{3}},
\end{align}
which however does not include any finite part for $\Lambda_{l}\geq2$.
Although this behavior might imply the deconfined hot QCD is probably
non-renormalizable in holography, it is possible to evaluate its correlation
function with the prescription in Section 3.3 if we take the boundary
value $\psi_{0}$ of $\psi$ carefully as

\begin{equation}
\psi_{0}=\lim_{Z\rightarrow+\epsilon^{-1}}\epsilon^{\frac{1}{6}}\psi\rightarrow\epsilon^{\frac{1}{6}}\left(\begin{array}{c}
\psi_{R}\\
0
\end{array}\right)=\epsilon^{-\frac{2}{3}\Lambda_{l}+\frac{2}{3}}\left(\begin{array}{c}
A\\
0
\end{array}\right),\ \epsilon\rightarrow0
\end{equation}
Thus conjugate momentum for $\psi_{0}$ is

\begin{equation}
\Pi_{0}=-\frac{\delta S_{f,d}^{\mathrm{D8}}}{\delta\psi_{0}}=\mathcal{T}_{d}\left(2\pi T\right)\left(0,D^{\dagger}\right)\epsilon^{-\frac{2}{3}\Lambda_{l}+\frac{2}{3}},
\end{equation}
which reduces to finite Green function as $G_{R}\sim D/A$. Then let
us use the ratio given in (\ref{eq:58}) so that the Green function
can be written as,

\begin{equation}
G_{R}^{\left(1,2\right)}=\mathcal{T}_{d}\left(2\pi T\right)\lim_{Z\rightarrow+\infty}Z^{1/3}\xi_{1,2}\left(Z\right).
\end{equation}
And the equation for $\xi_{1,2}$ can be obtained from (\ref{eq:66})
as,

\begin{equation}
\xi_{1,2}^{\prime}=-\frac{4}{3}K^{-1/2}\Lambda_{l}\xi_{1,2}-\frac{h\cdot\mathrm{k}}{2\pi T}K^{-2/3}+\frac{\omega}{2\pi T}Z^{-1}K^{-1/6}+\left(\frac{\omega}{2\pi T}Z^{-1}K^{-1/6}+\frac{h\cdot\mathrm{k}}{2\pi T}K^{-2/3}\right)\xi_{1,2}^{2},\label{eq:72}
\end{equation}
which can be solved numerically by using the incoming wave boundary
condition $\xi_{1,2}\left(0\right)=\pm i$ on the horizon.

\subsection{The numerical analysis}

In this section, we evaluate numerically the retarded Green functions
by solving (\ref{eq:60}) (\ref{eq:72}) with incoming wave boundary
conditions.

\begin{figure}[h]
\begin{centering}
\includegraphics[scale=0.3]{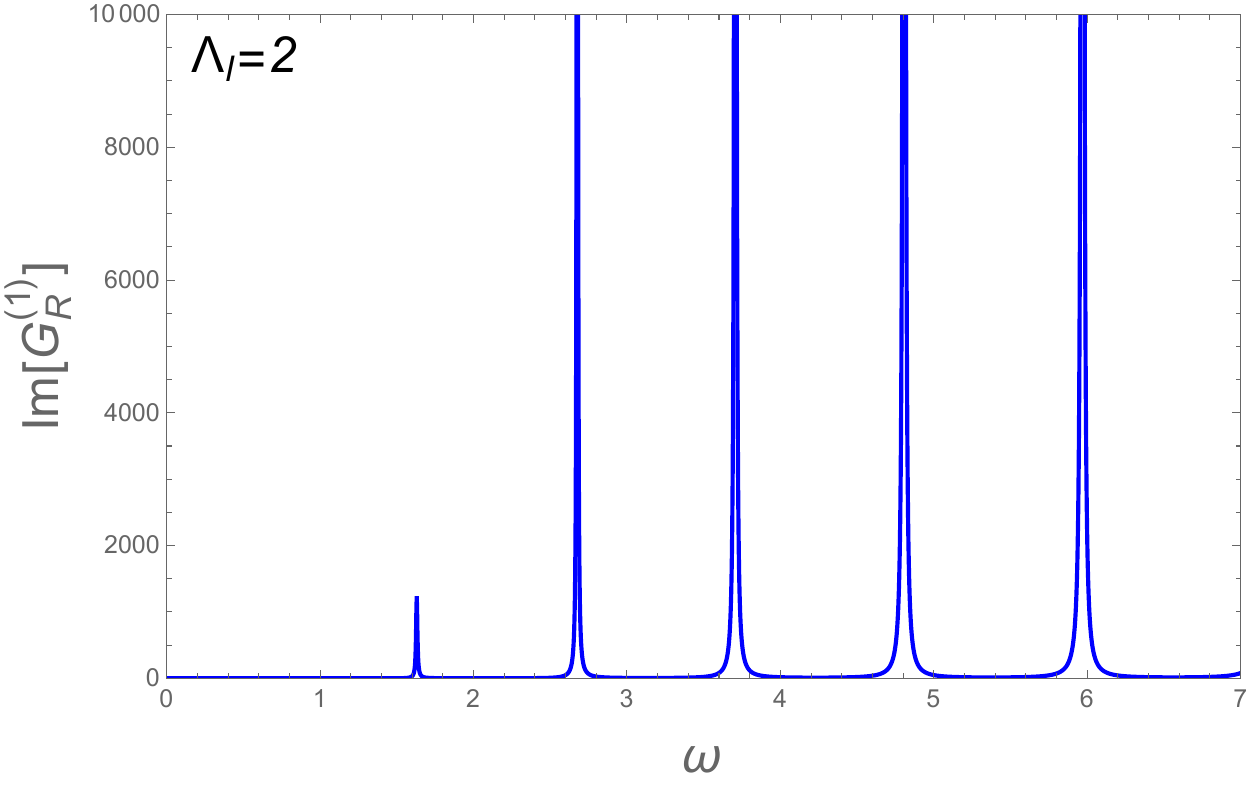}\includegraphics[scale=0.3]{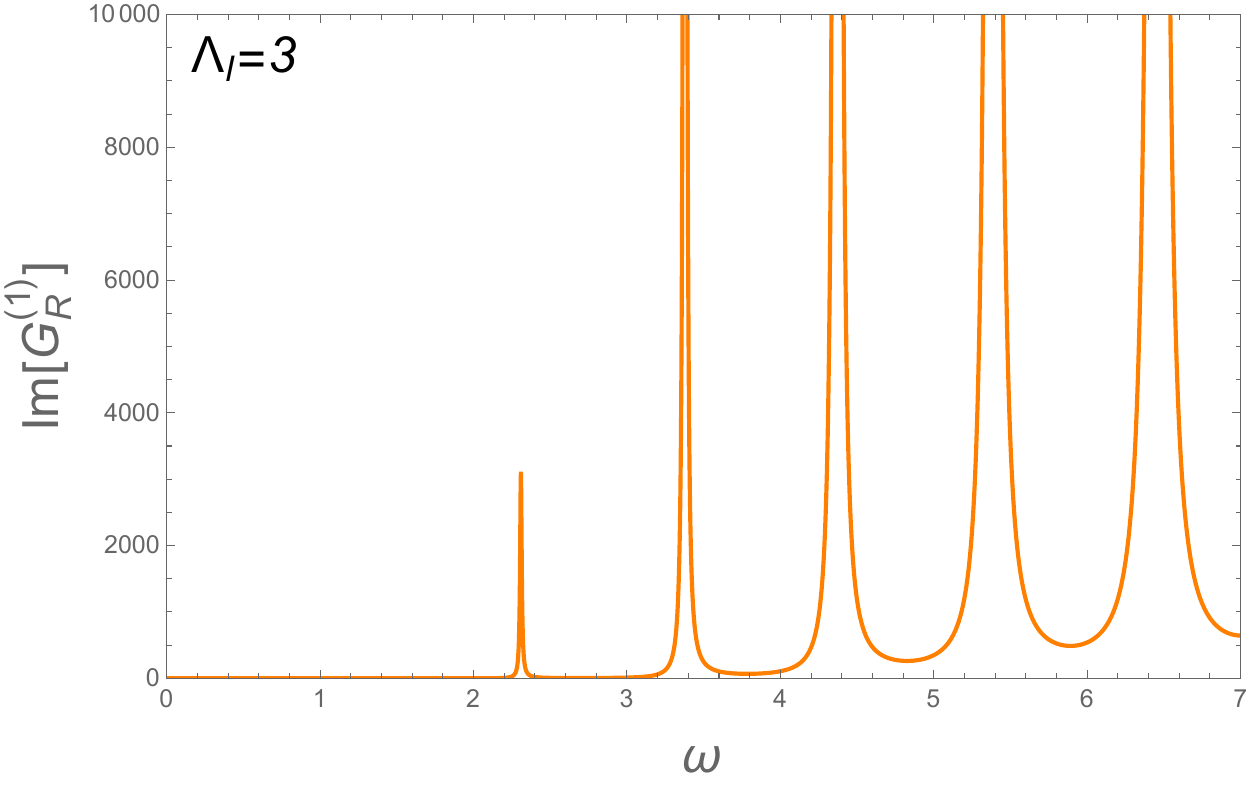}
\par\end{centering}
\begin{centering}
\includegraphics[scale=0.3]{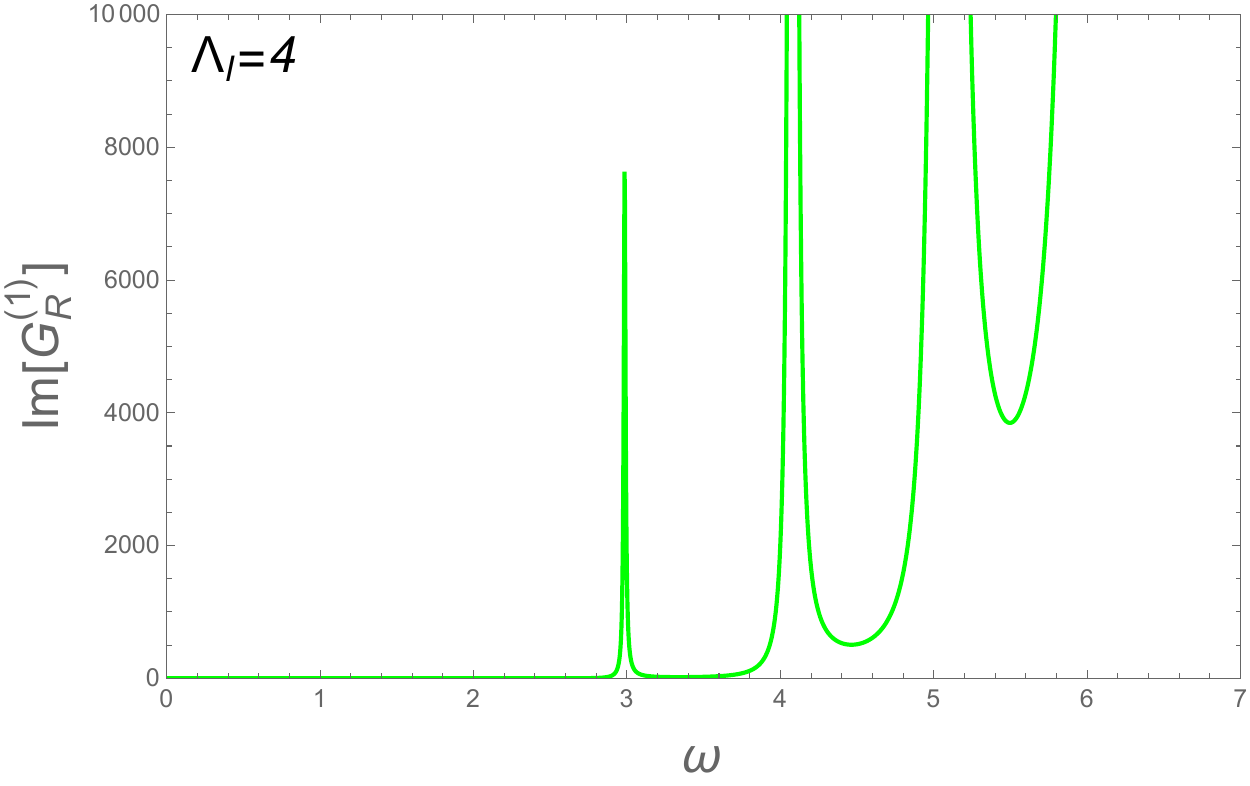}\includegraphics[scale=0.3]{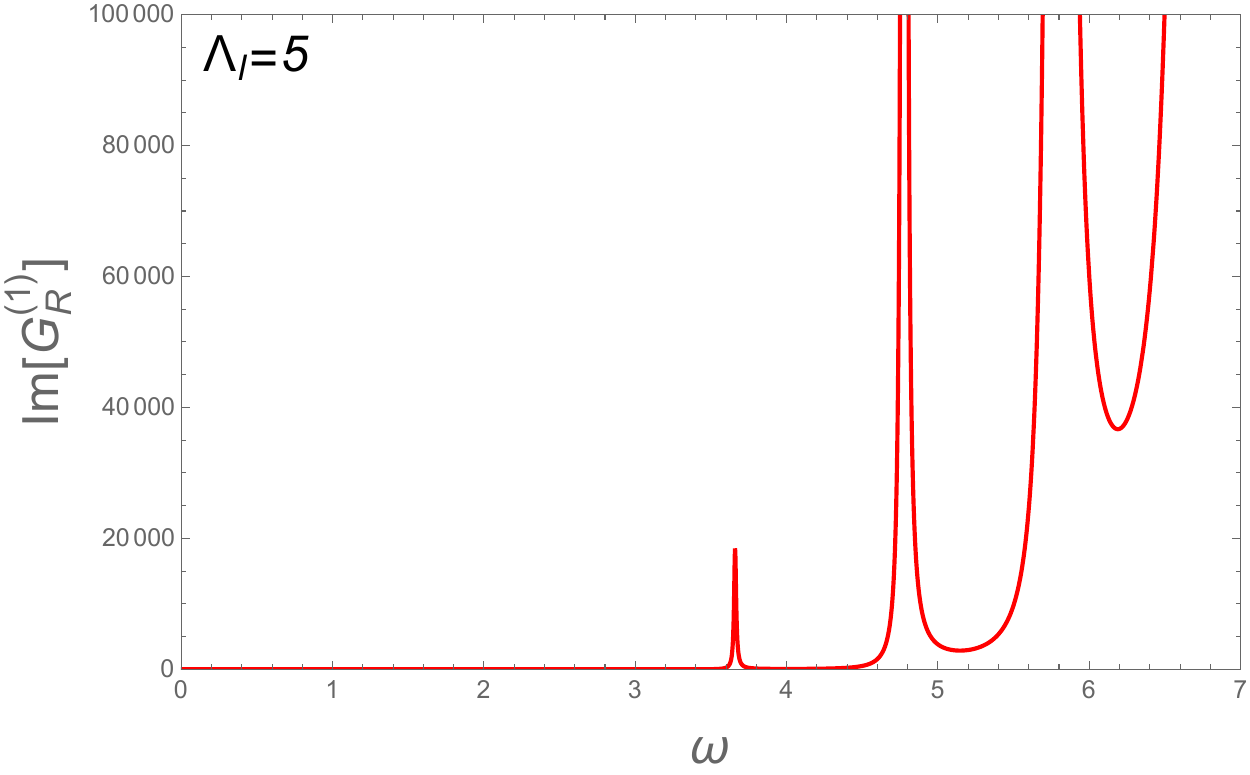}
\par\end{centering}
\caption{\label{fig:6} The imaginary part of confined retarded Green function
$G_{R}^{\left(1\right)}$ as the spectral function from the D4/D8
model with various $\Lambda_{l}$ for $\mathrm{k}=0$.}

\end{figure}

In confined phase, the results are collected in Figure \ref{fig:4}
and \ref{fig:5}. The behavior of the peaks in the Green function
displays discreteness as $\omega^{2}-k^{2}=M_{n}^{2}$ basically where
$M_{n}$ represents various constants. According to the general form
of the fermionic propagator
\begin{equation}
G_{R}\left(\omega,\vec{k}\right)=\frac{1}{ik_{\mu}\boldsymbol{\gamma}^{\mu}-M_{n}},\ M_{n}=m_{n}-\Sigma\left(k\right),\label{eq:73}
\end{equation}
here $m_{n},\Sigma\left(k\right)$ refers respectively to the eigenvalue
of the bare mass and self-energy, $M_{n}$ corresponds to the onshell
energy of the various bound states consisted of the flavored fermions
(quarks). And we believe this result is consistent with the property
of the bubble background (\ref{eq:22}) since its dual theory is expected
to be QCD with confinement states, e.g. meson, baryon states. We will
further discuss these bound states at the end of this manuscript.
Moreover, for $\omega>0$, the location of the peak in the Green function
can be read approximately once we set $\mathrm{k}=0$, so that the
peaks correspond to $\omega=m_{n}-\Sigma\left(k\right)$ as it is
illustrated in Figure \ref{fig:5}. In this sense, we find quantitatively,
for $\Lambda_{l}=2,l=0$, the bound energy (the position of the peak)
is located at $\omega\simeq1.6,2.7,3.7...$ in $G_{R}^{\left(1\right)}$
and at $\omega\simeq2.2,3.1,4.2...$ in $G_{R}^{\left(2\right)}$.
Furthermore, the location of the peaks also depends on the value of
$\Lambda_{l}$ as it is displayed in Figure \ref{fig:6}. 
 Accordingly, if we identify $h=\mp1$ as the parity of the bound
states, the onshell energy evaluated by the Green function agrees
consistently to the spectrum of the worldvolume fermion on the D8-branes
with various $\Lambda_{l}$ in this model as it is calculated in \cite{key-36}. 

\begin{figure}
\begin{centering}
\includegraphics[scale=0.5]{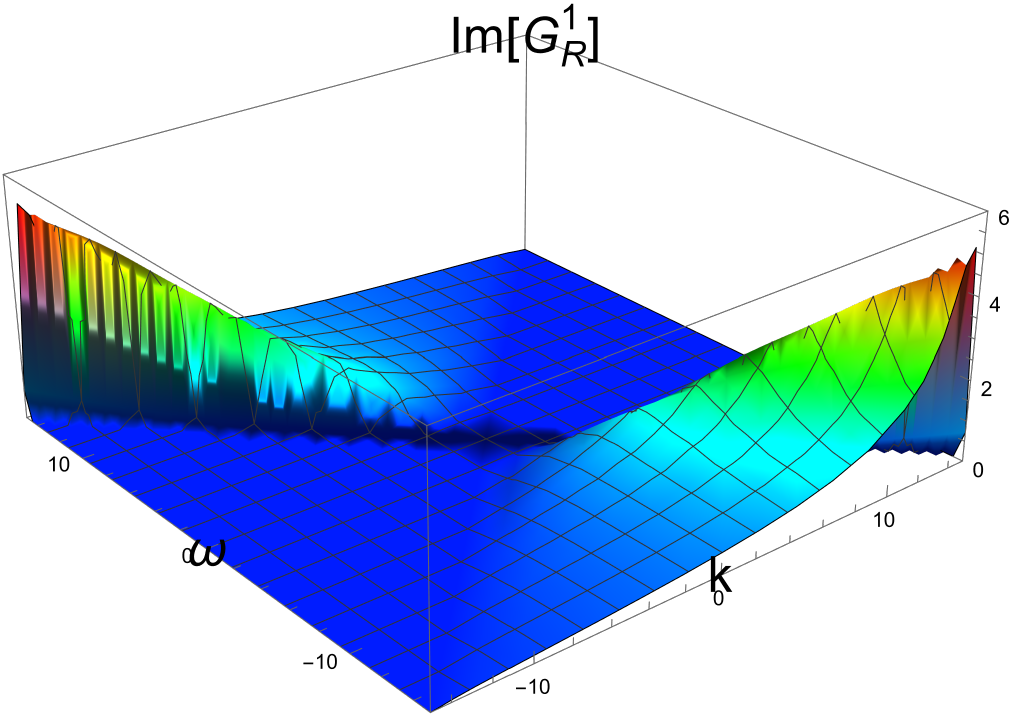}\includegraphics[scale=0.5]{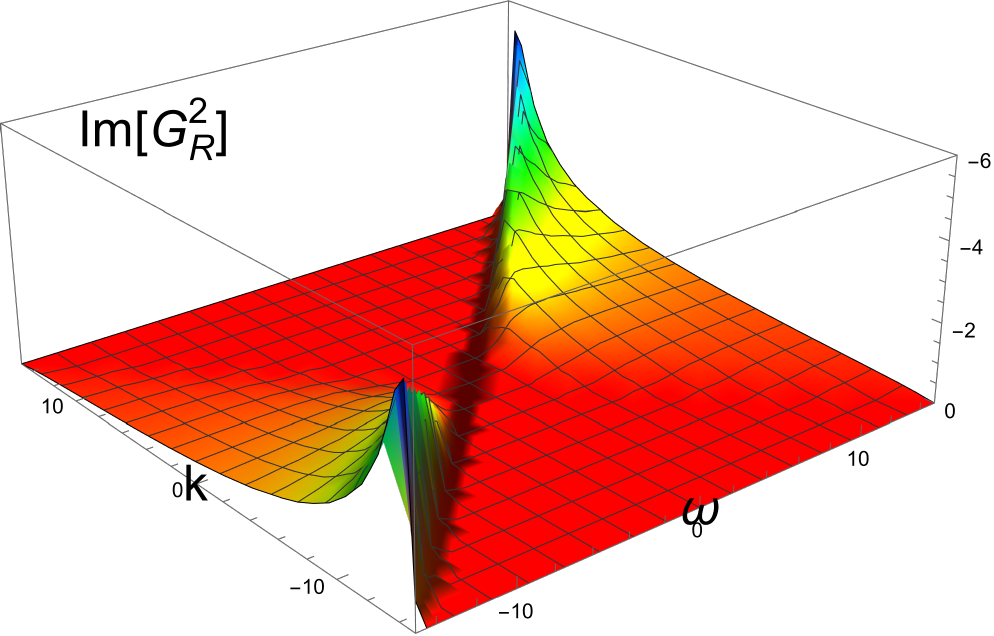}
\par\end{centering}
\caption{\label{fig:7} The 3d plot of the imaginary part of the Green function
$G_{R}^{\left(1,2\right)}$ from the black D4-brane background. The
parameters are chosen as $\Lambda_{l}=2,l=0,\mathcal{T}_{d}=1$ in
the unit of $2\pi T=1$.}

\end{figure}

\begin{figure}
\begin{centering}
\includegraphics[scale=0.35]{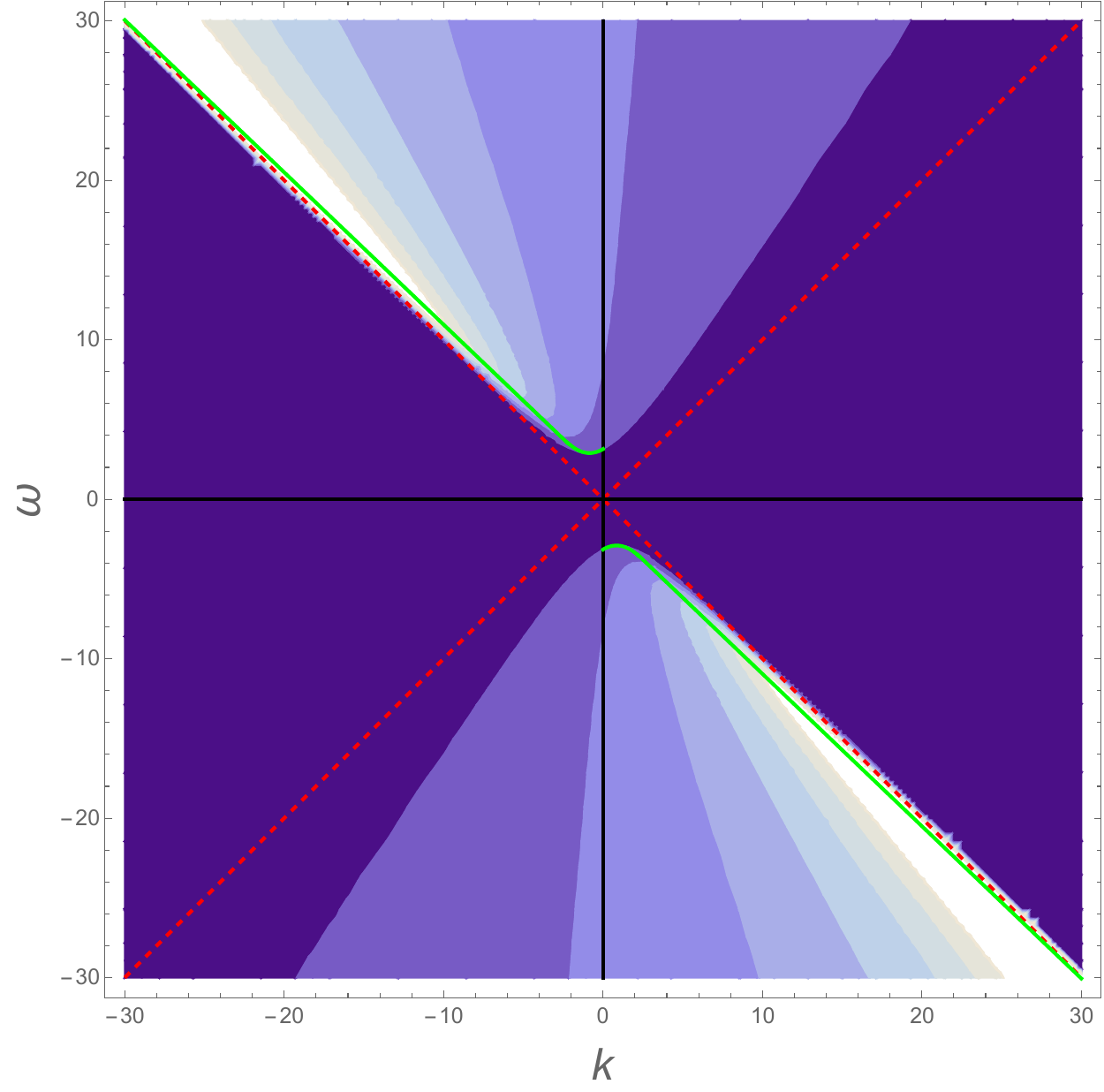}\includegraphics[scale=0.35]{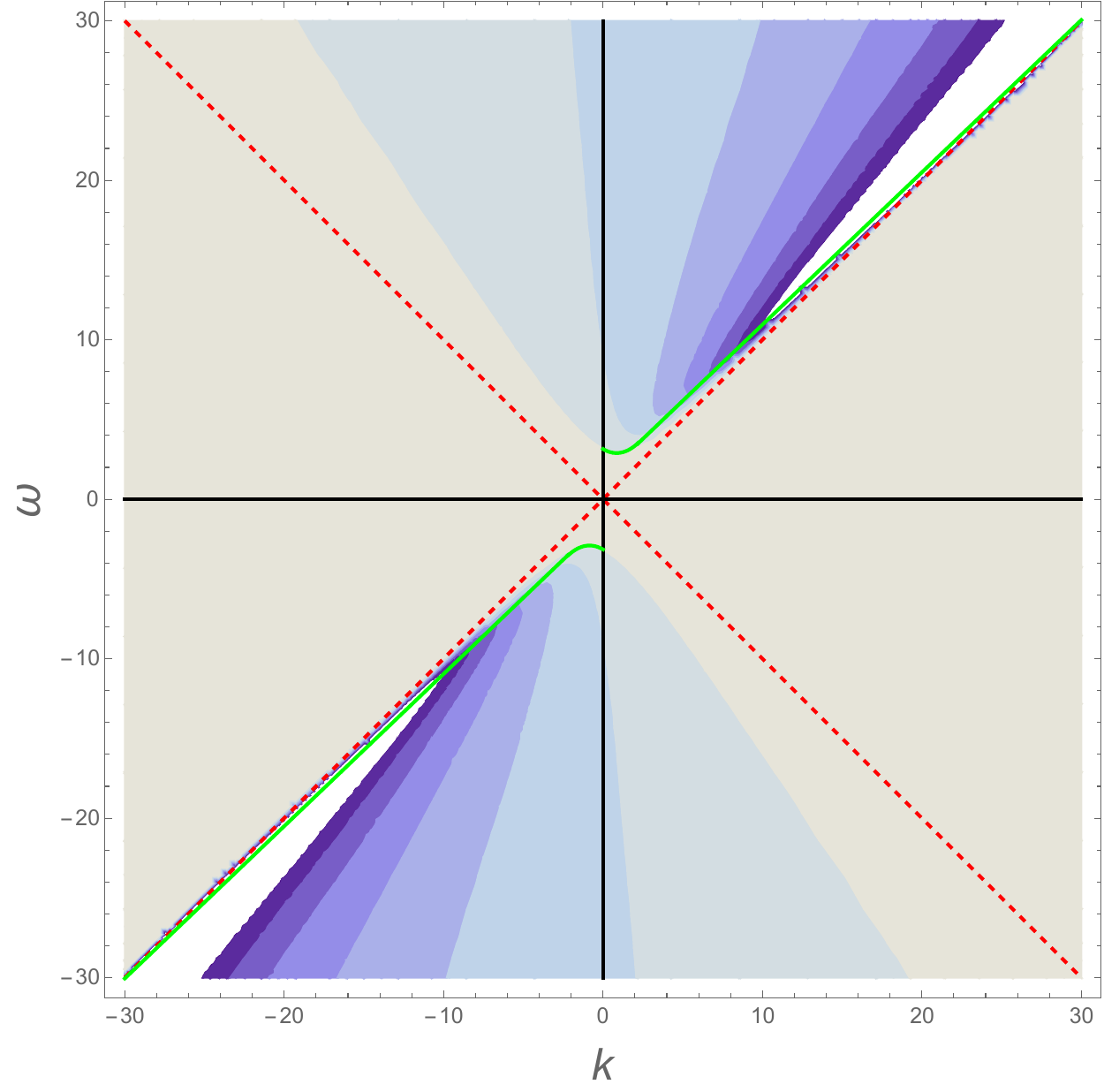}
\par\end{centering}
\caption{\label{fig:8} Density plot of the imaginary part of the Green function
$G_{R}^{\left(1,2\right)}$ from the black D4-brane background. The
white regions represent the protrusion or pit, i.e. the peaks, in
the Green functions which are fitted by green lines. The dashed lines
refer to $\omega=\pm\mathrm{k}$ as the light cone and the parameters
are chosen as $\Lambda_{l}=2,l=0,\mathcal{T}_{d}=1$ in the unit of
$2\pi T=1$.}

\end{figure}

Next, let us take a look at the Green function as the spectral function
obtained in the deconfined phase illustrated in Figure \ref{fig:7}
and \ref{fig:8} where the peaks in the Green function have been fitted
by green lines. 
 Since the relation of $G_{R}^{\left(1,2\right)}$ is given as $G_{R}^{\left(2\right)}\left(\omega,\mathrm{k}\right)=-G_{R}^{\left(1\right)}\left(\omega,-\mathrm{k}\right)$,
let us focus on $G_{R}^{\left(2\right)}\left(\omega,\mathrm{k}\right)$
for convenience. First, we can see the behaviors of the particle onshell
(peaks in the Green function) is very different from them in the confined
case which satisfies $\omega\simeq\mathrm{k}$ for $\mathrm{k}\gg1$
and $\omega\simeq\mathrm{k}^{2}+...$ for $\mathrm{k}\ll1$. While
we discuss the positive frequency mode, it is same to the negative
frequency mode. This behavior is in qualitative agreement with the
dispersion curves of fermions obtained by the hard thermal loop (HTL)
approximation in hot QCD, up to one-loop calculation, as \cite{key-44},

\begin{align}
\omega\left(\mathrm{k}\right) & \simeq m_{f}\pm\frac{1}{3}\mathrm{k}+\frac{1}{3m_{f}}\mathrm{k}^{2},\ \mathrm{k}\ll1;\nonumber \\
\omega\left(\mathrm{k}\right) & \simeq\mathrm{k},\ \mathrm{k}\gg1,\label{eq:74}
\end{align}
where $m_{f}$ is the effective mass of fermion generated by the medium
effect given by

\begin{equation}
m_{f}=\sqrt{\frac{C_{F}}{8}}g_{\mathrm{YM}}T.
\end{equation}
We note that chemical potential can also contribute to mass of fermion
while it is not turned on in the current holographic setup and $C_{F}$
is suggested to be $C_{F}=4/3$ for fundamental quarks. According
to our numerical calculations presented in Figure \ref{fig:8}, the
effective mass of fermion is evaluated to be $m_{f}=\omega\left(\mathrm{k}=0\right)\simeq2.9\times\left(2\pi T\right)$
which is expected to be the non-perturbative results predicted by
holography. However, we must keep in mind that the holographic approach
is valid in strong coupling region i.e. the 't Hooft coupling constant
$\lambda=g_{\mathrm{YM}}^{2}N_{c}$ is finite and satisfies $\lambda\gg1$
for $N_{c}\rightarrow\infty$. Hence the consistent interpretation
here should be that, in QFT side, the high order contribution of Yang-Mills
coupling constant $g_{\mathrm{YM}}$ by HTL perturbation is suppressed
in the large $N_{c}$ limit since in this limit we have $g_{\mathrm{YM}}\ll1$.
Second, our current numerical evaluation of the the dispersion curves
implies the minus sign should be picked up in (\ref{eq:74}) for $\mathrm{k}\ll1$
while there are two branches given by HTL approximation in (\ref{eq:74}).
And following the discussion in \cite{key-30,key-31}, this branch
could correspond to the dispersion curves of the fermionic plasmino
at high temperature. Nonetheless, since the numerical calculation
always displays a width of the peaks in the spectral function, the
dispersion curves of quarks (which is expected to be $\omega\left(\mathrm{k}\right)\simeq m_{f}+\frac{1}{3}\mathrm{k}+\frac{1}{3m_{f}}\mathrm{k}^{2}$
for $\mathrm{k}\ll1$) may also be included in the spectral function.
However, distinguishing exactly the dispersion curves of fundamental
quark from plasmino would very depend on the trick of the numerical
calculation we chose. In this sense, it would be less necessary to
further compare our current numerical results with the HTL approximation.
Overall, we believe our results, based on the holography with the
principle in string theory, reveal mostly the fundamental properties
of both confinement and deconfinement in QCD, while they are very
different from some bottom-up approaches or phenomenological approaches
in this model as \cite{key-30,key-31} where the features of QCD confinement
is less clear. 

\section{Approach to the D3/D7 model}

In order to further test the prescription in Section 3.3, let us attempt
to use it in the D3/D7 approach in this section, which is a holographic
version of 1+2 dimensional QCD. Similarly as the case of the D4/D8
approach, in Section 5.1, we perform the Step 1 in the prescription
to obtain a 4d effective fermionic action with respect to the bubble
and black D3 background. In Section 4.2, we perform Step 2 and Step
3 which is to solve the equations of motion for the bulk field, define
its boundary value and derive its onshell action in order to evaluate
the correlation function.

\subsection{The 4d action for the bulk spinor}

We start with the SUGRA solutions (\ref{eq:4}) and (\ref{eq:2})
in the case of $p=3$ for the IIB SUGRA. So in the near horizon limit,
the bubble D3-brane background is summarized as\footnote{In the IIB SUGRA, the R-R 5-form is usually defined to be self-dual
which is replaced as $F_{5}\rightarrow\frac{1}{2}\left(F_{5}+\star F_{5}\right)$.},

\begin{align}
ds_{c}^{2} & =\frac{r^{2}}{R^{2}}\left[\eta_{\alpha\beta}dx^{\alpha}dx^{\beta}+f\left(r\right)\left(dx^{3}\right)^{2}\right]+\frac{R^{2}}{r^{2}}\left[\frac{dr^{2}}{f\left(r\right)}+r^{2}d\Omega_{5}^{2}\right],\nonumber \\
f\left(r\right) & =1-\frac{r_{KK}^{4}}{r^{4}},\!F_{5}=dC_{4}=4R^{4}g_{s}^{-1}\epsilon_{5},\ \alpha,\beta=0,1,2,\label{eq:76}
\end{align}
which corresponds to the confined phase of the dual theory and the
black D3-brane background is given as,

\begin{align}
ds_{d}^{2} & =\frac{r^{2}}{R^{2}}\left[-f_{T}\left(r\right)dt^{2}+\left(dx^{1}\right)^{2}+\left(dx^{2}\right)^{2}+\left(dx^{3}\right)^{2}\right]+\frac{R^{2}}{r^{2}}\left[\frac{dr^{2}}{f\left(r\right)}+r^{2}d\Omega_{5}^{2}\right],\nonumber \\
f_{T}\left(r\right) & =1-\frac{r_{H}^{4}}{r^{4}},\!F_{5}=dC_{4}=4R^{4}g_{s}^{-1}\epsilon_{5},\label{eq:77}
\end{align}
which corresponds to the deconfined phase of the dual theory. Here
we use $\epsilon_{5}$ to refer to the volume form of a unit $S^{5}$
and $R^{4}=4\pi g_{s}N_{c}l_{s}^{4}$. Note that the dilaton field
vanishes for $p=3$ due to (\ref{eq:2}) and since $x^{3}$ is compactified
on a circle, the dual field theory is effectively the 1+2 dimensional
Yang-Mills theory below the energy scale $M_{KK}=2\pi\beta_{3}^{-1}$
where $\beta_{3}$ refers to the period of $x^{3}$ \cite{key-14,key-38}.

Then let us introduce the probe D7-branes as flavors embedded into
the background (\ref{eq:76}) (\ref{eq:77}) in order to investigate
its worldvolume fermions. According to Table \ref{tab:1}, the D7-branes
are located at $x^{3,9}=\mathrm{const}$, hence we impose the coordinate
transformation,

\begin{equation}
r=\frac{r_{KK,H}}{\sqrt{2}}\left(\frac{1}{\zeta^{2}}+\zeta^{2}\right)^{1/2},\zeta^{2}=\rho^{2}+u^{2},u\equiv x^{9},
\end{equation}
to the radial and $S^{5}$ part in (\ref{eq:76}) (\ref{eq:77}),
then they can be rewritten as,

\begin{equation}
\frac{R^{2}}{r^{2}}\left[\frac{dr^{2}}{f\left(r\right)}+r^{2}d\Omega_{5}^{2}\right]=\frac{R^{2}}{\zeta^{2}}\left(d\rho^{2}+\rho^{2}d\Omega_{4}^{2}+du^{2}\right),
\end{equation}
where $\rho^{2}=\sum_{m=4}^{8}x^{m}x_{m}$ and $d\Omega_{4}^{2}$
refers to the metric on $S^{4}$ whose solid angle is defined by parametrizing
$x^{4,5...8}$. We note that due to $r\geq r_{KK,H}$, it reduces
to two branches for $\zeta$ equivalently as $0<\zeta\leq1$ and $\zeta\geq1$.
In our setup, we chose the branch of $\zeta\geq1$. Therefore the
induced metric on the flavor D7-branes reads,

\begin{align}
ds_{\mathrm{D7},c}^{2} & =\frac{r^{2}}{R^{2}}\eta_{\alpha\beta}dx^{\alpha}dx^{\beta}+\frac{R^{2}}{\zeta^{2}}\left(d\rho^{2}+\rho^{2}d\Omega_{4}^{2}\right),\ \alpha,\beta=0,1,2,\nonumber \\
ds_{\mathrm{D7},d}^{2} & =\frac{r^{2}}{R^{2}}\left[-f_{T}\left(r\right)dt^{2}+\left(dx^{1}\right)^{2}+\left(dx^{2}\right)^{2}\right]+\frac{R^{2}}{\zeta^{2}}\left(d\rho^{2}+\rho^{2}d\Omega_{4}^{2}\right).\label{eq:80}
\end{align}
We note that $u=L$ as a constant represents the separation of the
D3- and D7-branes and it is proportional to the bare mass of fundamental
quarks in the D3/D7 approach \cite{key-14} as the vacuum expectation
value (VEV) of a $\left(3,7\right)$ string. For chirally symmetric
dual theory, we may simply set $L=0$ which implies the fundamental
quarks are massless. Keeping these in mind, the action for the worldvolume
fermions on the D7-brane is given by (\ref{eq:11}) (\ref{eq:13})
as,

\begin{equation}
S_{f}^{\mathrm{D7}}=\frac{iT_{7}}{2}\int d^{8}x\sqrt{-g}\bar{\Psi}P_{-}\left(\Gamma^{\alpha}\nabla_{\alpha}-\frac{1}{2\cdot8\cdot5!}F_{KLMNP}\Gamma^{\alpha}\Gamma^{KLMNP}\Gamma_{\alpha}\right)\Psi,\label{eq:81}
\end{equation}
where we have imposed $\bar{\gamma}\bar{\Psi}=\bar{\Psi}$ and $F_{KLMNP}$
refers to the component of the R-R 5-form presented in (\ref{eq:76})
(\ref{eq:77}). After some straightforward calculations, the action
(\ref{eq:81}) can be written as,

\begin{align}
S_{f,c}^{\mathrm{D7}}= & \frac{i}{2}T_{7}R^{2}\int d^{3}xd\rho d\Omega_{4}\bar{\Psi}P_{-}\bigg(\frac{\rho^{4}r^{2}R}{\zeta^{5}}\gamma^{\alpha}\partial_{\alpha}+\frac{3\rho^{5}r^{3}}{2\zeta^{6}}\sqrt{f}\gamma^{\rho}+\frac{\rho^{4}r^{3}}{\zeta^{4}R}\gamma^{\rho}\partial_{\rho}\nonumber \\
 & +\frac{\rho^{3}r^{3}}{\zeta^{4}R}\gamma^{m}D_{m}+\frac{\rho^{4}r^{3}}{\zeta^{5}R}\bigg)\Psi,\label{eq:82}\\
S_{f,d}^{\mathrm{D7}}= & \frac{i}{2}T_{7}R^{2}\int d^{3}xd\rho d\Omega_{4}\bar{\Psi}P_{-}\bigg[\frac{\rho^{4}r^{2}R}{\zeta^{5}}\gamma^{0}\partial_{0}+\frac{\rho^{4}\sqrt{f_{T}}r^{3}}{2\zeta^{5}R}\left(\frac{\zeta f^{\prime}}{2f}+\frac{3\rho\sqrt{f_{T}}}{\zeta}\right)\gamma^{4}\nonumber \\
 & +\frac{\rho^{4}\sqrt{f_{T}}r^{2}R}{\zeta^{5}}\gamma^{\alpha}\partial_{\alpha}+\frac{\rho^{4}\sqrt{f_{T}}r^{3}}{\zeta^{4}R}\gamma^{\rho}\partial_{\rho}+\frac{\rho^{3}\sqrt{f_{T}}r^{3}}{\zeta^{4}R}\gamma^{m}D_{m}+\frac{\rho^{4}\sqrt{f_{T}}r^{3}}{\zeta^{5}R}\bigg]\Psi,\label{eq:83}
\end{align}
where we have used $\gamma^{0123\rho}\Psi=\Psi$ and

\begin{equation}
P_{-}=\frac{1}{2}\left(1-\Gamma_{\mathrm{D7}}\right),
\end{equation}
with respect to the bubble (\ref{eq:76}) and black brane (\ref{eq:77})
background. Follow the decomposition presented in (\ref{eq:34}) -
(\ref{eq:36}) with the Eigen equation on $S^{4}$ (\ref{eq:37})
and integrate the $S^{4}$ part, the action (\ref{eq:82}) can be
further written as a 4d form as,

\begin{align}
S_{f,c}^{\mathrm{D7}}= & i\mathcal{T}\int d^{3}xd\rho\bar{\psi}\left[\frac{1}{\zeta}-\frac{1}{\rho}\Lambda_{l}+\frac{R^{2}}{r\zeta}\boldsymbol{\gamma}^{\alpha}\partial_{\alpha}+\left(\frac{2\rho}{\zeta^{2}}-\frac{2}{\rho}\right)\boldsymbol{\gamma}+\boldsymbol{\gamma}\partial_{\rho}\right]\psi,\label{eq:85}
\end{align}
by rescaling $\psi\rightarrow\left(2\pi\alpha^{\prime}\right)\psi\frac{\zeta^{2}}{\rho^{2}}\frac{1}{r^{3/2}}$
and action (\ref{eq:83}) can be written as,

\begin{align}
S_{f,d}^{\mathrm{D7}}= & i\mathcal{T}\int d^{3}xd\rho\bar{\psi}\bigg[\frac{1}{\sqrt{f_{T}}}\frac{R}{r\zeta}\boldsymbol{\gamma}^{0}\partial_{0}+\frac{R}{r\zeta}\boldsymbol{\gamma}^{\alpha}\partial_{\alpha}+\frac{1}{R}\boldsymbol{\gamma}\partial_{\rho}+\left(\frac{3\rho\sqrt{f_{T}}}{2R\zeta^{2}}+\frac{f_{T}^{\prime}}{4f_{T}R}\right)\boldsymbol{\gamma}\nonumber \\
 & +\frac{1}{R}\left(\frac{1}{\zeta}-\frac{f_{T}^{\prime}}{4f_{T}}\right)-\frac{1}{R\rho}\left(\Lambda_{l}+2\right)+\frac{\rho}{R\zeta^{2}}\left(\frac{5}{2}-\frac{3}{2}\sqrt{f_{T}}\right)\bigg]\psi,\label{eq:86}
\end{align}
by rescaling $\psi\rightarrow\left(2\pi\alpha^{\prime}\right)\psi\frac{\zeta^{2}}{\rho^{2}r^{3/2}f_{T}^{1/4}}$
where

\begin{equation}
\mathcal{T}=\frac{1}{2}T_{7}R^{2}\Omega_{4}\left(2\pi\alpha^{\prime}\right)^{2}.
\end{equation}
We will use these 4d fermionic actions (\ref{eq:85}) (\ref{eq:86})
to compute the two-point correlation function in the 2+1 dimensional
dual theory.

\subsection{Use the prescription}

Let us use the prescription in Section 3.3 to study the two-point
Green function for the 1+2 dimensional QCD in the D3/D7 model.

\subsubsection*{Confined phase}

For the confined phase, the equation of motion for the worldvolume
$\psi$ can be obtained by varying (\ref{eq:85}) as, 

\begin{equation}
\left[\frac{1}{\xi}-\frac{1}{\rho}\Lambda_{l}+\frac{R^{2}}{r\xi}\boldsymbol{\gamma}^{\alpha}\partial_{\alpha}+\left(\frac{2\rho}{\xi^{2}}-\frac{2}{\rho}\right)\boldsymbol{\gamma}+\boldsymbol{\gamma}\partial_{\rho}\right]\psi=0.
\end{equation}
Using the ansatz given in (\ref{eq:43}), we can obtain the coupled
equations as,

\begin{align}
\left(\frac{1}{\xi}-\frac{1}{\rho}\Lambda_{l}+\frac{2\rho}{\xi^{2}}-\frac{2}{\rho}+\partial_{\rho}\right)\psi_{R}+i\frac{R^{2}}{r\xi}\sigma^{\alpha}\partial_{\alpha}\psi_{L} & =0,\nonumber \\
i\frac{R^{2}}{r\xi}\bar{\sigma}^{\alpha}\partial_{\alpha}\psi_{R}+\left(\frac{1}{\xi}-\frac{1}{\rho}\Lambda_{l}-\frac{2\rho}{\xi^{2}}+\frac{2}{\rho}-\partial_{\rho}\right)\psi_{L} & =0.\label{eq:89}
\end{align}
Setting $k_{\mu}=\left(-\omega,\mathrm{k},0,0\right)$, the equations
presented in (\ref{eq:89}) can be written as decoupled second order
differential equations, which, at boundary $r\rightarrow\infty$ ($\rho\rightarrow\infty$),
take the asymptotic form as,

\begin{align}
\psi_{R}^{\prime\prime}+\frac{2}{\rho}\psi_{R}^{\prime}-\frac{\left(\Lambda_{l}^{2}+3\Lambda_{l}+2\right)}{\rho^{2}}\psi_{R} & =0,\nonumber \\
\psi_{L}^{\prime\prime}+\frac{2}{\rho}\psi_{L}^{\prime}-\frac{\left(\Lambda_{l}^{2}+\Lambda_{l}\right)}{\rho^{2}}\psi_{L} & =0.
\end{align}
And they can be solved exactly as,

\begin{align}
\psi_{R} & =A\rho^{1+\Lambda_{l}}+B\rho^{-2-\Lambda_{l}},\nonumber \\
\psi_{L} & =C\rho^{\Lambda_{l}}+D\rho^{-1-\Lambda_{l}},
\end{align}
where $A,B,C,D$ are integration constant. As we have outlined $\Lambda_{l}\geq2$,
so the boundary value $\psi_{0}$ of $\psi$ should be defined as,

\begin{equation}
\psi_{0}=\lim_{\rho\rightarrow\infty}\rho^{-1-\Lambda_{l}}\psi=\rho^{-1-\Lambda_{l}}\left(\begin{array}{c}
\psi_{R}\\
0
\end{array}\right)=\left(\begin{array}{c}
A\\
0
\end{array}\right).
\end{equation}
Then the onshell action (\ref{eq:85}) will include a finite part
as,

\begin{align}
S_{f,c}^{\mathrm{D7}} & \supseteq-\mathcal{T}\int d^{3}x\left[\left(\psi_{L}^{\dagger}\psi_{R}\right)|_{\rho\rightarrow\infty}+h.c+...\right]\nonumber \\
 & =-\mathcal{T}\int d^{3}x\left[C^{\dagger}A\rho^{2\Lambda_{l}+1}+D^{\dagger}A+h.c\right]|_{\rho\rightarrow\infty},\label{eq:93}
\end{align}
thus it request for a holographic counterterm as,

\begin{equation}
S_{ct}=\mathcal{T}\int d^{3}x\left[C^{\dagger}A\rho^{2\Lambda_{l}+1}+h.c\right]|_{\rho\rightarrow\infty}.
\end{equation}
The existing finite part in (\ref{eq:93}) is due to the residual
isometry of $\mathrm{AdS}_{5}$ since the D3-brane solution (\ref{eq:76})
(\ref{eq:77}) as bulk geometry is $\mathrm{AdS}_{5}\times S^{5}$
asymptotically. Follow the discussion in Section 3, 4, the retarded
Green function is given by,

\begin{equation}
\mathcal{T}D=G_{R}\left(\omega,\vec{k}\right)A.
\end{equation}
Using the representation (\ref{eq:57}) and introducing the ratio
(\ref{eq:58}), the Green function can be written as,

\begin{equation}
G_{R}^{\left(1,2\right)}=\lim_{\rho\rightarrow\infty}\rho^{2\Lambda_{l}+2}\xi_{1,2},
\end{equation}
where the equation for $\xi_{1,2}$ can be obtained from (\ref{eq:89})
as,

\begin{equation}
\xi_{1,2}^{\prime}=\frac{R^{2}}{r\zeta}\left(\omega-h\cdot\mathrm{k}\right)+2\left(\frac{1}{\zeta}-\frac{1}{\rho}\Lambda_{l}\right)\xi_{1}+\frac{R^{2}}{r\zeta}\left(\omega+h\cdot\mathrm{k}\right)\xi_{1,2}^{2},\label{eq:97}
\end{equation}
which will be evaluated numerically with the incoming wave boundary
condition and $h=\pm1$ for $\xi_{1,2}$ respectively. 
\begin{figure}[th]
\begin{centering}
\includegraphics[scale=0.27]{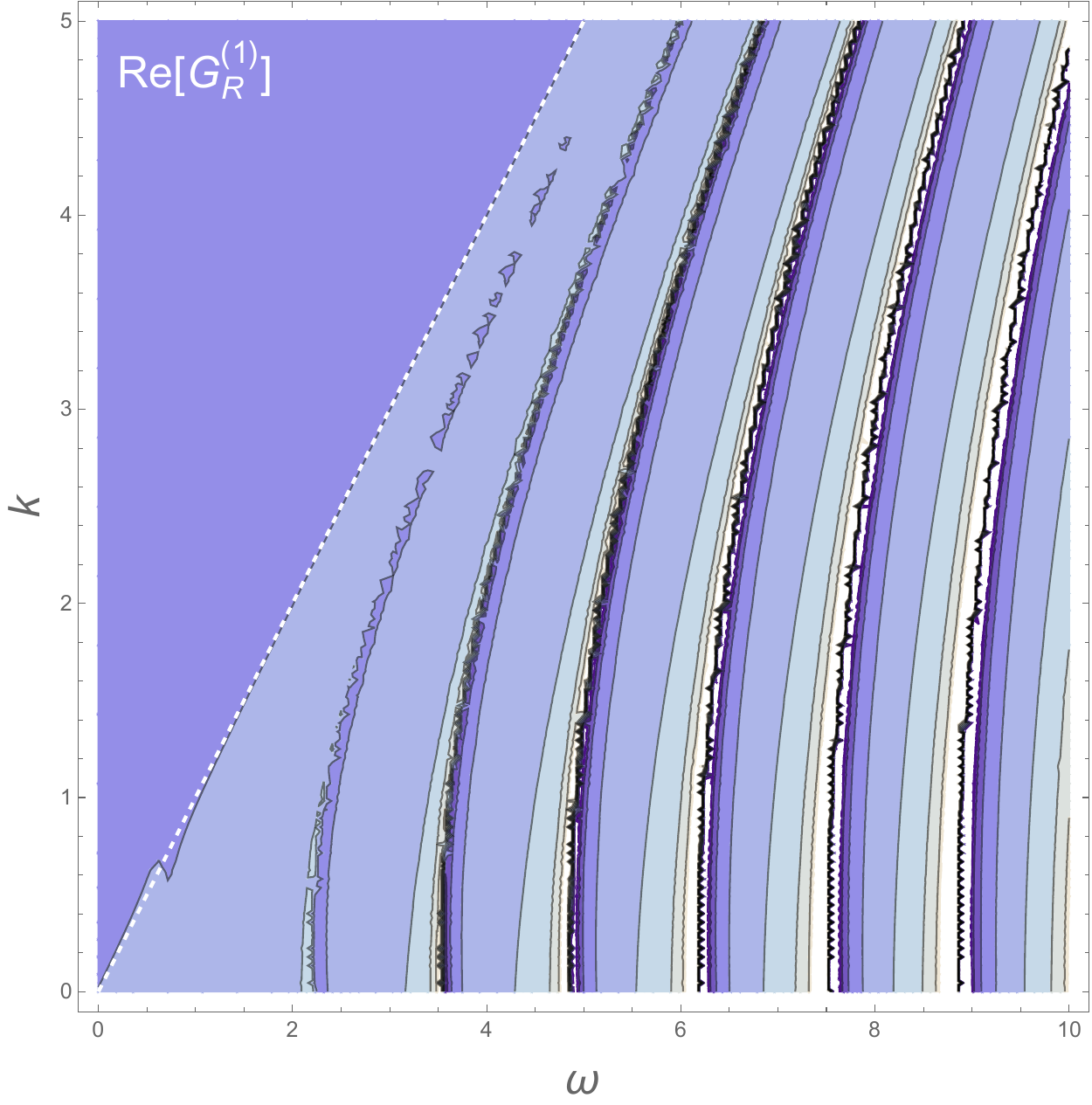}\includegraphics[scale=0.27]{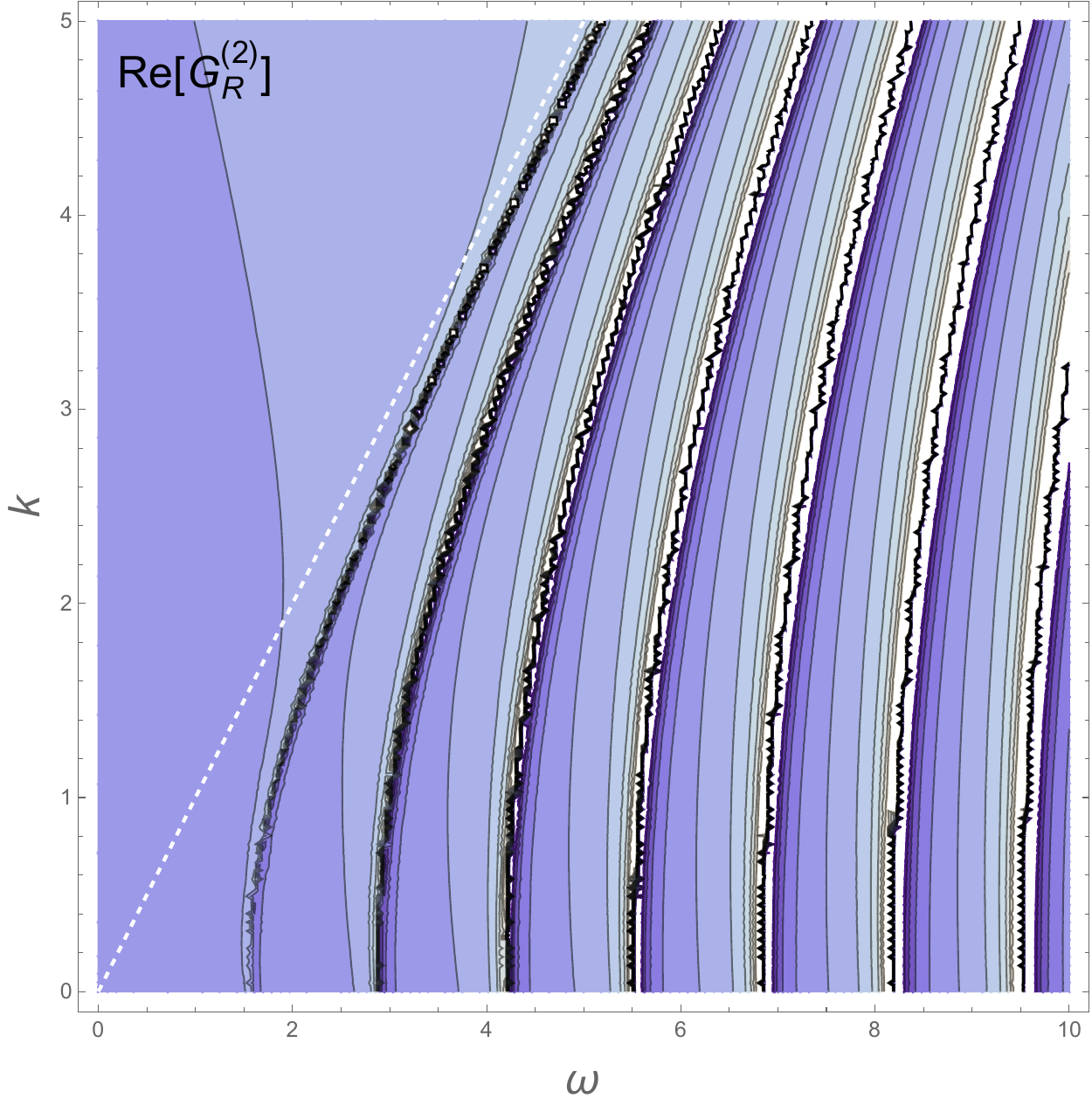}
\par\end{centering}
\begin{centering}
\includegraphics[scale=0.27]{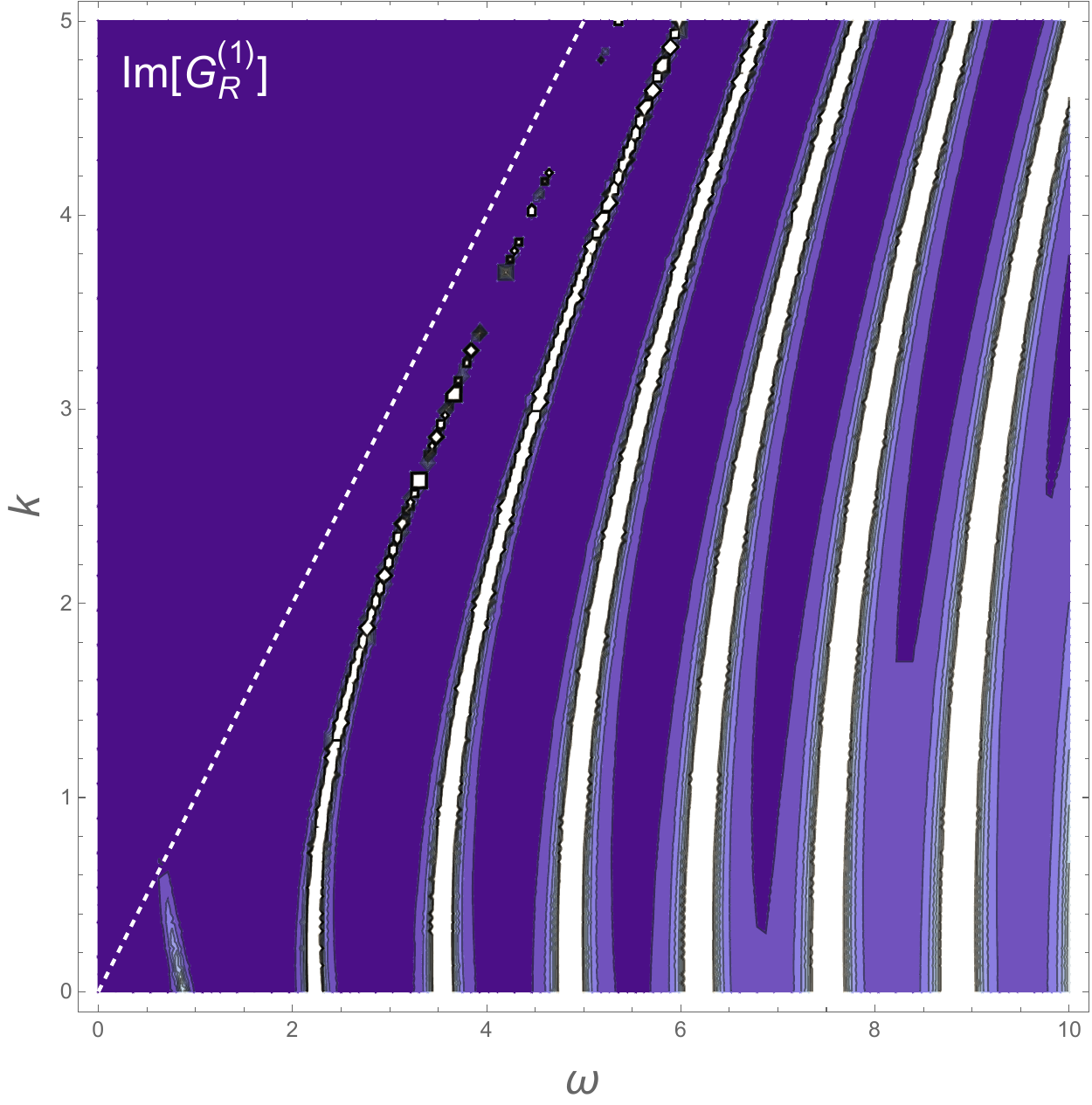}\includegraphics[scale=0.27]{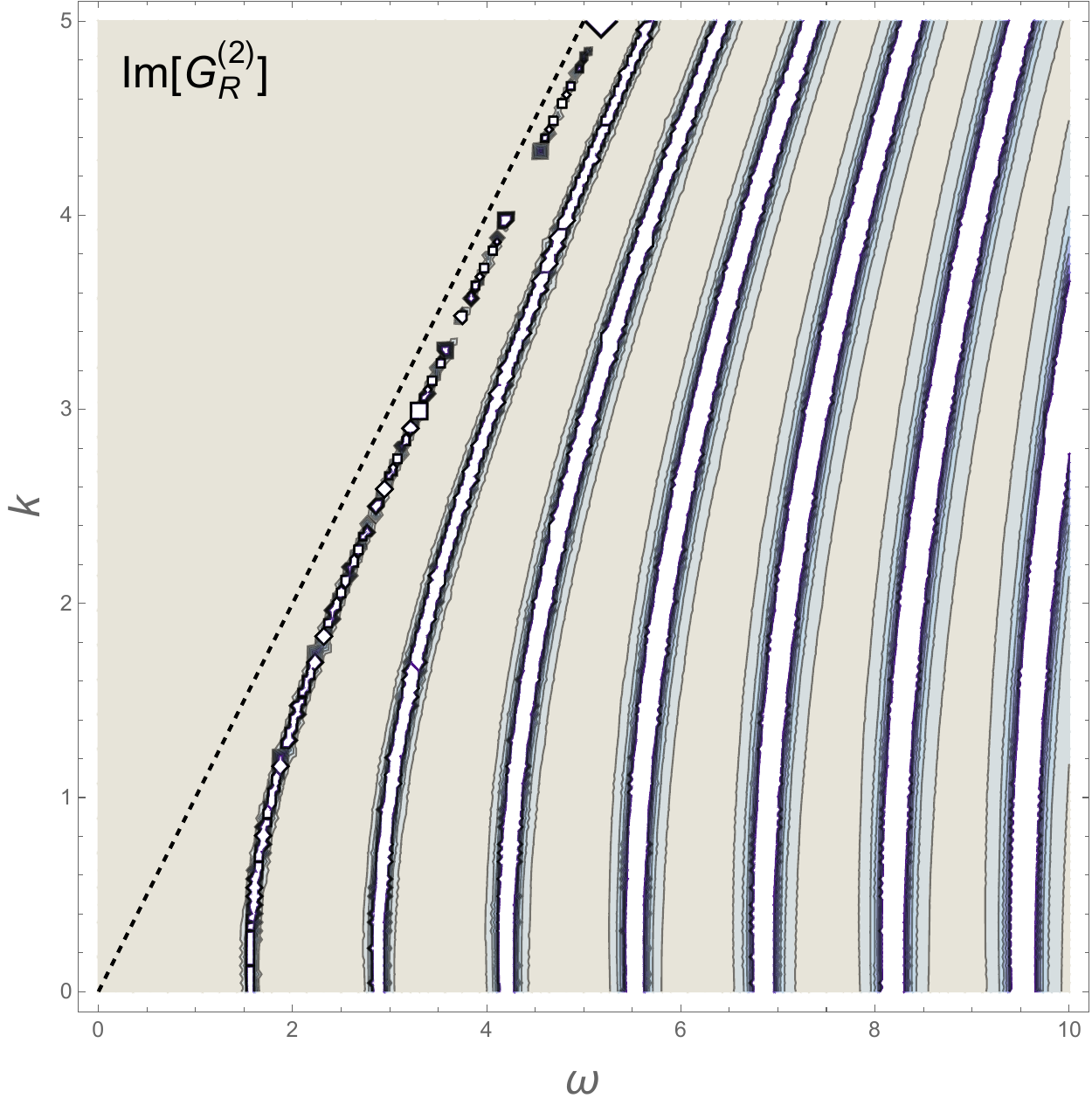}
\par\end{centering}
\caption{\label{fig:9} Density plot of the confined retarded Green function
$G_{R}^{\left(1,2\right)}$ as the spectral function of the 1+2 dimensional
QCD from the D3/D7 model. The parameters are chosen as $M_{KK}=1,\Lambda_{l}=2,L=0$.
The white regions refer to the peaks in the Green function and dashed
lines refer to the light cones.}
\end{figure}
 
\begin{figure}[th]
\begin{centering}
\includegraphics[scale=0.3]{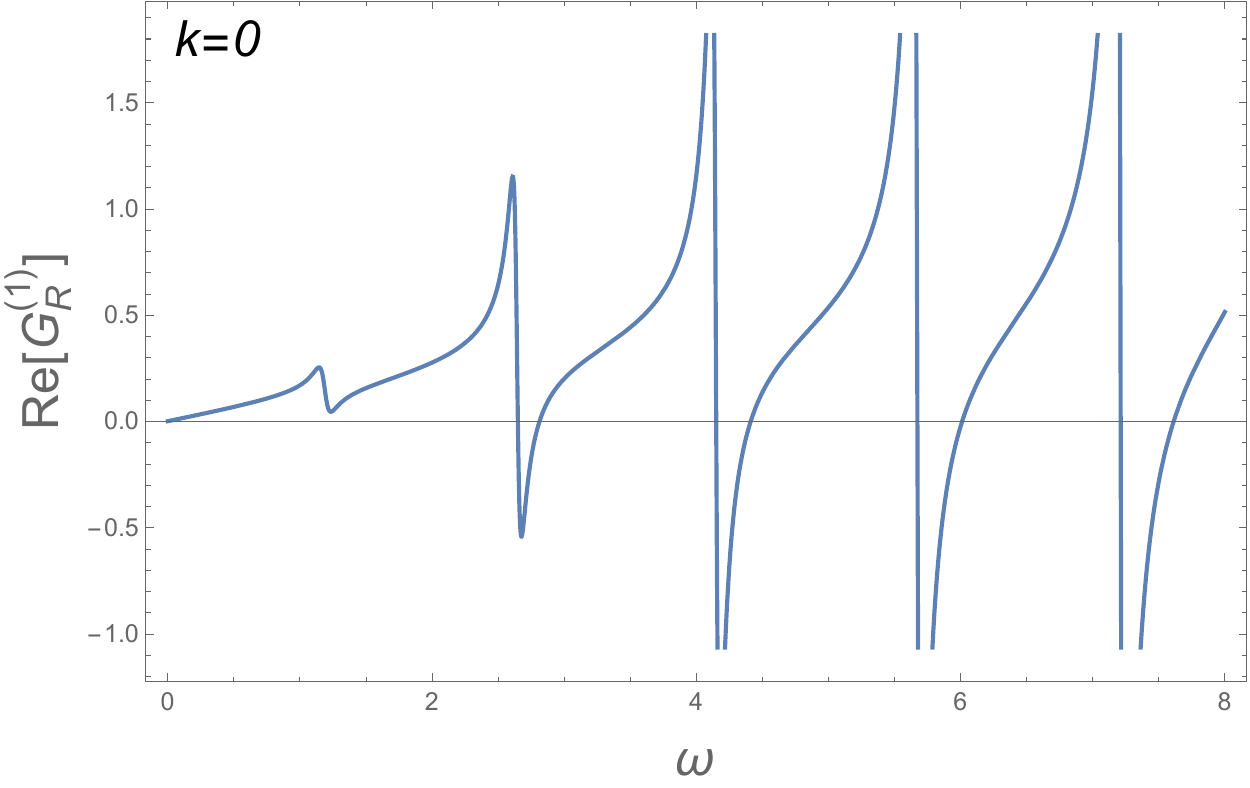}\includegraphics[scale=0.3]{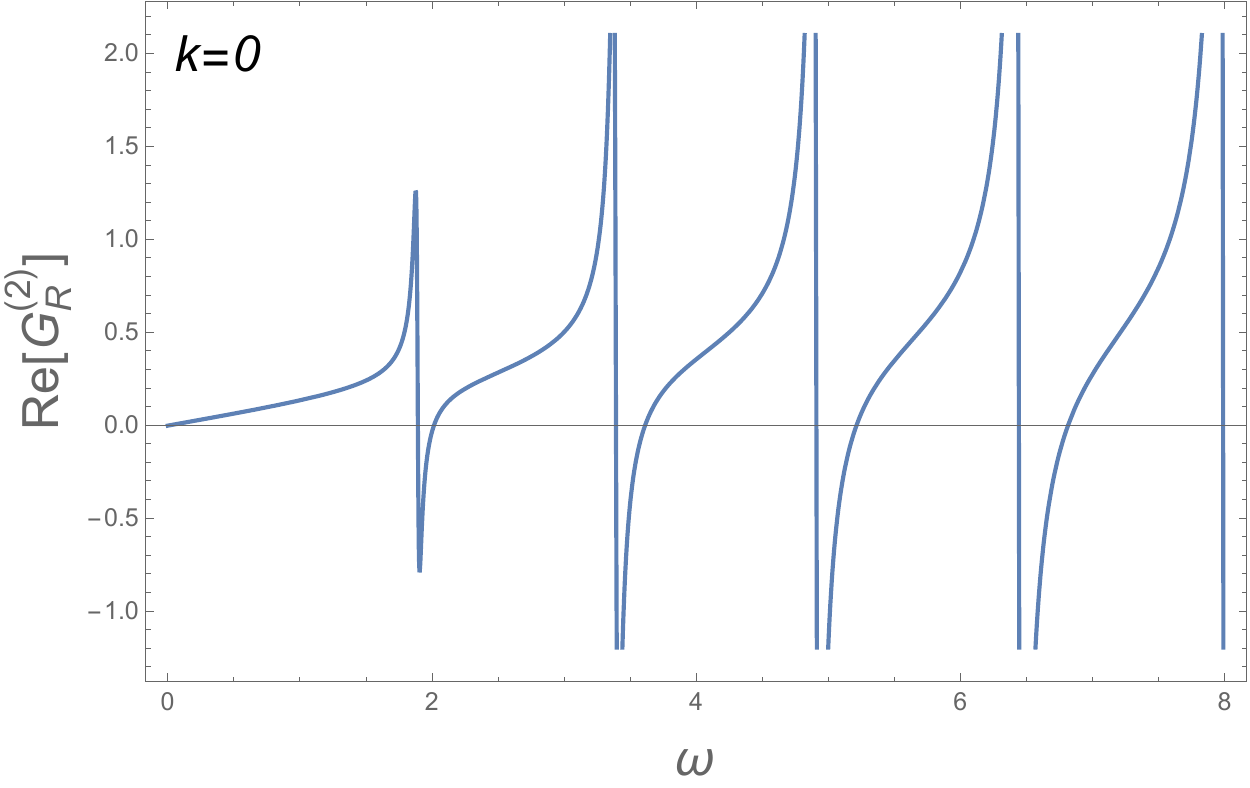}
\par\end{centering}
\begin{centering}
\includegraphics[scale=0.3]{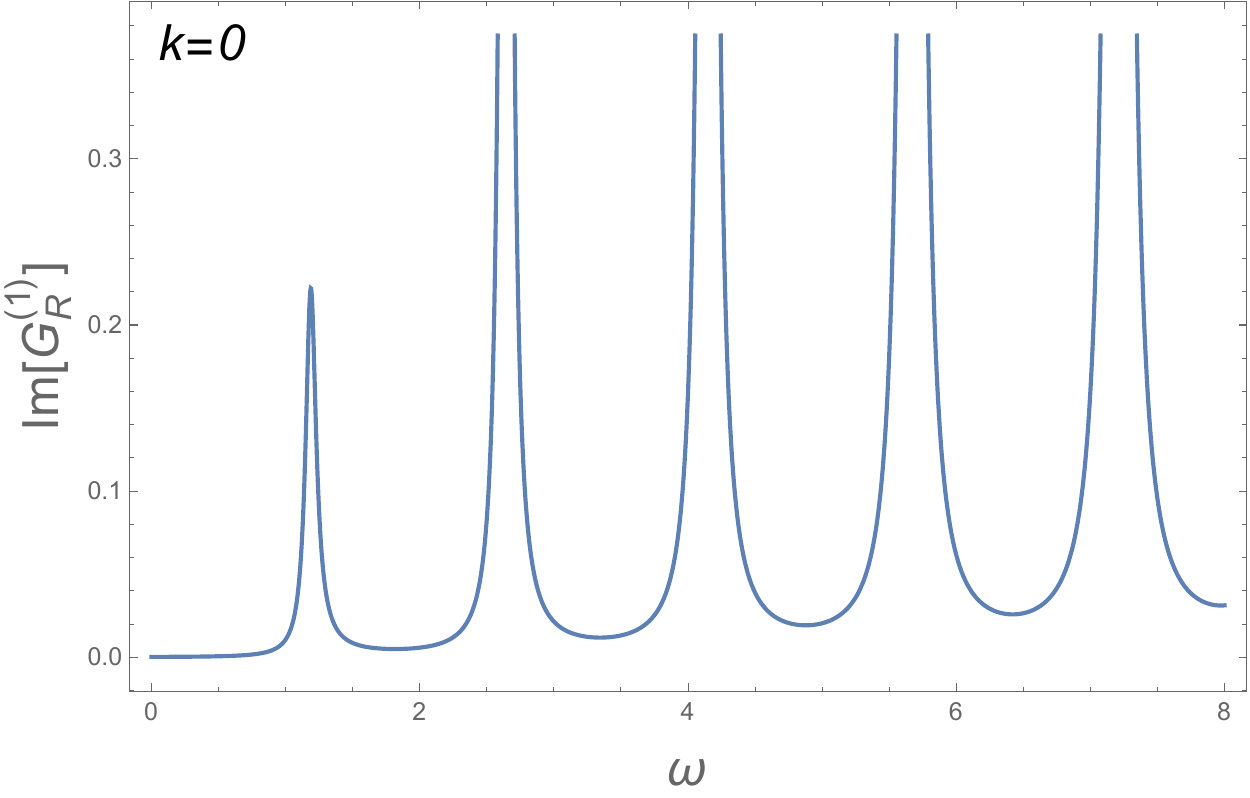}\includegraphics[scale=0.3]{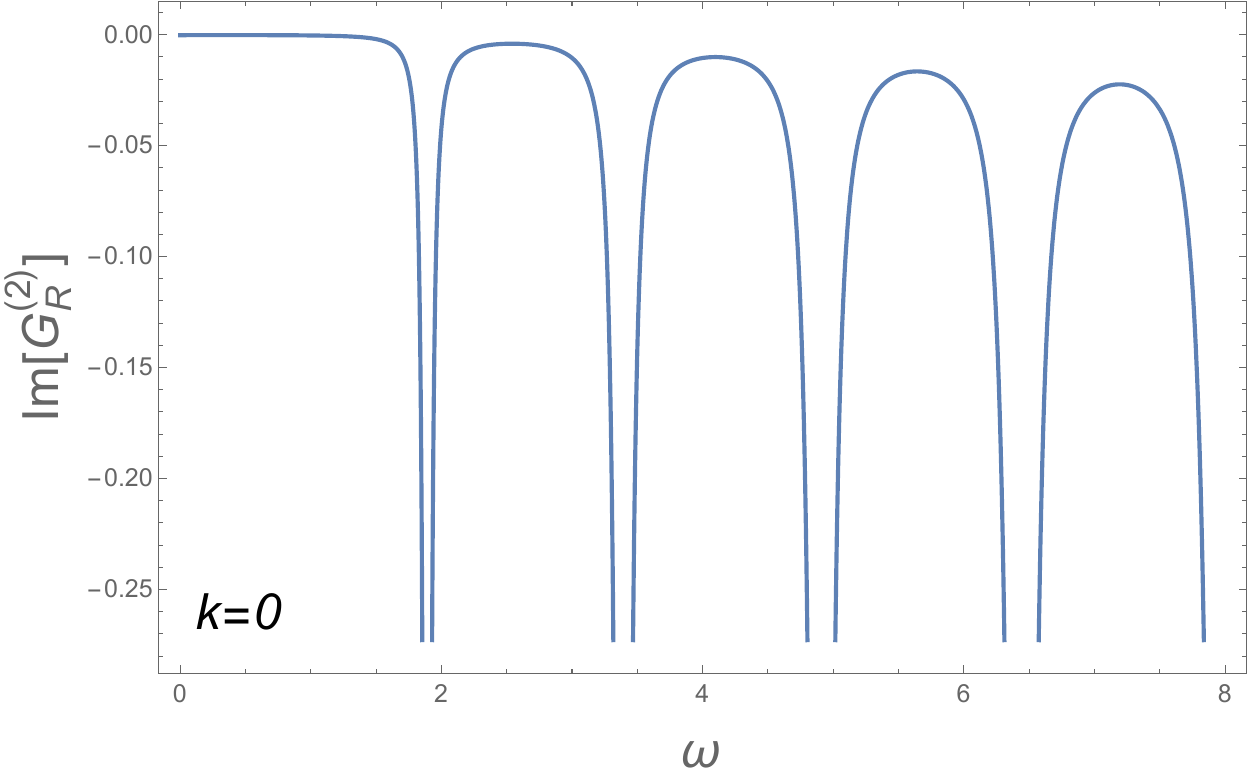}
\par\end{centering}
\caption{\label{fig:10} The confined retarded Green function $G_{R}^{\left(1,2\right)}$
as the spectral function of the 1+2 dimensional QCD from the D3/D7
approach. The parameters are set as $M_{KK}=1,\Lambda_{l}=2,L=0$.}
\end{figure}

\subsubsection*{Deconfined phase}

For the deconfined phase, the equation of motion for the worldvolume
$\psi$ is given by varying action (\ref{eq:86}) which is,

\begin{align}
\bigg[\frac{1}{\sqrt{f_{T}}}\frac{R}{r\zeta}\boldsymbol{\gamma}^{0}\partial_{0}+\frac{R}{r\zeta}\boldsymbol{\gamma}^{\alpha}\partial_{\alpha}+\frac{1}{R}\boldsymbol{\gamma}\partial_{\rho}+\left(\frac{3\rho\sqrt{f_{T}}}{2R\zeta^{2}}+\frac{f_{T}^{\prime}}{4f_{T}R}\right)\boldsymbol{\gamma}\nonumber \\
+\frac{1}{R}\left(\frac{1}{\zeta}-\frac{f_{T}^{\prime}}{4f_{T}}\right)-\frac{1}{R\rho}\left(\Lambda_{l}+2\right)+\frac{\rho}{R\zeta^{2}}\left(\frac{5}{2}-\frac{3}{2}\sqrt{f_{T}}\right)\bigg]\psi & =0.
\end{align}
Using the ansatz (\ref{eq:43}), the coupled equations for $\psi_{L,R}$
are,

\begin{align}
\bigg[\frac{1}{R}\partial_{\rho}+\frac{3\rho\sqrt{f_{T}}}{2R\zeta^{2}}+\frac{f_{T}^{\prime}}{4f_{T}R}+\frac{1}{R}\left(\frac{1}{\zeta}-\frac{f_{T}^{\prime}}{4f_{T}}\right)-\frac{1}{R\rho}\left(\Lambda_{l}+2\right)\nonumber \\
+\frac{\rho}{R\zeta^{2}}\left(\frac{5}{2}-\frac{3}{2}\sqrt{f_{T}}\right)\bigg]\psi_{R}+i\frac{R}{r\zeta}\left(\frac{1}{\sqrt{f_{T}}}\partial_{0}+\sigma^{\alpha}\partial_{\alpha}\right)\psi_{L} & =0,\nonumber \\
\bigg[-\frac{1}{R}\partial_{\rho}-\frac{3\rho\sqrt{f_{T}}}{2R\zeta^{2}}-\frac{f_{T}^{\prime}}{4f_{T}R}+\frac{1}{R}\left(\frac{1}{\zeta}-\frac{f_{T}^{\prime}}{4f_{T}}\right)-\frac{1}{R\rho}\left(\Lambda_{l}+2\right)\nonumber \\
+\frac{\rho}{R\zeta^{2}}\left(\frac{5}{2}-\frac{3}{2}\sqrt{f_{T}}\right)\bigg]\psi_{L}+i\frac{R}{r\zeta}\left(\frac{1}{\sqrt{f_{T}}}\partial_{0}+\bar{\sigma}^{\alpha}\partial_{\alpha}\right)\psi_{R} & =0,\label{eq:99}
\end{align}
which can reduce to two second-order differential equations. At boundary
$r\rightarrow\infty$ ($\rho\rightarrow\infty$), they are,

\begin{align}
\psi_{R}^{\prime\prime}+\frac{5}{\rho}\psi_{R}^{\prime}-\frac{\left(5+2\Lambda_{l}\right)\left(2\Lambda_{l}-3\right)}{4\rho^{2}} & =0,\nonumber \\
\psi_{R}^{\prime\prime}+\frac{5}{\rho}\psi_{R}^{\prime}-\frac{\left(2\Lambda_{l}-5\right)\left(2\Lambda_{l}+3\right)}{4\rho^{2}} & =0,
\end{align}
which can be solved as,

\begin{align}
\psi_{R} & =\rho^{\Lambda_{l}-\frac{3}{2}}A+\rho^{-\Lambda_{l}-\frac{3}{2}}B,\nonumber \\
\psi_{L} & =\rho^{-\Lambda_{l}-\frac{3}{2}}C+\rho^{-\frac{5}{2}+\Lambda_{l}}D.
\end{align}
While the boundary value of $\psi$ remains to be $\psi_{R}$, the
onshell action (\ref{eq:86}) taking the form as,

\begin{align}
S_{f,c}^{\mathrm{D7}} & \supseteq-\mathcal{T}\int d^{3}x\left[\left(\psi_{L}^{\dagger}\psi_{R}\right)|_{\rho\rightarrow\infty}+h.c+...\right]\nonumber \\
 & =-\mathcal{T}\int d^{3}x\left[\rho^{2\Lambda_{l}-4}D^{\dagger}A+h.c\right]|_{\rho\rightarrow\infty}.
\end{align}
does not include any finite part which implies the deconfined phase
of 3d QCD is also non-renormalizable in holography. In order to obtain
a finite Green function, we can follow the same discussion as it is
in the D4/D8 model. Therefore, let us first take the boundary value
$\psi_{0}$ of $\psi$ carefully as,

\begin{equation}
\psi_{0}=\lim_{\rho\rightarrow\epsilon^{-1}}\epsilon^{\frac{1}{2}}\psi=\epsilon^{-\Lambda_{l}+2}\left(\begin{array}{c}
A\\
0
\end{array}\right),\epsilon\rightarrow0.
\end{equation}
Then the conjugate momentum $\Pi_{0}$ for $\psi_{0}$ is given as,

\begin{equation}
\Pi_{0}=-\frac{\delta S_{f,d}^{\mathrm{D8}}}{\delta\psi_{0}}=\mathcal{T}\left(0,D^{\dagger}\right)\epsilon^{-\Lambda_{l}+2}.
\end{equation}
Accordingly, it leads to a finite Green function as,

\begin{equation}
\bar{\Pi}_{0}=G_{R}\left(\omega,\vec{k}\right)\psi_{0},\ G_{R}^{\left(1,2\right)}=\lim_{\epsilon\rightarrow0}\epsilon^{-1}\xi_{1,2},
\end{equation}
where the equation for $\xi_{1,2}$ defined in (\ref{eq:58}) can
be obtained from (\ref{eq:99}), as

\begin{align}
\xi_{1,2}^{\prime}= & \frac{R^{2}}{r\zeta}\left(\frac{\omega}{\sqrt{f_{T}}}-h\cdot\mathrm{k}\right)\xi_{2}+\frac{R^{2}}{r\zeta}\left(\frac{\omega}{\sqrt{f_{T}}}+h\cdot\mathrm{k}\right)\xi_{1,2}^{2}\nonumber \\
 & +2\left[\left(\frac{1}{\zeta}-\frac{f_{T}^{\prime}}{4f_{T}}\right)-\frac{1}{\rho}\left(\Lambda_{l}+2\right)+\frac{\rho}{\zeta^{2}}\left(\frac{5}{2}-\frac{3}{2}\sqrt{f_{T}}\right)\right]\xi_{1,2}.\label{eq:106}
\end{align}
And we will evaluate the Green function with the incoming wave boundary
condition on the horizon and $h=\pm1$ for $\xi_{1,2}$ respectively. 

\subsection{The numerical analysis}

\begin{figure}
\begin{centering}
\includegraphics[scale=0.5]{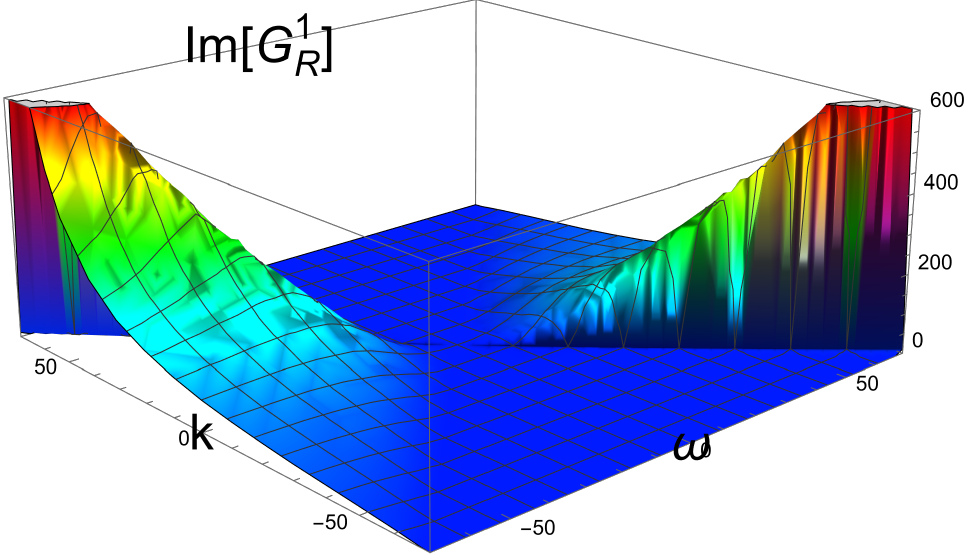}\includegraphics[scale=0.5]{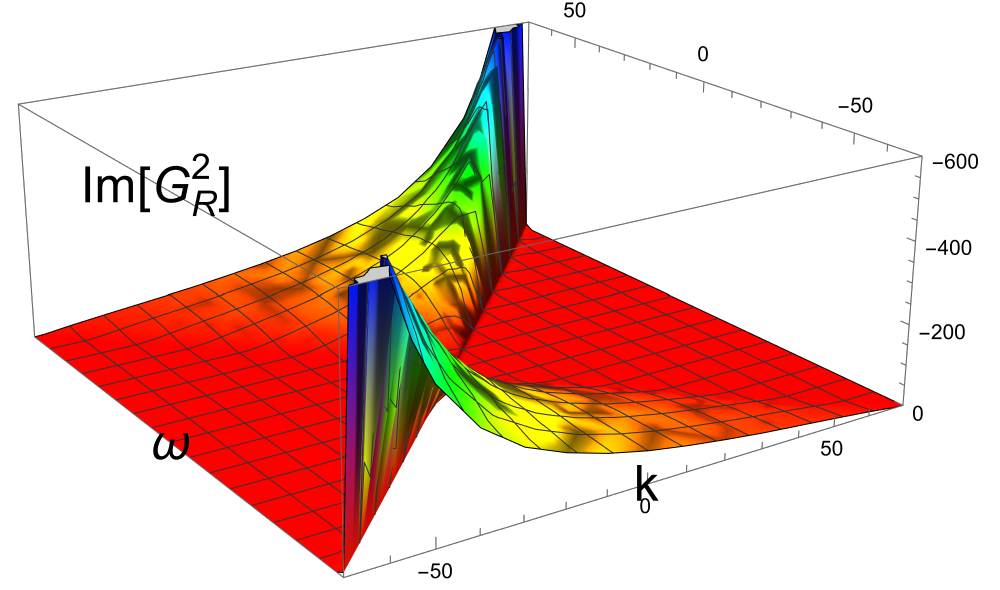}
\par\end{centering}
\caption{\label{fig:12} The 3d plot of the imaginary part of the Green function
$G_{R}^{\left(1,2\right)}$ from the black D3-brane background. The
parameters are chosen as $\Lambda_{l}=2,l=0,\mathcal{T}=1$ in the
unit of $2\pi T=1$.}
\end{figure}

\begin{figure}
\begin{centering}
\includegraphics[scale=0.35]{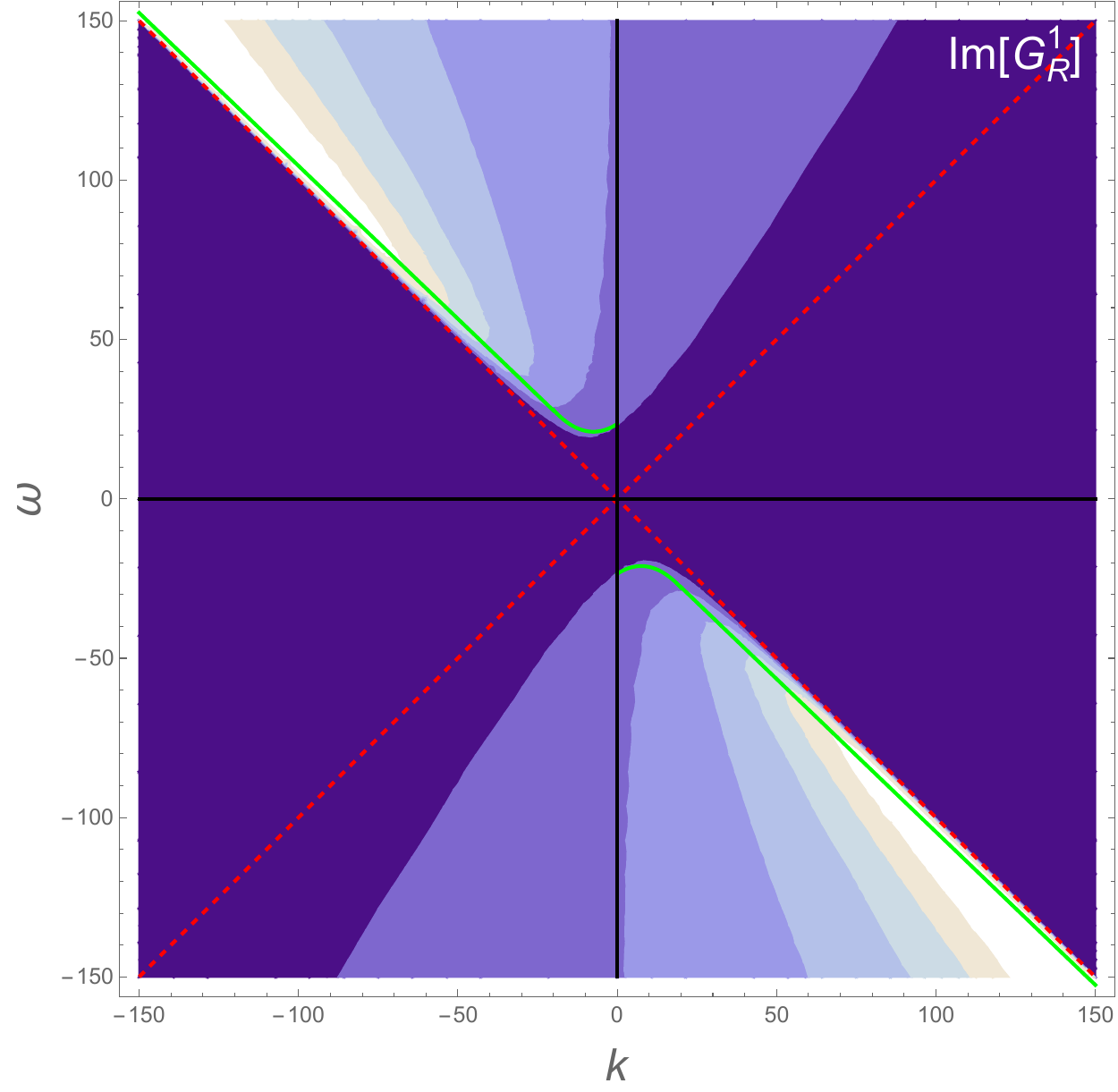}\includegraphics[scale=0.35]{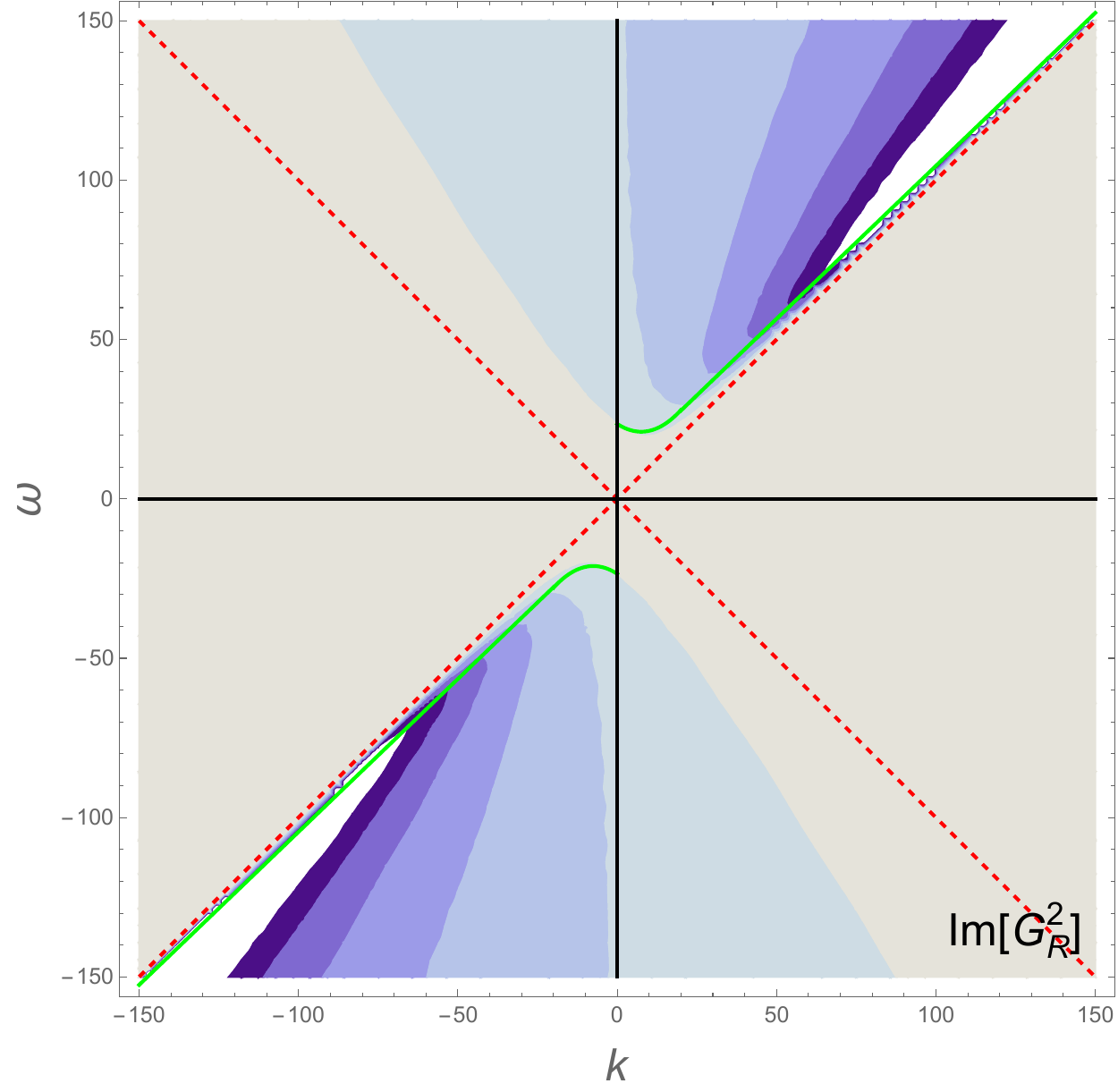}
\par\end{centering}
\caption{\label{fig:13} Density plot of the imaginary part of the Green function
$G_{R}^{\left(1,2\right)}$ from the black D3-brane background. The
white regions represent the protrusion or pit, i.e. the peaks, in
the Green functions which is fitted by green lines. The dashed lines
refer to $\omega=\pm\mathrm{k}$ as the light cone and the parameters
are chosen as $\Lambda_{l}=2,l=0,\mathcal{T}=1$ in the unit of $2\pi T=1$.}

\end{figure}

In this section, let us summarize the numerical results by solving
(\ref{eq:97}) and (\ref{eq:106}) in the D3/D7 approach. The behavior
of two-point Green function in the confined phase is illustrated in
Figure \ref{fig:9} which is basically similar to the approach in
the D4/D8 model. 
\begin{figure}[th]
\begin{centering}
\includegraphics[scale=0.32]{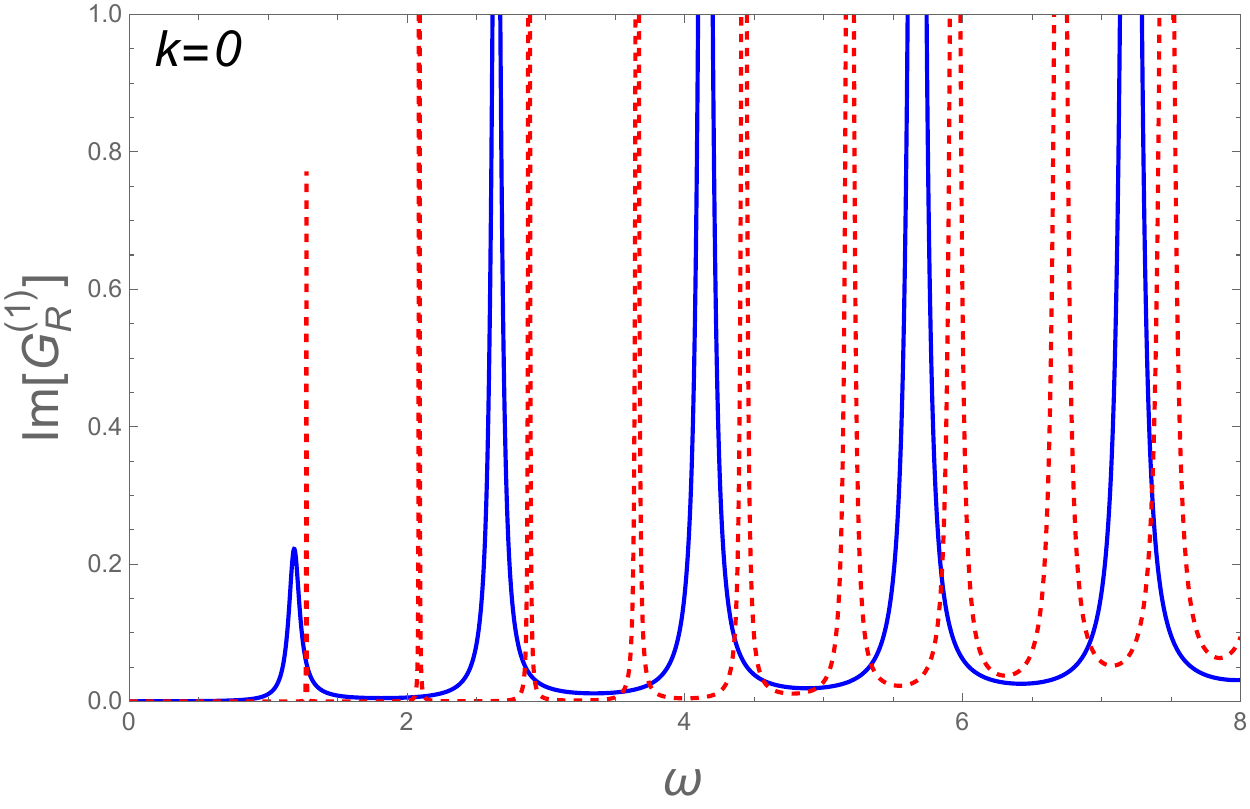}\includegraphics[scale=0.32]{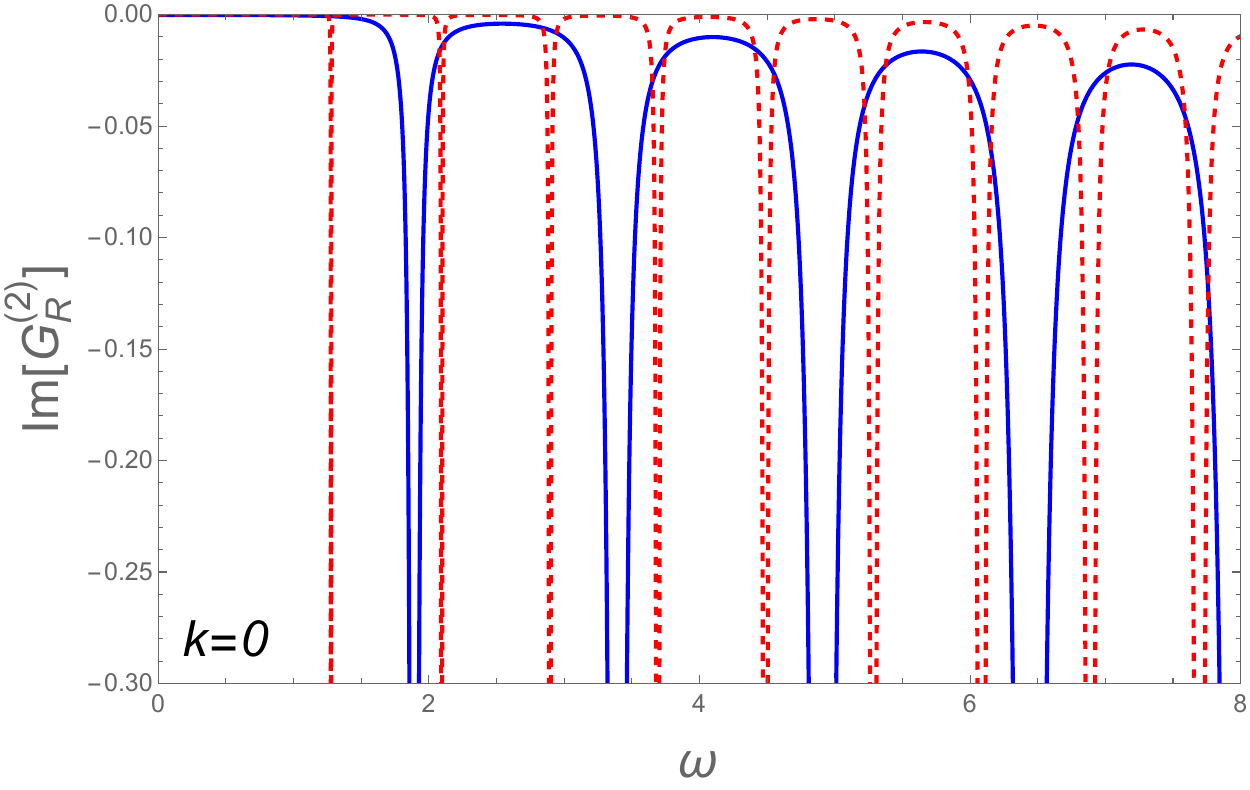}
\par\end{centering}
\caption{\label{fig:11} Imaginary part of the confined retarded Green function
$G_{R}^{\left(1,2\right)}$ with $L=0$ (solid blue line), $L=1$
(dashed red line).}
\end{figure}
 The peaks in the Green function display the discreteness representing
the onshell bound states as confinement in 3d QCD, as $\omega\simeq1.6,2.6,4.1...$
in $G_{R}^{\left(1\right)}$ and $\omega\simeq1.9,3.4,4.8...$ in
$G_{R}^{\left(2\right)}$ which is only quantitatively different from
the results in the D4/D8 model. To specify the position of the onshell
energy, we have plotted out the relation of the Green function $G_{R}$
as a function of $\omega$ which is illustrated in Figure \ref{fig:10}.
Note that, while we have set $L=0$ for the chirally symmetric theory,
the position of the peaks in the Green function depends on the quark
mass $L$ as it is illustrated in Figure \ref{fig:11}. The dependence
on $L$ in Green function is different from the D4/D8 approach, since
there is not a direction perpendicular to both D4- and D8-branes in
the D4/D8 model (which means the VEV of a $\left(4,8\right)$ string
is vanished i.e. the bare mass of fundamental quark is vanished.).
Furthermore, our numerical calculation reveals that the onshell mass
of the bound fermion is suppressed by the quark bare mass due to the
correction of the self-energy $\Sigma\left(k\right)$ in (\ref{eq:73}).
In the deconfined phase, we plot out the imaginary part of the Green
function as the spectral function in Figure \ref{fig:12} and \ref{fig:13}
in which the onshell condition fitted by green lines is in qualitative
agreement with the dispersion curves obtained by the hard thermal
loop approximation presented in (\ref{eq:74}). The effective mass
of fermion in hot medium is evaluated as $m_{f}\simeq4.8\left(2\pi T\right)$
by the D3/D7 model which also implies the high order contribution
in HTL approximation might be suppressed in the large $N_{c}$ limit.
 Altogether, the D3/D7 approach also reveals mostly the fundamental
properties of QCD as it is expected.

\section{Summary and discussion}

In this work, we study the correlation function of the flavored fermion
in the $\mathrm{D}_{p}/\mathrm{D}_{p+4}$ model ($p=4,3$) as the
holographic top-down approaches to QCD. Since our concern is the fermionic
correlator, the bulk spinor is identified to the worldvolume fermion
created by the $\left(p+4,p+4\right)$ string on the $\mathrm{D}_{p+4}$-brane.
Then we pick up the action for the worldvolume fermion on the D-brane
which is obtained by T-duality in string theory, and generalize the
prescription for two-point correlation function in AdS/CFT dictionary
into general D-brane background with respect to the flavored fermion
in the dual theory. Afterwards we apply our viewpoint to the case
of $p=4,3$ respectively. The numerical calculation and result show
that the Green functions in both D4/D8 and D3/D7 approaches reveal
the discrete peaks with the bubble background which represents the
bound states as confinement in QCD. In particular, the various bound
energies agree basically with the numerical evaluation in \cite{key-36}
for $p=4$ quantitatively. With the black brane background, the onshell
condition given by the Green function agrees qualitatively with the
dispersion curves obtained by the hard thermal loop approximation
which displays the behavior of the deconfined fermion in hot medium.
Overall our approaches to the D4/D8 and D3/D7 model illustrate the
fundamental properties of QCD in holography which would therefore
be remarkable.

Besides, let us give some comments about this work. First, although
the onshell condition in the deconfined phase is in qualitative agreement
with several bottom-up approaches involving minimal coupled fermion
e.g. \cite{key-30,key-31}, the behavior of Green function in confined
phase is totally different. The reasons are mostly that, as the dependence
of the inner sphere $S^{4}$ is neglected and the probe brane is absent
in the bottom-up approaches, the action for the bulk fermion would
take a different form so that the asymptotic behavior of fermion,
related to the Green function, is also different. And notice that
the action (\ref{eq:11}) for worldvolume fermion on a D-brane is
not minimal coupled in general, thus it is not surprised that the
results in our top-down approach could be different from those in
the bottom-up approaches. However, the property of confinement in
QCD is less clear in the previous works with bottom-up models or with
minimal coupled fermion.

Second, in the deconfined phase, while our analysis is consistent
with the hard thermal loop approximation, we must keep in mind that
the QFT approach is only valid in weak coupled field theory. Therefore
our current result in the deconfined phase (which is valid in the
strong coupling region) may imply the high order contribution from
the hard thermal loop approximation at large $N_{c}$ limit is suppressed.
On the other hand, the hard thermal loop approximation implies the
chemical potential may also contribute to the effective thermal mass
of fermion while it is not included in this work. However, if the
chemical potential (relating to the gauge field potential on the worldvolume
of the $\mathrm{D}_{p+4}$-brane) is taken into account in this setup,
its equation of motion coupling to bulk fermion would be totally different
from the bottom-up approaches with minimal coupled fermion according
to action (\ref{eq:11}) - (\ref{eq:13}), because the bulk fermion
is identified to be the fundamental representation of a $U\left(N\right)$
group in most bottom-up approaches while it is not in our top-down
approach. Accordingly, we believe the Green function with non-vanished
chemical potential in our top-down approach will display a very different
behavior from it in the bottom-up approaches and we will leave this
for the future work.

\begin{figure}
\begin{centering}
\includegraphics[scale=0.35]{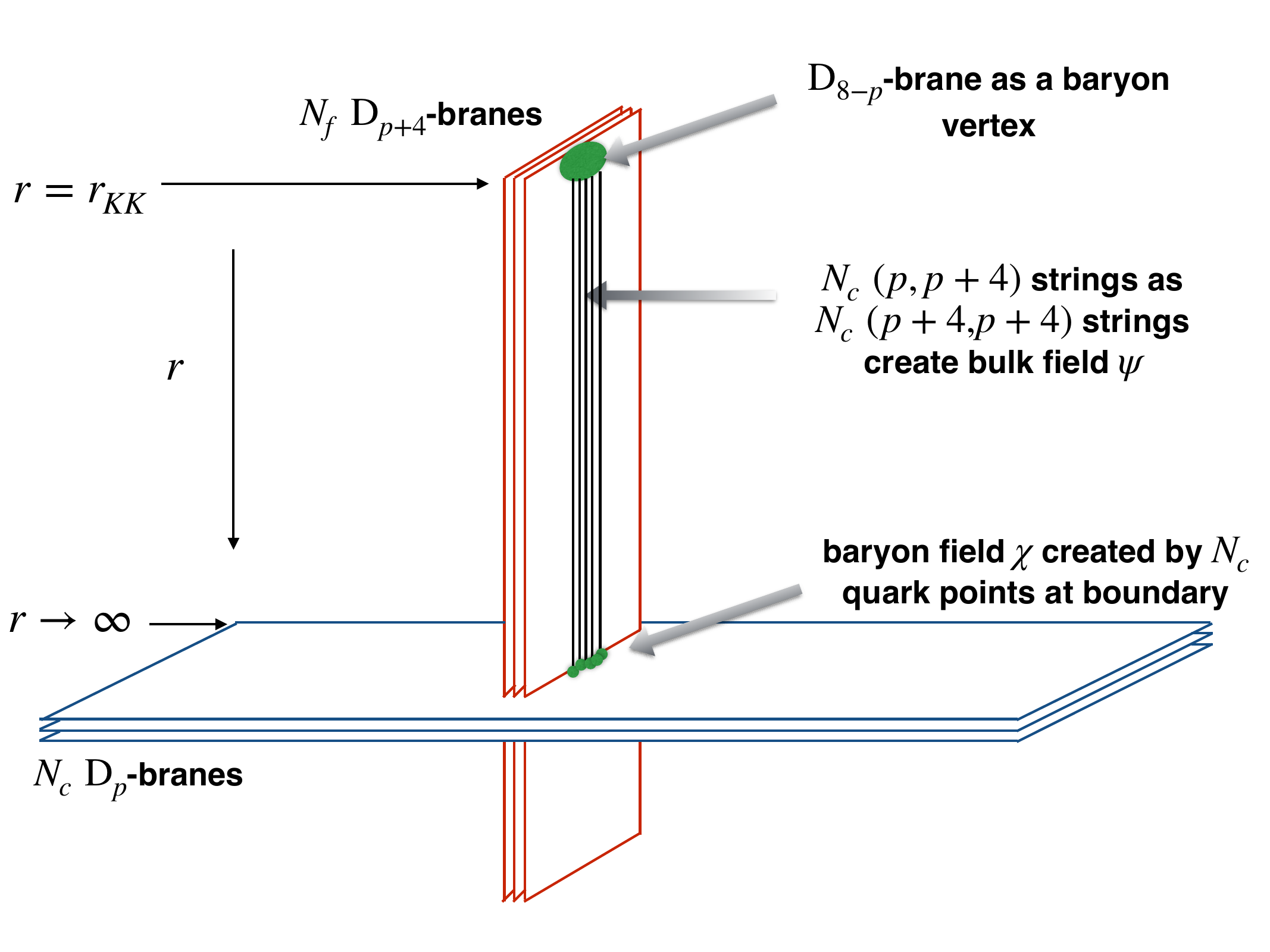}
\par\end{centering}
\caption{\textcolor{blue}{\label{fig:14} }The D-brane configuration of $\mathrm{D}_{p}/\mathrm{D}_{p+4}$
system with a baryon vertex. The bulk field $\psi$ is created by
$N_{c}$ $\left(p,p+4\right)$ strings as quarks and the dual field
$\chi$ at boundary is produced by $N_{c}$ quark points as a baryon
field.}

\end{figure}

Last but not least, let us discuss the interpretation of the dual
operator $\chi$ to the bulk fermion in terms of hadron physics and
outline how to interpret it as a baryon field in the confined phase.
As we have specified that to interpret the dual field $\chi$ as baryon
is more close to the realistic QCD, so it is necessary to let $\chi$
produced by multiple fundamental quarks takes baryon number in the
side of string theory. Fortunately, this can be achieved by following
Witten's \cite{key-45}, that is to introduce a $\mathrm{D}_{8-p}$-brane
wrapped on $S^{8-p}$ \cite{key-45,key-46,key-47} as a baryon vertex
located at $r=r_{KK}$ in the bubble background. Since the $N_{c}$
fundamental quarks created by the $\left(p,p+4\right)$ string must
always take baryon numbers, there must be $N_{c}$ $\left(p,p+4\right)$
strings as fundamental quarks connect the wrapped $\mathrm{D}_{8-p}$-brane
and stretch to the bulk boundary totally inside the $\mathrm{D}_{p+4}$-branes
thus they are flavored, colored, baryonic strings and are also $\left(p+4,p+4\right)$
strings. By taking into account a probe $\mathrm{D}_{p}$-brane at
the holographic boundary $r\rightarrow\infty$, the D-brane configuration
with a baryon vertex is illustrated in Figure \ref{fig:14}. 
In this configuration, the $\left(p,p+4\right)$ strings as fundamental
quarks are colored, flavored and take baryon numbers, hence the dual
field $\chi$ produced by multiple $N_{c}$ $\left(p,p+4\right)$
string points is a baryonic hadron field i.e. a baryon field. On the
other hand, since the $\left(p,p+4\right)$ strings are also $\left(p+4,p+4\right)$
strings, the bulk field $\psi$ dual to $\chi$ created by the multiple
quarks is also baryonic. So it provides a nicely holographic correspondence
with respect to baryon field. In this sense, the bound states as the
various peaks presented in the confined Green function refer to the
baryon spectrum in holography. And it could be possible to fit the
experimental data if we further identify the quantum number $l,s$
in (\ref{eq:37}) to the isospin and spin of the baryon. In this sense,
we can identify, for examples, the first three lowest states with
same parity presented in Figure \ref{fig:6} as proton, $N\left(1440\right)$
and $N\left(1710\right)$ for $l=1$ as \cite{key-48}. Thus our approach
with D4/D8 model gives the mass ratios of $N\left(1440\right)$ and
proton, $N\left(1710\right)$ and proton as $M_{N\left(1440\right)}/M_{proton}\simeq1.51,M_{N\left(1710\right)}/M_{proton}\simeq1.95,$
which are very close to the experimental data $M_{N\left(1440\right)}^{\mathrm{exp}}/M_{proton}^{\mathrm{exp}}\simeq1.53,M_{N\left(1440\right)}^{\mathrm{exp}}/M_{proton}^{\mathrm{exp}}\simeq1.82$
\cite{key-49}. And this viewpoint may also support that the open
strings on the D-brane behaves somehow as baryons in the $\mathrm{D}_{p}/\mathrm{D}_{p+4}$
model as it is discussed in \cite{key-37}. We further comment at
last, to interpret $\chi$ as quark field, i.e. produced by a single
$\left(p,p+4\right)$ string, may be possible in the deconfined background
since in hadron physics a free quark as a free color charge could
be observables theoretically in the deconfinement phase of QCD and
baryon will be dissolved in the deconfinement phase.

\section*{Acknowledgements}

We would like to thank Konstantinos Rigatos, Xin-li Sheng and Yan-qing
Zhao for helpful discussion. This work is supported by the National
Natural Science Foundation of China (NSFC) under Grant No. 12005033
and the Fundamental Research Funds for the Central Universities under
Grant No. 3132023198 and Grant No. 3132024198.

\end{document}